\documentclass[journal]{IEEEtran}
\usepackage{cite}
\usepackage{amsmath,amssymb,amsfonts}
\usepackage{algorithm}
\usepackage{graphicx}
\usepackage{textcomp}
\usepackage{xcolor}
\usepackage[hidelinks,bookmarks=false]{hyperref}
\usepackage[caption=false,font=footnotesize]{subfig}
\usepackage{comment}
\usepackage{tikz}
\usepackage{varwidth}
\usepackage{enumitem}
\usepackage{algpseudocode}
\usepackage{float}
\usepackage{threeparttable}
\usepackage[export]{adjustbox}
\usepackage{makecell}
\usepackage{cleveref}
\usepackage{dblfloatfix}
\usepackage{balance}

\usepackage{pifont}

\newcommand{\cmark}{\ding{51}} 
\newcommand{\xmark}{\ding{55}} 
\newcommand{\pmark}{\(\sim\)} 
\newcommand{\no}{\textcolor{gray}{\tiny$\cdot$}}

\usetikzlibrary{arrows.meta, positioning, shapes.geometric, calc}

\def\BibTeX{{\rm B\kern-.05em{\sc i\kern-.025em b}\kern-.08em T\kern-.1667em\lower.7ex\hbox{E}\kern-.125emX}}

\usepackage{booktabs}
\usepackage{tabularx}
\usepackage{array}
\newcolumntype{Y}{>{\raggedright\arraybackslash}X}
\newcolumntype{L}[1]{>{\raggedright\arraybackslash}p{#1}}
\newcolumntype{C}[1]{>{\centering\arraybackslash}p{#1}}
\usepackage{rotating}
\usepackage{booktabs}
\usepackage{multirow}
\usepackage{threeparttable}
\usepackage{pifont}
\usepackage{longtable}
\usepackage{threeparttablex}
\usepackage{soul} 
\usepackage[normalem]{ulem}
\usepackage{tikz}
\usetikzlibrary{arrows.meta, positioning}

\usepackage{tikz}
\usetikzlibrary{arrows.meta,positioning,decorations.pathreplacing}

\def\false{false}

\def\finalversion{false} 
\def\finalcolor{false} 
\definecolor{darkgreen}{RGB}{0,200,0}

\ifx\finalcolor\false
    \def\hatemcolor{blue}
    \def\carloscolor{olive}
    \def\alonsocolor{darkgreen}
    \def\juanRcolor{yellow}
\else
    
    \def\carloscolor{black}
    \def\alonsocolor{black}
    \def\juanRcolor{black}
\fi

\newcommand{\TODO}[1]{\ifx\finalversion\false{\textcolor{red}{(TODO: #1)}}\fi}

\newcommand{\defineauthorcomments}[2]{%
    \ifx\finalversion\false
        \expandafter\newcommand\csname #1\endcsname[1]{%
            {\bf \textcolor{#2}{##1}}%
        }
        \expandafter\newcommand\csname #1H\endcsname[1]{%
            \hl{##1}%
        }
        \expandafter\newcommand\csname #1C\endcsname[2]{%
            \csname #1H\endcsname{##1} %
            \csname #1\endcsname{##2}%
        }
        \expandafter\newcommand\csname #1CC\endcsname[2]{%
            \csname #1H\endcsname{##1} %
            \csname #1\endcsname{##2}%
        }
        \expandafter\newcommand\csname #1CO\endcsname[2]{%
            {##1} {\bf \textcolor{#2}{+#1: ##2+}}%
        }
        \expandafter\newcommand\csname #1CM\endcsname[3]{%
            \csname #1\endcsname{##3} %
            \csname #1CO\endcsname{##1}{##2}%
        }
        \expandafter\newcommand\csname #1ADD\endcsname[1]{%
            \csname #1\endcsname{##1}%
        }
        \expandafter\newcommand\csname #1RM\endcsname[1]{%
            \textcolor{gray}{\sout{##1}}%
        }
        \expandafter\newcommand\csname #1RP\endcsname[2]{%
            \csname #1\endcsname{##2} %
            \csname #1RM\endcsname{##1}%
        }
        \expandafter\newcommand\csname #1RPD\endcsname[2]{%
            ##1 %
            \csname #1\endcsname{##1 $\rightarrow$ ##2?}%
        }
        \expandafter\newcommand\csname #1MOD\endcsname[2]{%
            \csname #1RM\endcsname{##1}%
            \csname #1ADD\endcsname{##2}%
        }
    \else
        \expandafter\newcommand\csname #1\endcsname[1]{%
            ##1%
        }
        \expandafter\newcommand\csname #1H\endcsname[1]{%
            ##1%
        }
        \expandafter\newcommand\csname #1C\endcsname[2]{%
            ##1%
        }
        \expandafter\newcommand\csname #1CC\endcsname[2]{%
            ##1%
        }
        \expandafter\newcommand\csname #1CO\endcsname[2]{}
        \expandafter\newcommand\csname #1CM\endcsname[3]{%
            \csname #1ADD\endcsname{##3}%
        }
        \expandafter\newcommand\csname #1ADD\endcsname[1]{%
            \textcolor{#2}{##1}%
        }
        \expandafter\newcommand\csname #1RM\endcsname[1]{}
        \expandafter\newcommand\csname #1RP\endcsname[2]{%
            \csname #1ADD\endcsname{##2}%
        }
        \expandafter\newcommand\csname #1RPD\endcsname[2]{%
            ##1%
        }
        \expandafter\newcommand\csname #1MOD\endcsname[2]{%
            \csname #1ADD\endcsname{##2}%
        }
    \fi
}

\defineauthorcomments{AP}{\alonsocolor}
\defineauthorcomments{CA}{\carloscolor}
\defineauthorcomments{JR}{\juanRcolor}
\defineauthorcomments{HA}{\hatemcolor}

\newif\ifshowtodos
\showtodostrue

\definecolor{TaxBlue}{HTML}{2F6BFF}
\definecolor{TaxTeal}{HTML}{008C8C}
\definecolor{TaxGreen}{HTML}{2E8B57}
\definecolor{TaxPurple}{HTML}{7E57C2}
\definecolor{TaxOrange}{HTML}{D9822B}
\definecolor{TaxRed}{HTML}{B54747}

\begin{document}

\title{Wireless Foundation Models: \\
State-of-the-Art and Open Challenges}


\author{%
Alonso M. Pacheco Huachaca, Juan J. Rodriguez Rodriguez, Ahmed Aboulfotouh, \\ Nelson~L.~S.~da~Fonseca, Carlos A. Astudillo, and Hatem Abou-Zeid
\thanks{%
This work was partially funded by the Government of Canada through the Emerging Leaders in the Americas Program (ELAP); the INCT of Intelligent Communications Networks and the Internet of Things (ICoNIoT) via CNPq (405940/2022-0) and CAPES (88887.954253/2024-00).}
\thanks{%
A. Aboulfotouh and H. Abou-Zeid is with the Department of Electrical and Software Engineering, University of Calgary, Calgary, AB, Canada.\\
A.M.~Pacheco, J.J.~Rodriguez, N.L.S.~da~Fonseca, and C.A.~Astudillo are with the Institute of Computing, University of Campinas, SP, Brazil. 
}
}

\maketitle

\begin{abstract}

Wireless foundation models (WFMs) have emerged as a promising approach for learning reusable representations from large-scale wireless data and adapting them to downstream tasks. However, the rapidly growing literature remains fragmented across modalities, pretraining objectives, architectures, adaptation strategies, and evaluation protocols, making it difficult to assess progress toward broadly transferable models. This survey provides a systematic analysis of WFMs for physical-layer applications. We first introduce the main WFM design components, including pretraining, backbone architectures, and downstream adaptation. We then organize the literature into five physical-layer task families: signal recognition and demodulation, channel representation learning, RF sensing and localization, beam management, and spectrum sensing and monitoring, while separately examining multi-task PHY models. Across these categories, we analyze how existing models are pretrained, adapted, and evaluated, with particular attention to downstream task diversity and the distinction between in-distribution, partial-shift, and out-of-distribution transfer. Our analysis shows that current WFMs provide increasing evidence of reusable wireless representations, but this evidence varies considerably across task families and evaluation settings. Differences in datasets, modalities, architectures, pretraining objectives, adaptation protocols, and distribution shifts make it difficult to determine which design choices drive transfer and generalization. We conclude by identifying open directions for improving data availability, evaluation rigor, generalization, efficient adaptation, and real-world deployment, providing a unified framework for understanding the current WFM landscape and the requirements for developing more reusable foundation models for future physical-layer wireless systems.

\begin{IEEEkeywords}
Foundation models, wireless foundation models, deep learning, self-supervised learning, pretraining, transfer learning, network architecture, physical layer,  wireless communications, survey.
\end{IEEEkeywords}

\end{abstract}

\section{Introduction}
\label{sec:introduction}

The evolution toward sixth-generation (6G) wireless networks introduces new challenges related to massive connectivity, ultra-low latency, efficient spectrum use, sensing capabilities, and adaptation to diverse environments. These requirements are driving a new generation of communication systems expected to be more intelligent, efficient, adaptive, and reliable~\cite{dang2020whatshould6g, wang2023road6gvisionsrequirements}. The inherent complexity of 6G environments requires artificial intelligence (AI) to move beyond its traditional role as an optimization tool, as commonly seen in 5G, toward becoming a native component of wireless system design~\cite{yu2022roleofdeeplearning}.

Deep learning has achieved strong results in many wireless tasks, especially at the physical layer, including signal understanding, channel modeling, sensing and localization, beam management, and spectrum monitoring~\cite{zhao2024deeplearningphy,huang2020deepelerningphy5g}. However, many existing models are still designed for specific tasks and specific settings. They are often trained on fixed datasets, channel conditions, devices, or environments, which limits their ability to work well under the diverse conditions expected in 6G networks~\cite{doha2025deeplearningwireless,fontaine2024towardswireless}. This limitation becomes more important because wireless data can change significantly across frequencies, devices, antennas, mobility patterns, interference conditions, and physical environments~\cite{zhu2025wirelesslargeaimodel}.



Wireless Foundation models (WFMs) have emerged as a promising paradigm to address these limitations. Instead of training a separate model for each task or scenario, WFMs aim to learn reusable wireless representations that can be adapted to different tasks and deployment conditions. This paradigm is especially attractive for wireless systems because large amounts of wireless data can be collected or simulated, while labels are often expensive, limited, or tied to particular scenarios~\cite{aboulfotouh20256gwavesfm,chen2024bigaimodels6g}. By reusing pretrained representations, WFMs can reduce the dependence on task-specific pipelines, lower the amount of labeled data needed for adaptation, and shorten the time required to adjust models to changing wireless conditions. This is particularly relevant for 6G networks, where devices, channels, interference, mobility, and deployment environments are expected to vary continuously.

Despite this growing interest, the current WFM literature remains dispersed across different wireless data types, learning objectives, architectures, tasks, and adaptation strategies. This diversity highlights the potential of WFMs, but also makes it necessary to understand the motivations behind these models, the strengths of different design choices, and the main patterns emerging across the field. In particular, it is important to clarify which physical-layer tasks are being addressed, what types of wireless data are being used, how models are pretrained and adapted, and what evidence is provided for transfer across tasks, datasets, and deployment conditions. Moreover, current surveys often focus on general AI models, large language models, or broad wireless intelligence, while the design and evaluation of WFMs for physical-layer wireless tasks remain less systematically organized~\cite{xu2024largemultimodalmodelslmms}.

To address this gap, this survey studies WFMs from a model-design and evaluation perspective. We focus on physical-layer tasks and organize the literature into five major families: signal recognition and demodulation, channel representation learning, RF sensing and localization, beam management, and spectrum sensing and monitoring. Across these task families, we examine how WFMs are built, what type of wireless information they learn, what datasets are used for pretraining and downstream evaluation, how they are adapted to downstream tasks, and how their transfer is evaluated. This organization allows us to identify recurring trends, limitations, and open directions for developing more reusable, efficient, and wireless-specific foundation models.



\begin{table*}[!t]
\centering
\footnotesize
\setlength{\tabcolsep}{3pt}
\renewcommand{\arraystretch}{0.95}
\caption{List of abbreviations and shorthand terms used in the survey.}
\label{tab:abbreviations}

\begin{tabular}{@{}p{1.20cm}p{4.30cm}p{1.20cm}p{4.30cm}p{1.20cm}p{4.30cm}@{}}
\toprule
\textbf{Abbrev.} & \textbf{Meaning}
& \textbf{Abbrev.} & \textbf{Meaning}
& \textbf{Abbrev.} & \textbf{Meaning} \\
\midrule

AMC & \makecell[l]{Automatic modulation\\classification}
& CER & \makecell[l]{Channel estimation\\and recovery}
& CIR & \makecell[l]{Channel impulse\\response} \\

CSI & \makecell[l]{Channel state\\information}
& EnvR & \makecell[l]{Environment\\reconstruction}
& FDD & \makecell[l]{Frequency-division\\duplexing} \\

FMCW & \makecell[l]{Frequency-modulated\\continuous wave}
& I/Q & \makecell[l]{In-phase/quadrature\\signal}
& ISAC & \makecell[l]{Integrated sensing and\\communications} \\

\makecell[l]{LoS/\\NLoS} & \makecell[l]{Line-of-sight /\\non-line-of-sight}
& LSI & \makecell[l]{Localization and\\spatial inference}
& MC & Modulation classification \\

MIMO & \makecell[l]{Multiple-input\\multiple-output}
& OFDM & \makecell[l]{Orthogonal frequency-\\division multiplexing}
& OOD & Out-of-distribution setting \\

PEFT & Parameter-efficient fine-tuning
& Rx & Receiver / received signal
& SFM & \makecell[l]{Spectrum forecasting\\and monitoring} \\

SSL & Self-supervised learning
& TDD & Time-division duplexing
& Tx & Transmitter / transmitted signal \\

WFM & Wireless foundation model
& WTR & Wireless technology recognition
& Act & Activity recognition \\

AD & Anomaly detection
& AoA & Angle-of-arrival
& BM & Beam management \\

BS & Base station
& CDiag & Channel diagnostics
& CE & Channel estimation \\

CP & Channel prediction
& CR & Channel reconstruction
& Dem & Demodulation \\

DL & Downlink
& FB & CSI feedback
& FM & Foundation model \\

FT & Fine-tuning
& ID & In-distribution setting
& LoRA & Low-rank adaptation \\

LP & Linear probing
& MoE & Mixture-of-experts
& MT-FT & Multi-task fine-tuning \\

MU & Multi-user
& PHY & Physical layer
& Prc & Precoding \\

PSD & Power spectral density
& PT & Pretraining
& RF & Radio frequency \\

RFF & RF fingerprinting
& SID & Signal identification
& SNR & Signal-to-noise ratio \\

SS & Spectrum sensing
& SSM & State-space model
& ToA & Time-of-arrival \\

UL & Uplink
& ViT & Vision Transformer
& & \\

\midrule
\textbf{Short.} & \textbf{Meaning}
& \textbf{Short.} & \textbf{Meaning}
& \textbf{Short.} & \textbf{Meaning} \\
\midrule

chnl. & Channel
& sen. & Sensing
& cls. & Classification \\

corr. & Correction
& det. & Detection
& diag. & Diagnostics \\

est. & Estimation
& forecast. & Forecasting
& gen. & Generation \\

loc. & Localization
& mgmt. & Management
& pred. & Prediction \\

prec. & Precoding
& rec. & Recognition
& recon. & Reconstruction \\

seg. & Segmentation
& \makecell[l]{mask.\\rec.} & Masked reconstruction
& \makecell[l]{W.\\signal} & Wireless signal \\

\bottomrule
\end{tabular}
\end{table*}

\subsection{Survey Methodology}
\label{subsec:survey_methodology}

To ensure transparent and reproducible study selection, we follow a structured literature-selection methodology informed by the PRISMA framework~\cite{page2021prisma}. The process comprises identification, screening, eligibility assessment, and inclusion. Candidate studies were screened by title and abstract, assessed in full text against the eligibility criteria, and grouped according to their downstream PHY tasks.


\subsubsection{Search Strategy and Sources}

The search was conducted using academic databases, publisher platforms, and preprint repositories, including \textit{IEEE Xplore}, \textit{ACM Digital Library}, \textit{ScienceDirect}, \textit{SpringerLink}, \textit{Web of Science}, \textit{Scopus}, and \textit{arXiv}. 

The search strategy combined terms related to foundation models and wireless physical-layer systems, such as ``wireless foundation model'', ``radio foundation model'', ``channel foundation model'', ``large wireless model'', ``spectrum foundation model'', ``physical-layer foundation model'', ``self-supervised radio pretraining'', and ``foundation model for wireless communications'', focusing on papers published from 2024 until June 2026.

\subsubsection{Inclusion and Exclusion Criteria}

Papers were included if they met the following criteria:
(i) they focused on physical-layer wireless problems;
(ii) they proposed a wireless foundation model built or pretrained with wireless-domain data, including signal, channel, or spectrum representations; and (iii) they reported evaluation on at least one concrete physical-layer downstream task.

Papers were excluded when they were outside the physical-layer wireless scope, focused mainly on generic LLM-based wireless applications, or addressed non-wireless foundation-model settings. Papers that only mentioned foundation models as future motivation, without model design, wireless-domain pretraining, or downstream evaluation, were also excluded.

\subsection{Motivation and Contributions }
\label{subsec:contributions_organization}

Existing surveys provide valuable perspectives on AI and foundation models for wireless networks~\cite{xu2024largemultimodalmodelslmms,jiang2025lamsurvey,zhang2026mmfm}. However, recent surveys differ in scope and emphasis, and there remains room for a task-centered and evaluation-oriented analysis of PHY-oriented WFMs across multiple task families.

The contributions of this work can be summarized as follows.
\begin{enumerate}[font={\itshape},label={C\arabic*.}]
    \item We introduce a structured taxonomy of WFM physical-layer tasks, organizing current models according to their functional roles within the communication stack.

    \item We analyze how foundation-model principles, including large-scale pretraining, reusable representations, adaptation, and transferability, are being incorporated into PHY-oriented wireless tasks and protocols.

    \item We provide a task-centered and model-level comparison of PHY WFMs, examining input modalities, pretraining objectives, backbone architectures, adaptation strategies, datasets, and distribution-shift settings.

    \item We identify key limitations and open research directions for developing scalable, generalizable, and deployable WFMs for next-generation wireless systems.

\end{enumerate}

This paper is organized as follows.
\Cref{sec:related-work} reviews related surveys and positions this work within the existing literature.
\Cref{sec:tutorial} introduces the main WFM concepts, including pretraining, backbone architectures, and adaptation strategies.
\Cref{sec:phy-wfm} presents the PHY task taxonomy and dataset overview, while
\Cref{sec:analysis_signal,sec:analysis_channel,sec:analysis_rfsensing,sec:analysis_beam,sec:analysis_spectrum} analyze WFMs across the main PHY task families.
\Cref{sec:multitask_phy_wfms} discusses multi-task PHY WFMs.
Finally, \Cref{sec:open_directions} presents open challenges and future directions, and \Cref{sec:conclusion} concludes the paper.

Table~\ref{tab:abbreviations} summarizes the main abbreviations and shorthand terms used throughout the survey, including common wireless concepts, foundation-model terminology, adaptation protocols, and task labels used in the comparative tables.

\section{Related Work}
\label{sec:related-work}

Research on foundation models for wireless communications is expanding rapidly. However, most existing surveys approach this space from a broad communications-AI perspective, typically emphasizing large AI models, AI-native wireless systems, or general 6G visions. In these works, foundation models are usually discussed as part of a larger transformation of communication systems, rather than as a clearly defined class of wireless-native pretrained models. Our survey takes a narrower and more methodological perspective: it focuses on wireless foundation models (WFMs) for physical-layer tasks, with particular attention to how they are built, pretrained, adapted, and evaluated.

This distinction is summarized in Table~\ref{tab:wfm_related_surveys}. As the table shows, prior surveys tend to be strong along some dimensions while remaining limited along others. Broad surveys are useful for understanding the role of large AI models in future communication systems, but they are less focused on wireless-native representations and model-level WFM comparison. More specialized surveys provide deeper insight into specific branches, such as channel modeling, prediction and control, or network traffic, but they do not consolidate PHY-oriented WFMs across multiple task families.

The surveys most closely aligned with our focus are those that move beyond generic AI-for-wireless discussions and engage directly with foundation-model development in wireless settings. The survey on channel foundation models~\cite{jiang2025cfm} is particularly relevant because it treats channel foundation models as a dedicated research direction and reviews SSL-based approaches for learning reusable channel representations. It is also closely related to our work in its focus on wireless-native pretraining and model-level discussion. However, its scope is centered on channel modeling, and therefore it does not cover the broader set of PHY tasks addressed by recent WFMs, such as signal recognition, RF sensing and localization, beam management, and spectrum sensing. A related but differently organized perspective appears in the survey on multimodal foundation models for prediction and control in wireless networks~\cite{zhang2026mmfm}. This work is relevant because it discusses foundation models in wireless settings, including multimodal data, datasets, and prediction/control applications. However, its analysis remains broader and more application-oriented, whereas our survey places the model design itself at the center, with emphasis on how WFMs are pretrained, adapted, and evaluated in PHY-oriented tasks. These differences in scope and analytical emphasis are summarized in Table~\ref{tab:wfm_related_surveys}.


A second group of related works covers adjacent parts of this landscape. The systematic review on network traffic foundation models~\cite{perezjove2026ntfm} analyzes pretraining, downstream tasks, datasets, and transfer in network traffic, but focuses on network-layer rather than PHY data and tasks. The survey on learning-based wireless PHY approaches~\cite{mengistu2026phyfm} spans conventional machine learning to foundation-model developments, but does not center its analysis on WFM design, adaptation, and transfer across PHY task families. Broader surveys on telecom foundation models and large AI models for communications~\cite{zanouda2024tfm,jiang2025lamsurvey,zhu2025wirelesslargeaimodel} provide useful context on how FMs, LAMs, and WLAMs may shape future communication systems. These works discuss applications, deployment challenges, and future opportunities, and some of them introduce or review wireless-oriented FM directions. However, they are not primarily organized around PHY-native pretrained models or a model-level comparison of WFMs.

More recently, Liang et al.~\cite{liang2026foundation} survey foundation models across the wireless system, spanning PHY processing, resource management, and network autonomy. Their PHY discussion covers representative adapted and wireless-native models across several applications, while the survey also considers broader directions such as agentic models and wireless world modeling. Our survey instead adopts a specialized, task-centered PHY perspective, using PHY task families as the unit of analysis to systematically compare WFM pretraining, adaptation, and evaluation under different distribution shifts.




\begin{table*}[t]
\centering
\scriptsize
\setlength{\tabcolsep}{2.5pt}
\renewcommand{\arraystretch}{1.07}
\begin{threeparttable}
\caption{Comparison of surveys related to wireless foundation models.}
\label{tab:wfm_related_surveys}

\begin{tabular}{p{5.5cm} c *{6}{c} c *{5}{c}}
\toprule
\multirow{2}{*}{\textbf{Survey}} 
& \multirow{2}{*}{\textbf{Year}} 
& \multicolumn{6}{c}{\textbf{Comparison Dimensions}}
& \multirow{2}{*}{\textbf{Tutorial}}
& \multicolumn{5}{c}{\textbf{PHY Task Coverage}} \\
\cmidrule(lr){3-8}
\cmidrule(lr){10-14}
& 
& \makecell{\textbf{FM}\\\textbf{spec.}} 
& \makecell{\textbf{W.}\\\textbf{native}} 
& \makecell{\textbf{Model}\\\textbf{level}} 
& \textbf{PT} 
& \textbf{Adapt.}
& \makecell{\textbf{Eval.}\\\textbf{shift}}
& 
& \textbf{SID/Dem} 
& \makecell{\textbf{Channel}\\\textbf{R.}} 
& \makecell{\textbf{RF Sens.}\\\textbf{/Loc.}} 
& \textbf{Beam} 
& \textbf{Spectrum} \\
\midrule

Telecom Foundation Models Survey~\cite{zanouda2024tfm}
& 2024
& \cmark & \pmark & \xmark & \pmark & \pmark & \xmark
& \pmark
& \xmark & \xmark & \xmark & \xmark & \xmark \\

Channel Foundation Models Survey~\cite{jiang2025cfm}
& 2025
& \cmark & \cmark & \cmark & \cmark & \pmark & \pmark
& \cmark
& \xmark & \cmark & \xmark & \xmark & \xmark \\

Large AI Models Survey~\cite{jiang2025lamsurvey}
& 2025
& \pmark & \xmark & \pmark & \pmark & \pmark & \pmark
& \pmark
& \pmark & \pmark & \pmark & \pmark & \pmark \\

Wireless Large AI Models Survey~\cite{zhu2025wirelesslargeaimodel}
& 2025
& \pmark & \xmark & \pmark & \pmark & \pmark & \pmark
& \pmark
& \pmark & \pmark & \pmark & \pmark & \pmark \\

Network Traffic FM Review~\cite{perezjove2026ntfm}
& 2026
& \cmark & \cmark & \cmark & \cmark & \cmark & \pmark
& \xmark
& \xmark & \xmark & \xmark & \xmark & \xmark \\

Wireless PHY-to-FM Survey~\cite{mengistu2026phyfm}
& 2026
& \pmark & \cmark & \xmark & \xmark & \xmark & \xmark
& \pmark
& \pmark & \pmark  & \pmark & \xmark & \xmark \\

Multimodal FM Survey~\cite{zhang2026mmfm}
& 2026
& \cmark & \cmark & \pmark & \cmark & \cmark & \pmark
& \cmark
& \xmark & \pmark & \xmark & \pmark & \xmark \\

PHY-to-Network FM Survey~\cite{liang2026foundation}
& 2026
& \cmark & \cmark & \pmark & \cmark & \cmark & \pmark
& \pmark
& \pmark & \cmark & \pmark & \cmark & \pmark \\



\textbf{Ours}
& 2026
& \cmark & \cmark & \cmark & \cmark & \cmark & \cmark
& \cmark
& \cmark & \cmark & \cmark & \cmark & \cmark \\

\bottomrule
\end{tabular}

\begin{tablenotes}[flushleft]
\footnotesize
\item \textbf{Marks:} \cmark\ yes; \pmark\ partial; \xmark\ no.
\item \textbf{Model-level:} Explicit analysis of individual WFM methodologies and architectural choices.
\item \textbf{Eval.-shift:} Whether the survey systematically distinguishes transfer evidence across in-distribution, partial-shift, and out-of-distribution settings.
\item \textbf{Tutorial:} Whether the survey provides a conceptual or methodological overview, including definitions, pipelines, or pretraining/adaptation strategies.
\end{tablenotes}
\end{threeparttable}
\end{table*}

    
    
    

    
{



\definecolor{rqone}{RGB}{30, 60, 110}
\definecolor{rqfour}{RGB}{140, 90, 20}
\definecolor{dscolor}{RGB}{70, 80, 90}

\begin{figure}[b]
\centering
\resizebox{\linewidth}{!}{%
\begin{tikzpicture}[
    >=Stealth,
    font=\sffamily\small,
    encoder/.style={
        draw=rqone!70, thick, fill=rqone,
        rounded corners=4pt, align=center,
        text width=2.3cm, minimum height=2.4cm,
        text=white, font=\sffamily\small\bfseries
    },
    head/.style={
        draw=rqfour!70, thick, fill=rqfour!10,
        rounded corners=3pt, align=center,
        text width=1.3cm, minimum height=0.6cm,
        font=\scriptsize\bfseries, text=rqfour!90!black
    },
    loss/.style={
        circle, draw=dscolor!50, fill=dscolor!8,
        inner sep=0pt, minimum size=0.6cm,
        font=\scriptsize\itshape, text=dscolor
    },
    arrow/.style={->, thick, draw=black!70},
    meta_label/.style={font=\scriptsize\color{dscolor!70}}
]

\node[align=center, text=rqone] (x) {$x \in \mathcal{X}$\\ \scriptsize Observation};

\node[encoder, right=0.6cm of x] (f) {Shared Model\\[0.25cm] \Large $f_\theta$};
\node[meta_label, below=0.15cm of f] {Pretrained on $\mathcal{D}_{\mathrm{u}}$};

\node[align=center, right=0.6cm of f, text=rqone] (z) {$\mathbf{z} \in \mathbb{R}^{D_{\mathrm{enc}}}$\\ \scriptsize Representation};

\draw[arrow, rqone!70] (x) -- (f);
\draw[arrow, rqone!70] (f) -- (z);

\coordinate[right=0.4cm of z] (split);
\draw[thick, draw=dscolor!60] (z) -- (split);

\node[head, above right=0.95cm and 0.5cm of split] (phi1) {Head $\phi_1$};
\node[head, right=0.5cm of split] (phi2) {Head $\phi_2$};
\node[head, below right=0.95cm and 0.5cm of split] (phiT) {Head $\phi_{|\mathcal{T}|}$};
\node[font=\bfseries, yshift=0.1cm, text=rqfour!60] at ($(phi2)!0.5!(phiT)$) {\vdots};

\draw[arrow, rqfour!70] (split) |- (phi1.west);
\draw[arrow, rqfour!70] (split) -- (phi2.west);
\draw[arrow, rqfour!70] (split) |- (phiT.west);

\node[loss, right=1.2cm of phi1] (l1) {$\ell_1$};
\node[loss, right=1.2cm of phi2] (l2) {$\ell_2$};
\node[loss, right=1.2cm of phiT] (lT) {$\ell_{|\mathcal{T}|}$};

\draw[arrow, dscolor!60] (phi1) -- node[above, align=center, font=\scriptsize, text=dscolor] {$\hat{y}_1 \in \mathcal{Y}_1$} (l1);
\draw[arrow, dscolor!60] (phi2) -- node[above, align=center, font=\scriptsize, text=dscolor] {$\hat{y}_2 \in \mathcal{Y}_2$} (l2);
\draw[arrow, dscolor!60] (phiT) -- node[above, align=center, font=\scriptsize, text=dscolor] {$\hat{y}_{|\mathcal{T}|} \in \mathcal{Y}_{|\mathcal{T}|}$} (lT);

\node[meta_label, above=0.2cm of l1] (y1) {True $y_1$};
\draw[arrow, draw=dscolor!40] (y1) -- (l1);
\node[meta_label, above=0.2cm of l2] (y2) {True $y_2$};
\draw[arrow, draw=dscolor!40] (y2) -- (l2);
\node[meta_label, below=0.2cm of lT] (yT) {True $y_{|\mathcal{T}|}$};
\draw[arrow, draw=dscolor!40] (yT) -- (lT);

\node[above=1.1cm of f, font=\footnotesize\bfseries, text=rqone] (h1) {Representation Learning};

\coordinate (adapt_zone_start) at ($(split)+(0.2, 0)$);
\coordinate (adapt_zone_end) at (l2.east);
\coordinate (h2_pos) at ($(adapt_zone_start)!0.5!(adapt_zone_end)$);

\node[font=\footnotesize\bfseries, text=rqfour] at (h2_pos |- h1) {Supervised Adaptation};

\draw[dashed, rqone!15, thick] ($(z.east)!0.5!(split) + (0,-2.1)$) -- ($(z.east)!0.5!(split) + (0,3.0)$);

\end{tikzpicture}%
}
\caption{An input observation $x$ is mapped to a transferable latent representation $\mathbf{z}$ via a shared, pretrained backbone $f_\theta$. Task-specific heads $\phi_t$ provide lightweight adaptation, producing predictions $\hat{y}_t$ that are optimized against true labels $y_t$ via the multi-task objective defined in Equation~\ref{eq:wfm_general_objective}.}
\label{fig:problem_formulation}
\end{figure}
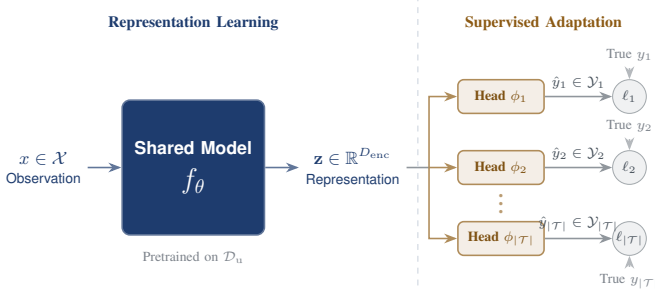
\section{Tutorial}
\label{sec:tutorial}

\subsection{What is a Wireless Foundation Model?}
\label{subsec:what_is_wfm}

A foundation model is a model pretrained on broad and diverse data, typically at scale, such that the learned representations can be adapted to downstream tasks beyond the objective used during pretraining~\cite{bommasani2021foundation,yazdnian2026multimodalfm}. While this broad definition is useful and widely adopted, in this survey we apply a stricter interpretation in order to distinguish foundation-model behavior from narrower forms of pretrained transfer. Specifically, we treat foundation-model behavior as requiring three elements: (\textit{i}) pretraining on broad and diverse data such that general structure is captured rather than a single task solved directly; (\textit{ii}) the pretrained representations can be reused across downstream tasks without requiring a separate pretraining pipeline for each task; and (\textit{iii}) downstream tasks that differ meaningfully in output structure, required capabilities, or application domain. 

\subsubsection{\textbf{Formal Definition}}
Consider a wireless system in which measurements collected from the radio environment are used to support a range of downstream tasks. These tasks may belong to communication, sensing, or localization, and the available observations may take different forms depending on the sensing, transmission, or processing stage at which they are acquired.

As shown in \autoref{fig:problem_formulation}, the objective is to build a shared wireless foundation model that learns general and reusable representations from wireless data and supports multiple downstream tasks through lightweight adaptation. Rather than training a separate model for each task, the goal is to learn a common backbone that captures structure useful across tasks and deployment settings while keeping task-specific overhead small.

\noindent \textbf{Inputs.} Let $x \in \mathcal{X}$ denote a wireless observation drawn from an input space $\mathcal{X}$. In general, $\mathcal{X}$ may include one or more wireless modality families, depending on the setting under consideration. 

\noindent \textbf{Shared model.} A parameterized model $f_\theta : \mathcal{X} \rightarrow \mathbb{R}^{D_{\mathrm{enc}}}$ maps each input to a latent representation $\mathbf{z} = f_\theta(x)$, where $\mathbf{z} \in \mathbb{R}^{D_{\mathrm{enc}}}$ is intended to preserve information useful across downstream tasks. Depending on the setting, $f_\theta$ may operate on a single modality family or may include input adapters that allow a shared backbone to process multiple modalities.

\noindent \textbf{Task heads.} For each downstream task $t \in \mathcal{T}$ with target space $\mathcal{Y}_t$, a task-specific head $\phi_t : \mathbb{R}^{D_{\mathrm{enc}}} \rightarrow \mathcal{Y}_t$ maps the shared representation to a prediction $\hat{y}_t = \phi_t(\mathbf{z})$. These heads provide the minimal adaptation needed to specialize the shared representation to each downstream objective.

\noindent \textbf{Self-supervised pretraining.} Before supervised adaptation, the encoder is pretrained on unlabeled wireless data drawn from $\mathcal{D}_{\mathrm{u}} \subset \mathcal{X}$. Pretraining encourages $f_\theta$ to learn robust and transferable structure directly from wireless observations, so that downstream tasks can be supported with limited labeled data and minimal task-specific modification.

\noindent \textbf{Learning goal.} The general supervised objective is
\begin{equation}
    \label{eq:wfm_general_objective}
    \min_{\theta,\{\phi_t\}} \sum_{t\in\mathcal{T}} \mathbb{E}_{(x,y_t)} \, \ell_t\big(\phi_t(f_\theta(x)), y_t\big),
\end{equation}
where $\ell_t$ is the loss function associated with task $t$. The goal is to achieve strong performance across tasks while keeping the backbone shared and the per-task adaptation lightweight.

\subsubsection{\textbf{Foundation Models vs SSL Pretraining}}
An important notion is that SSL pretraining alone is necessary but not sufficient to describe a model as a WFM. What matters is whether generalization to diverse tasks emerges from the pretraining.
This distinction is important because adaptation to multiple downstream tasks does not automatically imply foundation-model behavior. When the reported downstream tasks and data are very similar to the pretraining data, the result may reflect transfer within a narrow task family rather than broad generality. In such cases, the model may be closer to multi-task pretraining or domain adaptation, where the representation is adjusted to a new distributional context but does not clearly demonstrate new functional capabilities. In practical terms, we treat downstream tasks as more strongly supportive of strict foundation-model behavior when they differ meaningfully in what the model must do, rather than being only slight reformulations of the same task. 

We capture these notions by classifying downstream evaluation as in-distribution (ID), partial shift (Partial), or out-of-distribution (OOD). Results obtained under ID evaluation provide weaker support for strict foundation-model behavior, since they mainly show that the model remains useful under conditions similar to those seen during pretraining. In contrast, OOD evaluation provides stronger support when the downstream tasks differ clearly in dataset, scenario, acquisition conditions, or input modality. However, this support should also be interpreted by degree: when the shift is only minor, the result is better understood as \emph{partial} generalization rather than as a strong basis for considering the model a strict foundation model. Under this view, stricter foundation-model behavior is better supported when a model succeeds across meaningfully distinct downstream tasks and under meaningfully different data conditions.
In contrast, models that rely on large-scale or self-supervised pretraining but remain centered on a single task or on a narrow family of closely related tasks are better viewed as \emph{FM-like}, or task-specialized pretrained models.





\subsection{Pretraining Strategies}
\label{subsec:pretraining_strategies}

Pretraining is a key step in wireless foundation models because it provides the basis for learning useful representations that can be adapted to several tasks. Supervised pretraining can work well, but it depends on large labeled datasets, which are often difficult to obtain in wireless systems. Labels such as user positions, beam indices, activity classes, or spectrum occupancy may require controlled measurements, simulations, specialized hardware, or manual annotation~\cite{gizzini2022channelest,guo2022csifeedback,li2021wirelesssensing}. Unsupervised pretraining avoids labels, but it may not provide a clear enough learning signal for downstream adaptation.

For this reason, self-supervised learning (SSL) has become a central pretraining strategy for WFMs. SSL uses mechanisms such as augmentations, masking, temporal ordering, prediction targets, or reconstruction losses to create supervisory signals from unlabeled data~\cite{shwartzziv2023compresssl,balestriero2023cookbook,zhao2024comparisonreview}. This is well suited to wireless systems, where large amounts of raw I/Q samples, CSI, CIRs, spectrograms, radio maps, and sensing data can be collected or simulated, while task-specific labels remain costly and hard to transfer across devices, environments, and propagation
conditions~\cite{aboulfotouh2024ssradio,liu2024wifo,mashaal2025iqfm,chu2026wirelessjepa}. 



The following paragraphs provide an overview of the main self-supervised and pretraining strategies used in WFMs.

\subsubsection{\textbf{Contrastive Learning}}

Contrastive learning is an SSL strategy that trains a model by comparing different views of the data. Views from the same instance are treated as positive pairs, while views from different instances are treated as negative pairs. The objective is to bring positives closer in the representation space and push negatives apart. Figure~\ref{fig:group_pretraining_1} summarizes this pipeline. 
Its limitation is that the learned representation depends strongly on how the views and pairs are defined. Poorly chosen augmentations can remove task-relevant information, while false negatives can push apart samples that should actually remain close.

In general AI, contrastive learning appears in augmentation-based methods such as \textit{SimCLR}~\cite{chen2020simclr}, memory- or momentum-based variants such as \textit{MoCo}~\cite{he2020moco}, and multimodal alignment models such as \textit{CLIP}~\cite{radford2021clip}. In WFMs, the same principle is useful when different views of a wireless signal should remain aligned despite problematic variations. For example, \textit{IQFM} applies contrastive learning to I/Q data so that different views of the same signal are aligned in representation space, improving robustness across signal variations~\cite{mashaal2025iqfm}.

\subsubsection{\textbf{Masked Reconstruction}}


Masked reconstruction trains a model to recover masked parts of the input from visible context, providing a scalable self-supervised signal without manual labels. Figure~\ref{fig:group_pretraining_1} illustrates this pipeline. It is useful when the data contain strong local or global dependencies that allow missing regions to be inferred from context. The limitation lies in choosing the right masking level: insufficient masking may encourage low-level shortcuts, whereas excessive masking can make reconstruction overly difficult.

Masked modeling is well established in general AI through language and vision methods such as BERT and MAE, where masked tokens or image patches are recovered from visible context~\cite{devlin2019bert,he2022maskedautoencoders}. In WFMs, this objective is useful because wireless data contain dependencies across time, frequency, antennas, or space that can provide natural self-supervised signals. \textit{SSRadio} illustrates this with masked spectrogram reconstruction for spectrum-related tasks~\cite{aboulfotouh2024ssradio}, while \textit{6G WavesFM} extends masked modeling across spectrograms, CSI, and I/Q data for multiple PHY tasks~\cite{aboulfotouh20256gwavesfm}.

\subsubsection{\textbf{Latent Prediction}}

Latent prediction learns representations by predicting missing, future, or related parts of the data in an embedding space, operating on higher-level representations rather than reconstructing raw input samples as in masked autoencoders. It also differs from contrastive learning because it does not necessarily require negative samples and can reduce dependence on carefully engineered augmentations, especially when valid transformations are difficult to define. Figure~\ref{fig:group_pretraining_1} illustrates this pipeline.
The advantage of latent prediction is that it can encourage the model to capture semantic, temporal, or structural dependencies while avoiding excessive focus on raw-sample reconstruction. Its limitation is that the prediction target must be carefully designed; otherwise, the model may learn weak or uninformative representations instead of useful structure. 

In general AI, this idea appears in \textit{I-JEPA}, which predicts missing image regions in representation space to encourage learning of high-level structure rather than pixel-level reconstruction~\cite{assran2023ijepa}.
In WFMs, it is useful when the goal is to learn dependencies from I/Q samples, CSI, or spectrograms without reconstructing the exact waveform. \textit{WirelessJEPA} provides a representative example by adapting JEPA to wireless I/Q data through an antenna--time representation and spatio-temporal masking, allowing the model to learn dependencies across antennas and time~\cite{chu2026wirelessjepa}.

\begin{figure}[!h]
    \centering
    \includegraphics[width=0.95\linewidth,trim=0 5mm 0 0mm, clip]{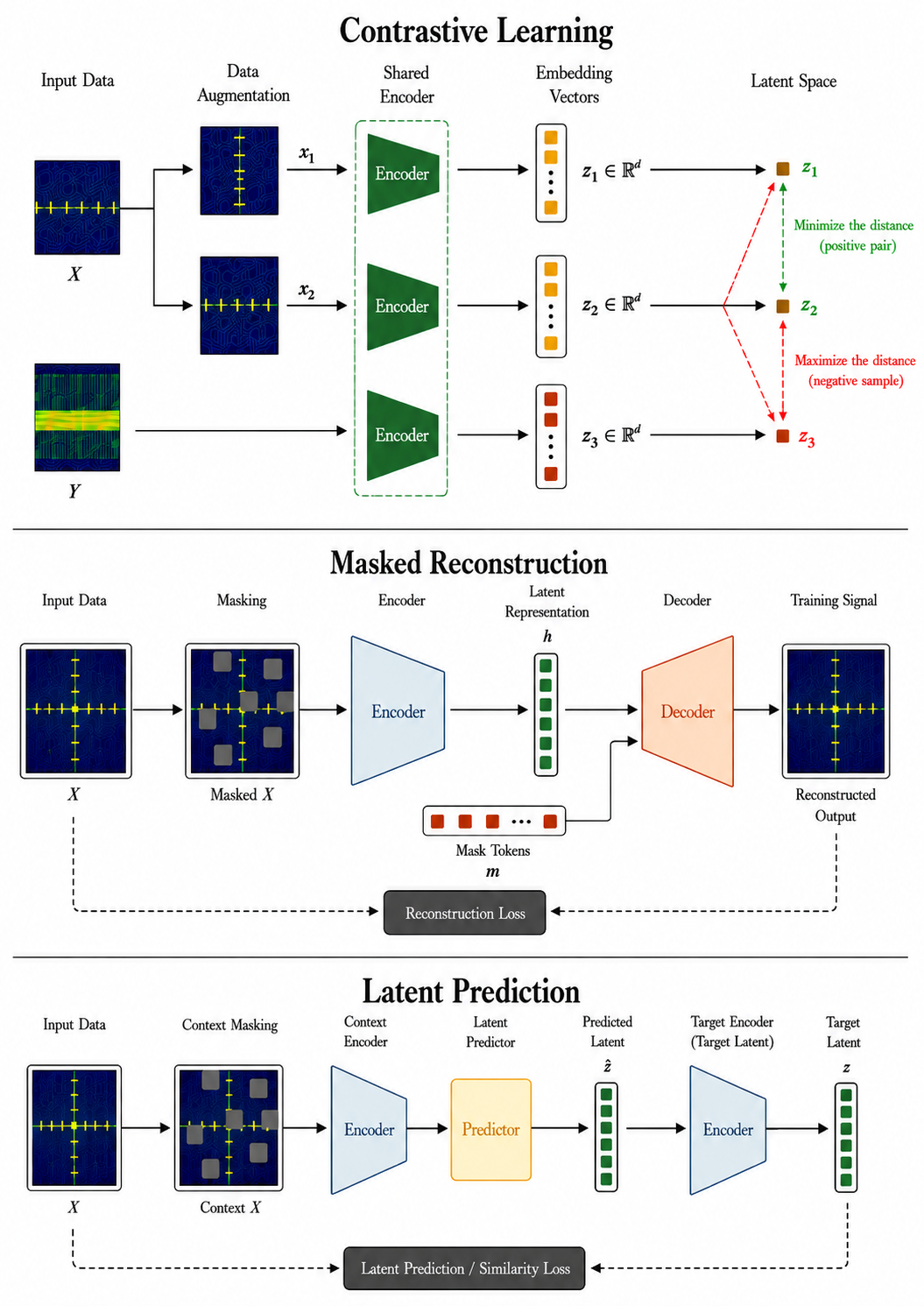}
    \caption{Representative WFM pretraining objective families.}
    \label{fig:group_pretraining_1}
\end{figure}

\subsubsection{\textbf{Generative}}

Generative pretraining learns to produce data-domain outputs, often by modeling how samples are generated, completed, or denoised. Unlike contrastive and latent-prediction objectives, it requires an interpretable output in the data domain rather than only organizing internal representations. It differs from masked reconstruction when the goal is not only to recover hidden parts of the same input, but to generate complete or refined data-domain outputs, often through autoregressive or denoising processes. Figure~\ref{fig:group_pretraining_2} illustrates the general generative pretraining pipeline. 
Its limitation is that generated outputs may be statistically realistic but not necessarily consistent with the underlying physical constraints of the domain, so additional validation is needed.

In general AI, generative learning includes autoregressive models, which generate data sequentially, and diffusion models, which generate samples through iterative denoising; DDPMs formalize the latter by learning to remove noise step by step~\cite{ho2020denoising}. In WFMs, this objective is useful when the goal is to synthesize or refine wireless-domain outputs. \textit{ChannelGPT} generates channel-related outputs conditioned on environmental context, helping capture how propagation patterns vary across scenarios~\cite{yu2024channelgpt}. \textit{WiFo-MUD} follows the diffusion branch by progressively refining noisy symbol estimates for multi-user demodulation, improving robustness across heterogeneous user and channel configurations~\cite{yang2026wifomud}.


\subsubsection{\textbf{Temporal Prediction}}

Temporal prediction learns from the temporal ordering of data by using past observations to predict future observable values. Unlike latent prediction, where the target is an embedding or hidden representation, temporal prediction directly supervises the model with future samples or measurements. It also differs from masked reconstruction because the model is not asked to recover missing parts of the current input, but to forecast what comes next from historical context. Figure~\ref{fig:group_pretraining_2} illustrates this pipeline. 
Its limitation is that the learned behavior may become tied to the temporal patterns and scenarios used during training, so generalization to different time scales must be explicitly evaluated.

In general AI, this appears through time-series foundation models such as \textit{Chronos}, which reformulates forecasting as a language-modeling problem by tokenizing time-series values and training Transformer-based models to predict future values~\cite{ansari2024chronos}. In WFMs, this objective is useful when future information can improve communication or network decisions: \textit{Multi-Task Prediction FM} trains a causal Transformer to forecast future values from historical wireless time series, covering channel prediction within a unified framework~\cite{sheng2025multitaskfm}.


\subsubsection{\textbf{Distillation-Based Learning}}

Distillation-based learning uses a teacher model to guide a student model, with supervision coming not from labels or positive--negative pairs but from the teacher's predictions, output distributions, or feature representations. Depending on the goal, distillation can be used either to learn stronger representations or to compress a large model into a smaller one. Figure~\ref{fig:group_pretraining_2} illustrates the general distillation-based learning pipeline. 
Its limitation is teacher dependence: the student may inherit the teacher's errors, or biases, so the quality of the teacher strongly affects the learned student.

In general AI, distillation-based learning appears both in self-distillation methods such as \textit{DINOv2}, which trains Vision Transformers with teacher--student augmented views for label-free representation learning~\cite{oquab2023dinov2}, and in classical knowledge distillation, where a smaller student reproduces a larger teacher for efficient deployment~\cite{hinton2015distilling}. In WFMs, this strategy suits models operating under computation, memory, or latency constraints: \textit{Tiny-WiFo} trains a lightweight student to imitate the larger WiFo model for channel prediction, preserving teacher behavior while reducing model size and inference cost~\cite{zhang2025tinywifo}.

\begin{figure}[!h]
    \centering
    \includegraphics[width=0.95\linewidth,trim=0 5mm 0 5mm, clip]{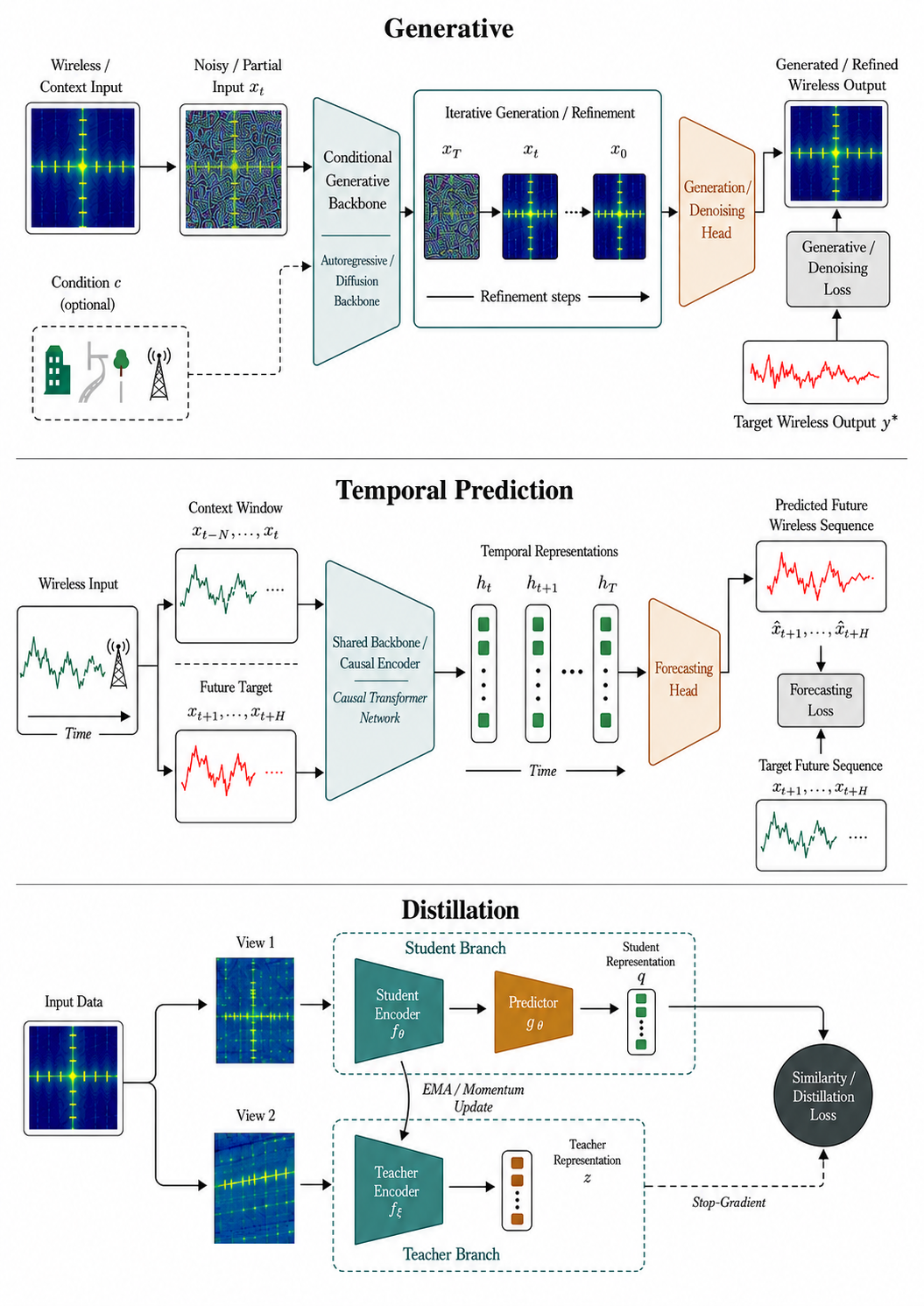}
    \caption{Representative WFM pretraining objective families.}
    \label{fig:group_pretraining_2}
\end{figure}




\subsubsection{\textbf{Hybrid}}

Hybrid objectives combine two or more training signals within the same pretraining framework. Instead of relying on a single objective, such as contrastive alignment, masked reconstruction, or prediction, hybrid methods jointly optimize complementary losses. The goal is to encourage the model to learn different types of structure at the same time. Figure~\ref{fig:group_pretraining_3} illustrates this pipeline.
Its limitation is that the training becomes harder to interpret and tune, since the final representation depends on how the losses are balanced and whether the objectives reinforce or conflict with each other.

In general AI, hybrid objectives combine complementary signals, such as reconstruction and contrastive learning, to learn both local structure and discriminative representations, as shown in methods such as \textit{CMAE} and \textit{RECON}~\cite{huang2024contrastivemasked,qi2023contrastreconstruct}. In WFMs, hybrid objectives are useful when one pretraining signal is insufficient to capture wireless structure. \textit{SpectrumFM}, for example, combines masked reconstruction with next-slot prediction to learn both local signal recovery and future spectrum dynamics~\cite{zhou2025spectrumfm}. Pilot WSensing FM uses a hybrid objective because wireless sensing requires representations that preserve radar signal structure, remain robust to view perturbations, and capture temporal motion dynamics~\cite{serbetci2025pilotwsensing}. 



\subsubsection{\textbf{Task-Driven Supervision}}

Task-driven supervision uses real downstream tasks to train a shared model. Instead of defining supervision only from the input structure, as in SSL, the model is guided by task information, such as labels, task descriptions, or task-specific losses. The objective is to align a common backbone with a subset of related functions, rather than only learn a generic representation disconnected from the target functions. Figure~\ref{fig:group_pretraining_3} illustrates this pipeline.
Its limitation is that the learned behavior may remain tied to the tasks, labels, or scenarios included during training, so transfer beyond that task set requires additional evidence.

In general AI, task-driven supervision is related to models that organize diverse tasks under a shared training format, such as \textit{T5}'s text-to-text formulation and \textit{FLAN}'s instruction tuning with task descriptions and examples~\cite{raffel2020t5,wei2022flan}. In WFMs, this strategy is useful because it guides the model toward PHY-relevant functions instead of learning only generic signal structure. \textit{MUSE-FM} uses short task descriptions to specify the target wireless function~\cite{zheng2025musefm}, while \textit{ICWLM} uses a few input--output examples in context so the same model can infer mappings such as precoding or channel prediction~\cite{wen2025icwlm}.



\begin{figure}[!h]
    \centering
    \includegraphics[width=0.95\linewidth,trim=0 10mm 0 10mm, clip]{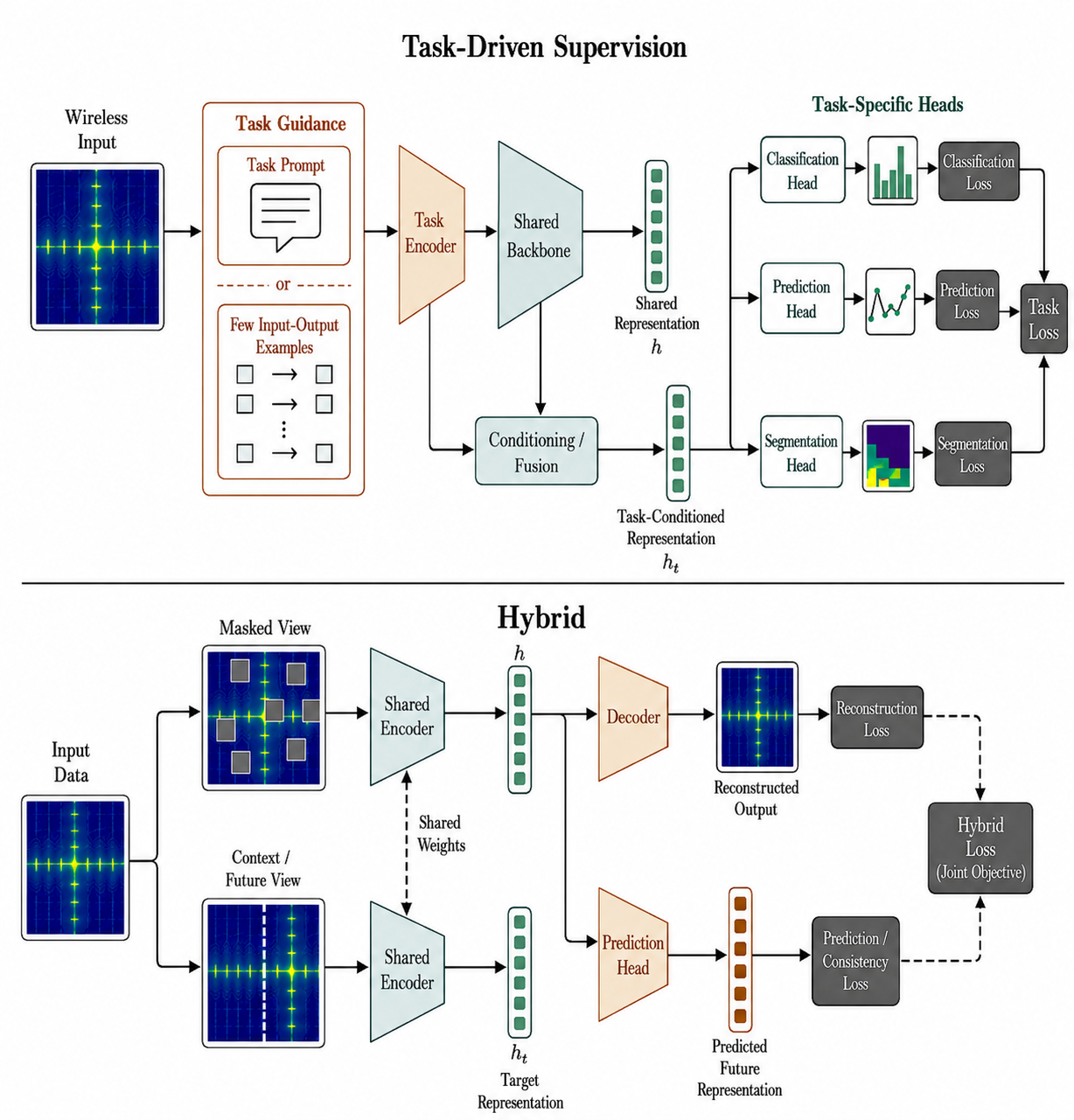}
    \caption{Representative WFM pretraining objective families.}
    \label{fig:group_pretraining_3}
\end{figure}

\subsection{Backbone Architectures}
\label{subsec:backbone_architectures}

Backbone architectures determine how wireless inputs are represented and processed before downstream adaptation. We summarize the main architectural families used in WFMs, emphasizing their implications for representation structure, scalability, and transfer.


\subsubsection{\textbf{CNN-based}}
CNN backbones exploit local structure in grid-like wireless representations such as I/Q windows, spectrograms, and CSI matrices, capturing correlations across time, frequency, antennas, or spatial dimensions~\cite{lecun1998gradient,he2016deep}. Within WFMs, \textit{SSRadio}, for example, combines convolutional and recurrent processing for masked spectrogram modeling and downstream spectrum tasks~\cite{aboulfotouh2024ssradio}.



\subsubsection{\textbf{Transformer-based}}
Transformers have become the dominant WFM backbone family because attention can model long-range dependencies across structured wireless inputs~\cite{vaswani2017attention,bommasani2021foundation}. Within WFMs, Transformer backbones appear with several recurring architectural extensions.

\textit{Vision Transformers (ViTs)} are a patch-based Transformer variant for images, dividing images into patches treated as tokens instead of processing pixel by pixel~\cite{dosovitskiy2021image}. In WFMs, this idea is useful when wireless data are represented as image-like structures, such as spectrograms, CSI maps, or radio maps. This explains why ViT models are often used with masked reconstruction objectives, where the model learns by recovering missing patches from the visible context~\cite{he2022maskedautoencoders}.

\textit{Mixture-of-Experts (MoE) Transformers} extend standard Transformers by adding multiple expert modules, while activating only a subset of them for each input~\cite{shazeer2017outrageously,fedus2022switch}. This allows the model to increase its capacity without using all parameters for every sample. In WFMs, this is relevant for heterogeneous wireless data, because different experts can specialize in different propagation conditions, scenarios, or signal patterns. \textit{LWM-Spectro} illustrates this design by using a MoE Transformer backbone for spectrogram-based modulation classification~\cite{kim2026lwmspectro}.

\textit{CNN--Transformer backbones} combine convolutional layers for local pattern extraction with attention layers for broader context modeling. This design appears in general vision models such as CoAtNet, which combines both to balance locality and efficiency~\cite{dai2021coatnet}. In WFMs, \textit{SpectrumFM}, for instance, combines CNNs and multi-head attention to model both local and global structure in I/Q spectrum data~\cite{zhou2025spectrumfm}.   

\textit{Dual-branch and multimodal Transformers} use separate encoders or branches to process different views or modalities before aligning or fusing their representations. Their main advantage is that they can preserve modality-specific information while still learning a shared representation space. Within WFMs, this is useful when the model must connect heterogeneous wireless inputs. \textit{CSI-CLIP}, for example, treats CSI and CIR as paired channel modalities and learns a shared representation through contrastive alignment~\cite{jiang2025csiclip}.

\textit{Geometry-aware attention models} add structural information to attention-based backbones. Instead of treating all input elements only as independent tokens, they use spatial, topological, or geometric relationships to guide how information is exchanged~\cite{velickovic2018graphattention}. In WFMs, this is useful for radio maps or spatially organized wireless data, where propagation depends on geometry and neighborhood structure. \textit{FM-RME} follows this direction by combining a geometry-aware feature extraction module with an attention-based network for multi-dimensional radio map estimation~\cite{yang2026fmrme}.

\textit{Diffusion Transformers} use a Transformer as the denoising or refinement network inside a diffusion-based generative process~\cite{ho2020denoising,peebles2023scalable}. In WFMs, this is useful when wireless signals or symbols must be refined through iterative denoising. \textit{WiFo-MUD} illustrates this architecture by using a diffusion Transformer backbone for conditional multi-user demodulation~\cite{yang2026wifomud}.

\subsubsection{\textbf{MLP-based}}
MLPs offer lightweight feature encoding for edge-oriented WFMs, where latency and parameter efficiency are prioritized. \textit{Lightweight Edge FM} uses patch-independent MLP encoders for wireless time series, reducing complexity relative to Transformer-based alternatives~\cite{cheraghinia2025lightweightedge}.


\subsubsection{\textbf{RNN-based}}
Recurrent neural networks (RNNs) process sequential data by updating a hidden state over time to capture temporal dependencies, with variants like LSTM improving the modeling of longer patterns~\cite{hochreiter1997long,graves2013speech}. In WFMs, recurrent backbones are useful for time-evolving wireless inputs, such as spectrum traces or temporal CSI, and can be combined with convolutions to capture both local structure and temporal evolution. 

\subsubsection{\textbf{SSMs}}
State-space models (SSMs) process sequences through structured dynamical representations rather than self-attention. Recent architectures such as Mamba use selective state-space mechanisms to model long sequences efficiently~\cite{gu2022efficiently,gu2023mamba}.  In WFMs, this is useful when wireless inputs are long or structured and attention-based models become computationally expensive. \textit{WiMamba} adapts Mamba as the main CSI encoder. Its bidirectional design is intended to capture relationships across the structured CSI input, where antenna and subcarrier dimensions do not follow a natural one-way temporal order~\cite{raviv2026wimamba}.

\noindent \textbf{Summary:} \noindent To provide a practical reading guide, Fig.~\ref{fig:backbone_family_map} summarizes how the main backbone families are related in WFMs. 

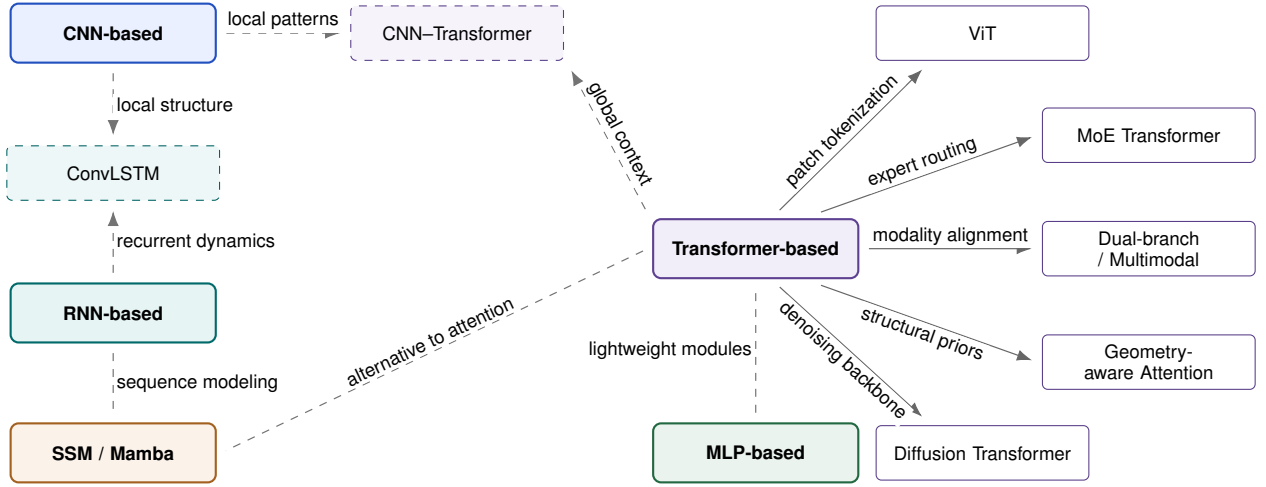
\begin{figure*}[t]
\centering
\scriptsize
\begin{tikzpicture}[
    font=\sffamily\scriptsize,
    family/.style args={#1}{
        rectangle,
        rounded corners=3pt,
        draw=#1!75!black,
        fill=#1!10,
        thick,
        align=center,
        text width=2.55cm,
        minimum height=0.78cm,
        font=\sffamily\bfseries\scriptsize,
        outer sep=2pt
    },
    variant/.style args={#1}{
        rectangle,
        rounded corners=2pt,
        draw=#1!70!black,
        fill=white,
        align=center,
        text width=2.65cm,
        minimum height=0.72cm,
        font=\sffamily\scriptsize,
        outer sep=2pt
    },
    bridge/.style args={#1}{
        rectangle,
        rounded corners=2pt,
        draw=#1!70!black,
        fill=#1!7,
        dashed,
        align=center,
        text width=2.65cm,
        minimum height=0.72cm,
        font=\sffamily\scriptsize,
        outer sep=2pt
    },
    arrow/.style={
        -{Latex[length=2mm]},
        draw=black!60,
        line width=0.45pt,
        shorten <=2pt,
        shorten >=2pt
    },
    softarrow/.style={
        -{Latex[length=2mm]},
        dashed,
        draw=black!55,
        line width=0.45pt,
        shorten <=2pt,
        shorten >=2pt
    },
    relation/.style={
        dashed,
        draw=black!55,
        line width=0.45pt,
        shorten <=2pt,
        shorten >=2pt
    },
    edgeLabel/.style={
        font=\sffamily\scriptsize,
        fill=white,
        inner sep=1.2pt
    }
]

\node[family=TaxBlue]   (cnn)      at (-6.8,  3.40) {CNN-based};
\node[bridge=TaxTeal]   (convlstm) at (-6.8,  1.55) {ConvLSTM};
\node[family=TaxTeal]   (rnn)      at (-6.8, -0.30) {RNN-based};
\node[family=TaxOrange] (ssm)      at (-6.8, -2.15) {SSM / Mamba};

\node[bridge=TaxPurple] (cnntrans) at (-2.25,  3.40) {CNN--Transformer};
\node[family=TaxPurple] (trans)    at ( 1.70,  0.55) {Transformer-based};
\node[family=TaxGreen]  (mlp)      at ( 1.70, -2.15) {MLP-based};

\node[variant=TaxPurple] (vit)   at ( 4.70,  3.40) {ViT};
\node[variant=TaxPurple] (moe)   at ( 6.90,  2.05) {MoE Transformer};
\node[variant=TaxPurple] (multi) at ( 6.90,  0.55) {Dual-branch / Multimodal};
\node[variant=TaxPurple] (geo)   at ( 6.90, -0.95) {Geometry-aware Attention};
\node[variant=TaxPurple] (diff)  at ( 4.70, -2.15) {Diffusion Transformer};

    
\draw[softarrow]
    (cnn.east) -- node[edgeLabel, above] {local patterns} (cnntrans.west);

\draw[softarrow]
    (trans.north west) -- node[edgeLabel, above, sloped] {global context} (cnntrans.south east);

\draw[softarrow]
    (cnn.south) -- node[edgeLabel, right] {local structure} (convlstm.north);

\draw[softarrow]
    (rnn.north) -- node[edgeLabel, right] {recurrent dynamics} (convlstm.south);

\draw[relation]
    (rnn.south) -- node[edgeLabel, right] {sequence modeling} (ssm.north);

\draw[relation]
    (trans.west) -- node[edgeLabel, above, sloped] {alternative to attention} (ssm.east);

\draw[relation]
    (mlp.north) -- node[edgeLabel, left] {lightweight modules} (trans.south);

\draw[arrow]
    (trans.60) -- node[edgeLabel, above, sloped] {patch tokenization} (vit.210);

\draw[arrow]
    (trans.30) -- node[edgeLabel, above, sloped] {expert routing} (moe.west);

\draw[arrow]
    (trans.0) -- node[edgeLabel, above] {modality alignment} (multi.west);

\draw[arrow]
    (trans.-30) -- node[edgeLabel, below, sloped] {structural priors} (geo.west);

\draw[arrow]
    (trans.-60) -- node[edgeLabel, below, sloped] {denoising backbone} (diff.150);

\end{tikzpicture}
\caption{Relationship map of backbone families used in WFMs. Solid arrows indicate variants derived from Transformer-based backbones. Dashed arrows pointing into dashed nodes indicate bridge architectures that combine backbone families, while dashed lines without arrowheads indicate conceptual, lightweight, or alternative design relations. Edge labels summarize the main reason for each connection.}
\label{fig:backbone_family_map}
\end{figure*}

\subsection{Adaptation Strategies}
\label{subsec:adaptation_strategies}

After pretraining, WFMs must be adapted to downstream PHY tasks, datasets, or deployment conditions. The main strategies differ in how much of the pretrained model is updated. Zero-shot evaluation uses the pretrained model without task-specific parameter updates, providing a strict test of direct transfer~\cite{radford2021clip,bommasani2021foundation}. Frozen fine-tuning and linear probing freeze the backbone and train only a task-specific head, making them useful for evaluating representation quality~\cite{chen2020simclr,he2022maskedautoencoders}. PEFT methods, such as adapters, prompts, or low-rank updates, provide an intermediate option for adapting larger WFMs with lower memory and training cost~\cite{houlsby2019parameter,hu2022lora}. Partial and full fine-tuning offer greater adaptation flexibility by updating selected layers or the complete model, but make it harder to separate representation quality from task-specific re-optimization. Multi-task fine-tuning adapts the model with several downstream objectives jointly, which can encourage shared representations but requires balancing potentially competing task signals~\cite{caruana1997multitask,ruder2017overview}. Figure~\ref{fig:adaptation_strategies} summarizes these strategies along the efficiency--flexibility axis.

\begin{figure}[!h]
    \centering
    \includegraphics[width=0.90\linewidth,trim=0 20mm 0 0mm, clip]{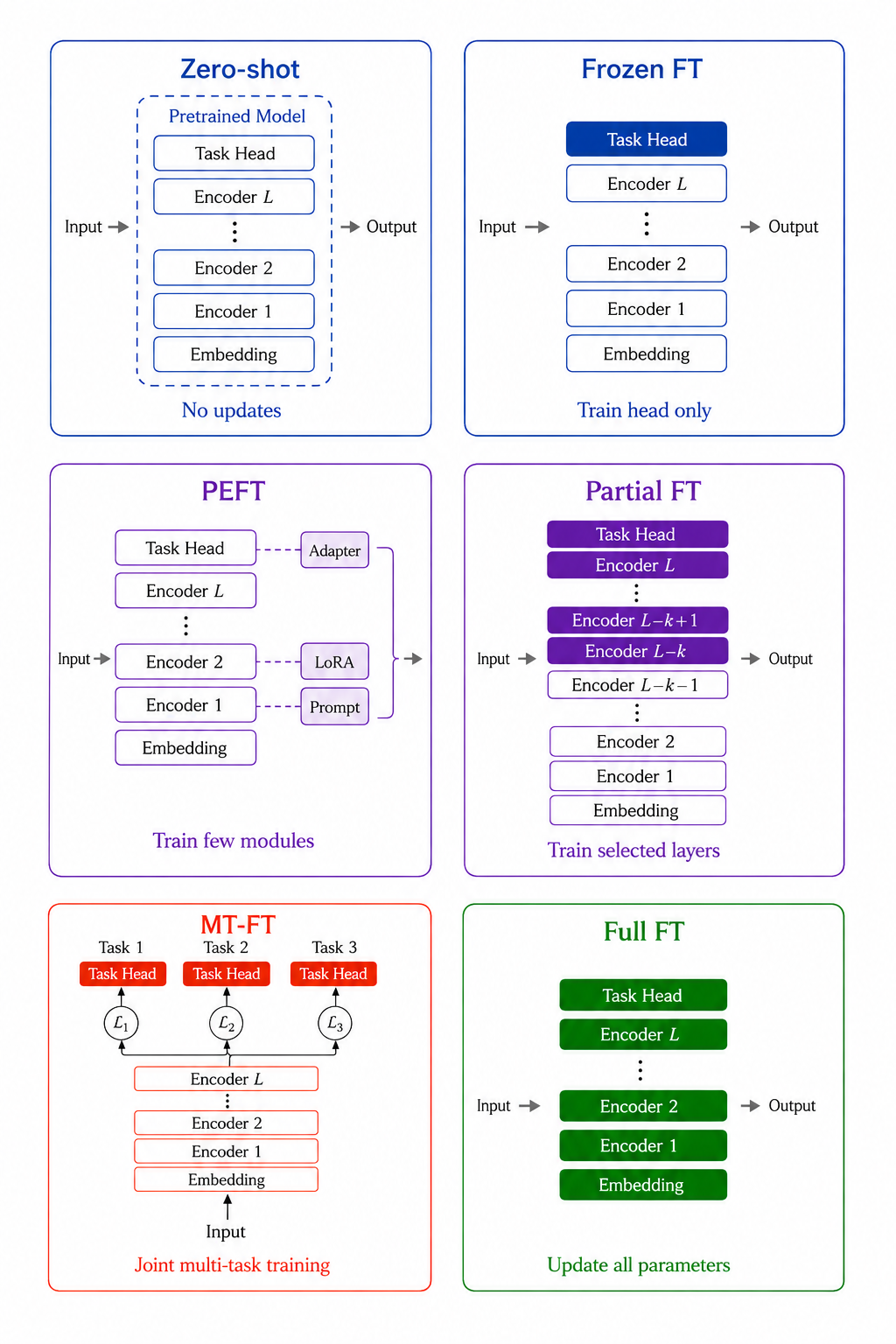}
    \caption{Overview of foundation-model adaptation strategies, ordered from lower update cost to higher adaptation flexibility.}
    \label{fig:adaptation_strategies}
\end{figure}

\section{WFMs for the PHY Layer: Taxonomy and Overview} 
\label{sec:phy-wfm}
The physical layer has become a major focus of the emerging wireless foundation-model literature, consistent with the broader adoption of AI for signal-level representations such as I/Q samples, CSI, and spectrograms \cite{huynh2023gaiphy,fontaine2024towardswireless,jiang2025lamsurvey,zhu2025wirelesslargeaimodel,chatzieleftheriou2026ai6g}. These models are evaluated across a diverse range of downstream PHY tasks.

To organize this literature, we introduce the task taxonomy in Section~\ref{subsec:phy_task_taxonomy_definition}. The taxonomy does not assign each WFM to a single function; instead, it provides a common structure for comparing how models are pretrained, adapted, and evaluated across PHY tasks. We use this structure to assess transfer evidence across task families, datasets, adaptation protocols, and distribution-shift settings.


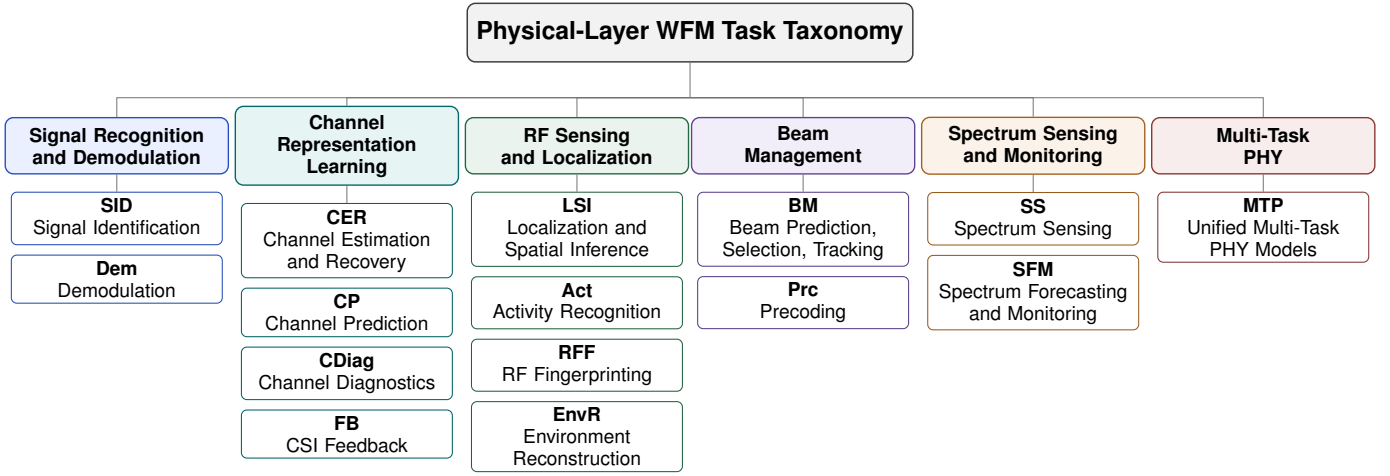
\begin{figure*}[t]
\centering
\resizebox{\textwidth}{!}{%
\begin{tikzpicture}[
    font=\sffamily\scriptsize,
    root/.style={
        draw=black!75,
        fill=black!5,
        rounded corners=3pt,
        line width=0.55pt,
        minimum width=5.1cm,
        minimum height=0.75cm,
        align=center,
        font=\sffamily\bfseries\small
    },
    cat/.style args={#1/#2}{
        draw=#1!75!black,
        fill=#1!10,
        rounded corners=3pt,
        line width=0.50pt,
        text width=#2,
        minimum height=0.72cm,
        align=center,
        inner sep=3pt,
        font=\sffamily\bfseries\scriptsize
    },
    task/.style args={#1/#2}{
        draw=#1!70!black,
        fill=white,
        rounded corners=2pt,
        line width=0.38pt,
        text width=#2,
        minimum height=0.52cm,
        align=center,
        inner sep=2.2pt
    },
    link/.style={
        draw=black!55,
        line width=0.35pt
    }
]

\node[root] (root) at (0,0) {Physical-Layer WFM Task Taxonomy};

\node[cat=TaxBlue/2.65cm]   (sig)  at (-7.35,-1.45) {Signal Recognition\\and Demodulation};
\node[cat=TaxTeal/2.65cm]   (chan) at (-4.40,-1.45) {Channel\\Representation Learning};
\node[cat=TaxGreen/2.65cm]  (rf)   at (-1.45,-1.45) {RF Sensing\\and Localization};
\node[cat=TaxPurple/2.65cm] (beam) at ( 1.45,-1.45) {Beam\\Management};
\node[cat=TaxOrange/2.65cm] (spec) at ( 4.40,-1.45) {Spectrum Sensing\\and Monitoring};
\node[cat=TaxRed/2.65cm]    (mt)   at ( 7.35,-1.45) {Multi-Task\\PHY};

\coordinate (hub) at ($(root.south)+(0,-0.45)$);
\draw[link] (root.south) -- (hub);
\foreach \x in {sig,chan,rf,beam,spec,mt}
    \draw[link] (hub) -| (\x.north);

\node[task=TaxBlue/2.55cm, below=2.2mm of sig] (sid)
    {\textbf{SID}\\Signal Identification};
\node[task=TaxBlue/2.55cm, below=1.2mm of sid] (dem)
    {\textbf{Dem}\\Demodulation};

\node[task=TaxTeal/2.55cm, below=2.2mm of chan] (cer)
    {\textbf{CER}\\Channel Estimation\\and Recovery};
\node[task=TaxTeal/2.55cm, below=1.2mm of cer] (cp)
    {\textbf{CP}\\Channel Prediction};
\node[task=TaxTeal/2.55cm, below=1.2mm of cp] (cdiag)
    {\textbf{CDiag}\\Channel Diagnostics};
\node[task=TaxTeal/2.55cm, below=1.2mm of cdiag] (fb)
    {\textbf{FB}\\CSI Feedback};

\node[task=TaxGreen/2.55cm, below=2.2mm of rf] (lsi)
    {\textbf{LSI}\\Localization and\\Spatial Inference};
\node[task=TaxGreen/2.55cm, below=1.2mm of lsi] (act)
    {\textbf{Act}\\Activity Recognition};
\node[task=TaxGreen/2.55cm, below=1.2mm of act] (rff)
    {\textbf{RFF}\\RF Fingerprinting};
\node[task=TaxGreen/2.55cm, below=1.2mm of rff] (envr)
    {\textbf{EnvR}\\Environment\\Reconstruction};

\node[task=TaxPurple/2.55cm, below=2.2mm of beam] (bm)
    {\textbf{BM}\\Beam Prediction,\\Selection, Tracking};
\node[task=TaxPurple/2.55cm, below=1.2mm of bm] (prc)
    {\textbf{Prc}\\Precoding};

\node[task=TaxOrange/2.55cm, below=2.2mm of spec] (ss)
    {\textbf{SS}\\Spectrum Sensing};
\node[task=TaxOrange/2.55cm, below=1.2mm of ss] (sfm)
    {\textbf{SFM}\\Spectrum Forecasting\\and Monitoring};

\node[task=TaxRed/2.55cm, below=2.2mm of mt] (mtp)
    {\textbf{MTP}\\Unified Multi-Task\\PHY Models};

\foreach \a/\b in {
    sig/sid,
    chan/cer,
    rf/lsi,
    beam/bm,
    spec/ss,
    mt/mtp}
    \draw[link] (\a.south) -- (\b.north);

\end{tikzpicture}%
}
\caption{Hierarchical taxonomy of physical-layer WFM tasks. Major categories group functionally related physical-layer problems, and the individual task-level labels provide compact categories for organizing the surveyed foundation model and self-supervised learning methods.}
\label{fig:phy_taxonomy}
\end{figure*}

\subsection{Taxonomy and Definitions of Physical Tasks}
\label{subsec:phy_task_taxonomy_definition}

To organize the physical-layer literature in a compact and consistent manner, we adopt a hierarchical taxonomy composed of five major categories: \emph{Signal Recognition and Demodulation}, \emph{Channel Representation Learning}, \emph{RF Sensing and Localization}, \emph{Beam Management}, and \emph{Spectrum Sensing and Monitoring}. 
In addition, we include a  \emph{Multi-Task PHY} category to highlight models that explicitly address multiple physical-layer tasks. 
Within this hierarchy, task-level categories specify the concrete downstream problems targeted by each model. This design provides a compact organizational structure while preserving the distinctions that are most relevant for analysis.


\subsubsection{Signal Recognition and Demodulation}
This category covers receiver-side tasks, including signal identification and demodulation, following prior work on wireless signal recognition and receiver-side processing.

\textbf{Signal Identification (SID).}
This group includes tasks that identify the type or nature of a received signal by recognizing discriminative patterns from raw I/Q samples, CSI, or spectrogram-like representations. It encompasses problems such as wireless technology recognition and modulation classification \cite{li2019wirelesssignal,kulin2018spectrumid}.

\textbf{Demodulation (Dem).}
Demodulation refers to the recovery of transmitted symbols or bits from the received waveform. Unlike signal identification tasks, which focus on recognizing the signal type, demodulation aims to interpret the information transmitted by the signal itself \cite{doha2025deeplearningwireless}.


\subsubsection{Channel Representation Learning} 
This category covers tasks that estimate, recover, predict, diagnose, or compress channel information. In our taxonomy, this category includes:

\textbf{Channel Estimation and Recovery (CER).} 
CER includes tasks that infer or reconstruct channel state information from incomplete or noisy observations, using mechanisms such as interpolation, denoising, and completion \cite{gizzini2022channelest}.

\textbf{Channel Prediction (CP).}
Channel prediction refers to forecasting future channel states or channel evolution over time. Unlike estimation or recovery, which focus on present or missing observations, prediction explicitly models temporal dynamics to anticipate future channel conditions \cite{kim2025channelprediction}.

\textbf{Channel Diagnostics (CDiag).}
Channel diagnostics includes tasks that infer descriptive properties or conditions of the channel rather than reconstructing it directly. This includes general \textit{CSI Analysis} and specific channel-condition inference tasks such as \textit{LoS/NLoS classification}, mobility-level recognition, and other propagation-state indicators derived from CSI or related channel representations \cite{li2017nloscsi,zeng2024ckm}.

\textbf{CSI Feedback (FB).}
CSI feedback is treated as a separate task because it focuses on the efficient compression, representation, and transmission of channel state information from receiver to transmitter, especially in FDD systems. Its objective is not channel inference itself, but bandwidth-efficient channel reporting for downstream communication decisions \cite{guo2022csifeedback}.


\subsubsection{RF Sensing and Localization}
This category includes tasks that use wireless signals as a sensing modality, rather than only a communication carrier, to infer physical, spatial, or behavioral information about users and environments \cite{li2021wirelesssensing,ma2019wificsi}. In our taxonomy, this category includes:

\textbf{Localization and Spatial Inference (LSI).}
This category combines \textit{Localization}, \textit{Angle-of-Arrival estimation}, and \textit{Ranging}, as all three tasks infer spatial properties from wireless measurements. Although they differ in output format, they share the common goal of extracting spatial structure from physical-layer observations \cite{li2021wirelesssensing,ma2019wificsi}.

\textbf{Activity Recognition (Act).}
Activity recognition includes tasks that infer human motion, behavior, or physical actions from wireless signals. Typical examples include gesture recognition, presence-aware sensing, and human activity classification from CSI or RF measurements \cite{li2021wirelesssensing,ma2019wificsi}.

\textbf{RF Fingerprinting (RFF).}
RF fingerprinting is treated separately because its objective is to identify or authenticate a device based on hardware-specific imperfections in its RF emissions. Unlike signal identification, which targets the communication format or protocol, RFF focuses on transmitter identity and device-specific characteristics \cite{jagannath2022rffsurvey}.

\textbf{Environment Reconstruction (EnvR).}
Environment reconstruction refers to tasks that infer a representation of the surrounding physical or radio environment from wireless observations. Examples include radio map estimation, scene reconstruction, and environment-aware modeling, where the goal is to recover structural information about the propagation space \cite{li2021wirelesssensing,zeng2024ckm}.

\subsubsection{Beam Management}
This category contains tasks related to directional transmission and reception strategies, especially in MIMO, mmWave, and massive MIMO systems. These tasks concern how to select, adapt, or optimize beams and precoders to maintain link quality and communication efficiency. In our taxonomy, this category includes:

\textbf{Beam Management.}
Beam management includes tasks such as beam prediction, beam selection, and beam tracking, all of which aim to support directional communication by choosing or adapting suitable beams under changing channel or mobility conditions \cite{xue2024beammanagement}.

\textbf{Precoding (Prc).}
Precoding is maintained as a separate task because it concerns the generation or optimization of transmit-side beamforming or spatial precoding vectors, typically in MIMO settings. While related to beam management, it is more directly tied to transmission design and signal processing \cite{brilhante2023beambeam}.

\subsubsection{Spectrum Sensing and Monitoring}
This category covers tasks that analyze spectral activity across time and frequency. In our compact formulation, this category includes:

\textbf{Spectrum Sensing.} 
This includes tasks that detect or localize spectral activity across time and frequency, covering occupancy detection and spectrum segmentation to identify active bands, occupied regions, or signal structures~\cite{yucek2009spectrumsensing}.

\textbf{Spectrum Forecasting and Monitoring.} 
This includes tasks that track or anticipate changes in spectral activity over time, covering spectrum forecasting and anomaly detection to model spectrum dynamics or identify abnormal RF patterns~\cite{kulin2018spectrumid}.


\subsubsection{Multi-Task PHY}
This category covers models explicitly evaluated on multiple physical-layer task categories, highlighting unified designs that support cross-task transfer, shared representations, or broader generalization.


\begin{table*}[t]
\centering
\scriptsize
\setlength{\tabcolsep}{3.8pt}
\renewcommand{\arraystretch}{1.1}
\caption{Taxonomic coverage matrix for physical-layer wireless foundation models.}
\label{tab:phy_compact_taxonomy}
\begin{tabular}{l c *{14}{c} c}
\toprule
\multirow{2}{*}{\textbf{Model}} & \multirow{2}{*}{\textbf{Year}} 
& \multicolumn{2}{c}{\textbf{Signal Recogn.}} 
& \multicolumn{4}{c}{\textbf{Channel Rep. Learning}} 
& \multicolumn{4}{c}{\textbf{RF Sens. \& Local.}} 
& \multicolumn{2}{c}{\textbf{Beam Mgmt.}} 
& \multicolumn{2}{c}{\textbf{Spectrum}} 
& \multirow{2}{*}{\textbf{Total}} \\
\cmidrule(lr){3-4}
\cmidrule(lr){5-8}
\cmidrule(lr){9-12}
\cmidrule(lr){13-14}
\cmidrule(lr){15-16}
& & \textbf{SID} & \textbf{Dem} 
& \textbf{CER} & \textbf{CP} & \textbf{CDiag} & \textbf{FB} 
& \textbf{LSI} & \textbf{Act} & \textbf{RFF} & \textbf{EnvR} 
& \textbf{BM} & \textbf{Prc} 
& \textbf{SS} & \textbf{SFM} & \\
\midrule

RadioFM~\cite{ott2024radiofm}
& 2024 & \no & \no & \no & \no & \no & \no & \cmark & \no & \no & \no & \no & \no & \no & \no & 1 \\

ChannelGPT~\cite{yu2024channelgpt}
& 2024 & \no & \no & \cmark & \cmark & \no & \no & \no & \no & \no & \no & \no & \no & \no & \no & 2 \\

SSRadio~\cite{aboulfotouh2024ssradio}
& 2024 & \no & \no & \no & \no & \no & \no & \no & \no & \no & \no & \no & \no & \cmark & \cmark & 2 \\

6G-RadioFM~\cite{aboulfotouh2024radio6gfm}
& 2024 & \no & \no & \no & \no & \no & \no & \no & \cmark & \no & \no & \no & \no & \cmark & \no & 2 \\

WiFo~\cite{liu2024wifo}
& 2024 & \no & \no & \no & \cmark & \no & \no & \no & \no & \no & \no & \no & \no & \no & \no & 1 \\

LWM~\cite{alikhani2024lwm}
& 2024 & \no & \no & \no & \no & \cmark & \no & \no & \no & \no & \no & \cmark & \no & \no & \no & 2 \\

6G WavesFM~\cite{aboulfotouh20256gwavesfm}
& 2025 & \cmark & \no & \cmark & \no & \no & \no & \cmark & \cmark & \no & \no & \no & \no & \no & \no & 4 \\

WirelessGPT~\cite{yang2025wirelessgpt}
& 2025 & \no & \no & \cmark & \cmark & \no & \no & \no & \cmark & \no & \cmark & \no & \no & \no & \no & 4 \\

IQFM~\cite{mashaal2025iqfm}
& 2025 & \cmark & \no & \no & \no & \no & \no & \cmark & \no & \cmark & \no & \cmark & \no & \no & \no & 4 \\

Unified FM~\cite{cheraghinia2025unifiedfm}
& 2025 & \cmark & \no & \no & \no & \cmark & \no & \cmark & \no & \no & \no & \no & \no & \no & \no & 3 \\

MUSE-FM~\cite{zheng2025musefm}
& 2025 & \no & \cmark & \cmark & \no & \no & \no & \cmark & \no & \no & \no & \no & \cmark & \no & \no & 4 \\

CSI-CLIP~\cite{jiang2025csiclip}
& 2025 & \no & \no & \no & \no & \cmark & \no & \cmark & \no & \no & \no & \cmark & \no & \no & \no & 3 \\

ICWLM~\cite{wen2025icwlm}
& 2025 & \no & \no & \no & \cmark & \no & \no & \no & \no & \no & \no & \no & \cmark & \no & \no & 2 \\

ContraWiMAE~\cite{guler2025wimae}
& 2025 & \no & \no & \cmark & \no & \cmark & \no & \no & \no & \no & \no & \cmark & \no & \no & \no & 3 \\

Filter-and-Attend~\cite{wang2025filterandattend}
& 2025 & \no & \no & \cmark & \cmark & \no & \no & \cmark & \no & \no & \no & \no & \no & \no & \no & 3 \\

LWLM~\cite{pan2025lwlm}
& 2025 & \no & \no & \no & \no & \no & \no & \cmark & \no & \no & \no & \no & \no & \no & \no & 1 \\

WiFo-CF~\cite{liu2025wifocf}
& 2025 & \no & \no & \no & \no & \no & \cmark & \cmark & \no & \no & \no & \no & \no & \no & \no & 2 \\

BERT4MIMO~\cite{catak2025bert4mimo}
& 2025 & \no & \no & \no & \cmark & \no & \no & \no & \no & \no & \no & \no & \no & \no & \no & 1 \\

MMIMO-Prc-FM~\cite{emery2025precodingfm}
& 2025 & \no & \no & \no & \no & \no & \no & \no & \no & \no & \no & \no & \cmark & \no & \no & 1 \\

SSRadioRep~\cite{kanu2025ssradiorep}
& 2025 & \cmark & \no & \no & \no & \no & \no & \cmark & \no & \no & \no & \no & \no & \no & \no & 2 \\

SpaRTran~\cite{ott2025spartran}
& 2025 & \no & \no & \no & \no & \no & \no & \cmark & \no & \no & \no & \no & \no & \no & \no & 1 \\

Tiny-WiFo~\cite{zhang2025tinywifo}
& 2025 & \no & \no & \no & \cmark & \no & \no & \no & \no & \no & \no & \no & \no & \no & \no & 1 \\

Lightweight Edge FM~\cite{cheraghinia2025lightweightedge}
& 2025 & \cmark & \no & \no & \no & \cmark & \no & \no & \no & \no & \no & \no & \no & \no & \no & 2 \\

WiCo-MG~\cite{han2025wicomg}
& 2025 & \no & \no & \cmark & \no & \no & \no & \no & \no & \no & \no & \no & \no & \no & \no & 1 \\

WiCo-PG~\cite{sun2025wicopg}
& 2025 & \no & \no & \no & \no & \no & \no & \no & \no & \no & \cmark & \no & \no & \no & \no & 1 \\

Multi-Task Prediction FM~\cite{sheng2025multitaskfm}
& 2025 & \no & \no & \no & \cmark & \no & \no & \cmark & \no & \no & \no & \no & \no & \no & \no & 2 \\

Pilot WSensing FM~\cite{serbetci2025pilotwsensing}
& 2025 & \no & \no & \no & \no & \no & \no & \no & \cmark & \no & \no & \no & \no & \no & \no & 1 \\


SpectrumFM~\cite{zhou2025spectrumfm}
& 2025 & \cmark & \no & \no & \no & \no & \no & \no & \no & \no & \no & \no & \no & \cmark & \cmark & 3 \\

Multimodal WFM~\cite{aboulfotouh2025multimodal}
& 2025 & \cmark & \no & \no & \no & \no & \no 
& \cmark & \cmark & \cmark & \no 
& \no & \no 
& \cmark & \no & 5 \\

WiFo-MUD~\cite{yang2026wifomud}
& 2026 & \no & \cmark & \no & \no & \no & \no & \no & \no & \no & \no & \no & \no & \no & \no & 1 \\

CSI-MAE~\cite{jiang2026csimae}
& 2026 & \no & \no & \cmark & \no & \no & \cmark & \cmark & \no & \no & \no & \no & \no & \no & \no & 3 \\

LWM-Spectro~\cite{kim2026lwmspectro}
& 2026 & \cmark & \no & \no & \no & \cmark & \no & \no & \no & \no & \no & \no & \no & \no & \no & 2 \\

WiFo-M$^2$~\cite{zhang2026wifom2}
& 2026 & \no & \no & \cmark & \cmark & \no & \no & \no & \no & \no & \no & \cmark & \no & \no & \no & 3 \\

WiFo-E~\cite{wen2026wifoe}
& 2026 & \no & \no & \no & \no & \no & \no & \no & \no & \no & \no & \no & \cmark & \no & \no & 1 \\

WirelessJEPA~\cite{chu2026wirelessjepa}
& 2026 & \cmark & \no & \no & \no & \no & \no & \cmark & \no & \cmark & \no & \no & \no & \no & \no & 3 \\

MMFM4WCS~\cite{yazdnian2026multimodalfm}
& 2026 & \no & \no & \cmark & \no & \no & \no & \cmark & \no & \no & \no & \no & \cmark & \no & \no & 3 \\

AM-FM~\cite{zhu2026amfm}
& 2026 & \no & \no & \no & \no & \no & \no & \cmark & \cmark & \no & \no & \no & \no & \no & \no & 2 \\

FM-RME~\cite{yang2026fmrme}
& 2026 & \no & \no & \no & \no & \no & \no & \no & \no & \no & \cmark & \no & \no & \no & \cmark & 2 \\

WiMamba~\cite{raviv2026wimamba}
& 2026 & \no & \no & \cmark & \no & \cmark & \no & \cmark & \no & \no & \no & \cmark & \no & \no & \no & 4 \\


HeterCSI~\cite{zhang2026hetercsi}
& 2026 & \no & \no & \cmark & \cmark & \no & \no
& \no & \no & \no & \no
& \no & \no
& \no & \no & 2 \\

LWM-Temporal~\cite{alikhani2026lwmtemporal}
& 2026 & \no & \no & \no & \cmark & \no & \no
& \no & \no & \no & \no
& \no & \no
& \no & \no & 1 \\

FARM~\cite{gao2026farm}
& 2026 & \no & \no & \no & \no & \no & \no
& \no & \no & \no & \cmark
& \no & \no
& \no & \cmark & 2 \\

WiFo-MiSAC~\cite{liu2026wifomisac}
& 2026 & \no & \no & \cmark & \cmark & \no & \no
& \cmark & \no & \no & \no
& \cmark & \no
& \no & \no & 4 \\

CSI-JEPA~\cite{luo2026csijepa}
& 2026 & \no & \no & \no & \no & \no & \no
& \cmark & \cmark & \no & \no
& \no & \no
& \no & \no & 2 \\

SiFo~\cite{zhao2026sifo}
& 2026 & \no & \no & \no & \no & \no & \cmark
& \no & \no & \no & \no
& \no & \no
& \no & \no & 1 \\

SPA-MAE~\cite{chen2026spamae}
& 2026 & \no & \no & \cmark & \no & \cmark & \no
& \cmark & \no & \no & \no
& \cmark & \no
& \no & \no & 4 \\

ComHymba~\cite{yang2026comhymba}
& 2026 & \no & \no & \cmark & \cmark & \cmark & \no
& \cmark & \no & \no & \cmark
& \cmark & \no
& \no & \no & 6 \\

LatentWave~\cite{mohamed2026latentwave}
& 2026 & \cmark & \no & \no & \no & \cmark & \no
& \cmark & \no & \no & \no
& \cmark & \no
& \no & \no & 4 \\

CSI-CLIP++~\cite{jiang2026csiclippp}
& 2026 & \no & \no & \no & \no & \cmark & \no
& \cmark & \no & \no & \no
& \cmark & \no
& \no & \no & 3 \\

\midrule
\textbf{Total} & -- & 10 & 2 & 15 & 13 & 10 & 3 & 24 & 8 & 3 & 5 & 11 & 5 & 4 & 4 & \textbf{117} \\
\bottomrule
\end{tabular}

\begin{tablenotes}[flushleft]
\footnotesize
\item \textbf{Abbreviations:} SID = Signal Identification; Dem = Demodulation; CER = Channel Estimation \& Recovery; CP = Channel Prediction; CDiag = Channel Diagnostics; FB = CSI Feedback; LSI = Localization \& Spatial Inference; Act = Activity Recognition; RFF = RF Fingerprinting; EnvR = Environment Reconstruction; BM = Beam Management; Prc = Precoding; SS = Spectrum Sensing, including spectrum segmentation; SFM = Spectrum Forecasting \& Monitoring, including anomaly detection and spectrum activity modeling.
\end{tablenotes}

\end{table*}

\subsection{Categorization and Analysis of WFMs by Task}
\label{subsec:phy_task_taxonomy_results}

Table~\ref{tab:phy_compact_taxonomy} summarizes the taxonomic coverage of physical-layer wireless foundation models from 2024 to 2026, considering a data cut-off of June 2026, using the compact task taxonomy introduced in this survey. 
Thus, the table highlights which physical-layer capabilities are explicitly covered by each model, including signal identification, demodulation, channel-oriented inference, RF sensing and localization, beam-related decisions, spectrum-oriented tasks, and multi-task PHY settings.

Several trends emerge from this comparison. First, the number of identified works grows markedly from only 6 papers in 2024 to 23 in 2025, corresponding to an increase of approximately 283\%. The current 2026 set already contains 20 papers, which represents 87.0\% of the 2025 count and more than three times the 2024 volume.

Second, the most frequently covered major categories are \emph{RF Sensing and Localization} and \emph{Channel Rep. Learning}, appearing in 30/49 (61.2\%) and 28/49 (57.1\%) models, respectively. In comparison, \emph{Beam Management}, \emph{Signal Recognition}, and \emph{Spectrum Sensing and Monitoring} appear less often, with 16/49 (32.7\%), 12/49 (24.5\%), and 6/49 (12.2\%) models, respectively. This suggests that the literature has so far prioritized channel-aware inference and spatial or sensing-oriented reasoning, while spectrum-oriented WFMs remain relatively limited. At the same time, 35 of the 49 models (71.4\%) cover more than one fine-grained task, showing that cross-task evaluation is already common in the WFM literature.

When each model is assigned to its dominant category, Channel Representation Learning is the most common focus (18/49, 36.7\%), followed by Multi-Task PHY (14/49, 28.6\%) and RF Sensing and Localization (9/49, 18.4\%). Signal Recognition, Spectrum, and Beam Management remain smaller groups, with 3, 3, and 2 models, respectively.

In order to discuss each PHY category in detail, we analyze the reviewed WFMs under a common methodological framework. Specifically, for each group of tasks, we examine: i) the input modality used to represent the wireless signal or channel, ii) the pretraining objective adopted to learn transferable representations, iii) the backbone architecture used to model such representations, iv) the design of pretraining and fine-tuning datasets, v) the downstream adaptation strategy, and vi) the extent to which the resulting model is evaluated under out-of-distribution conditions. This shared framework enables a more consistent comparison across PHY task categories, even when individual WFMs support multiple downstream tasks.

\subsection{Datasets}
\label{subsec:phy_datasets}



Table~\ref{tab:wfm_unified_datasets} organizes the datasets used by reviewed PHY WFMs according to task family, usage stage, signal modality, realism, accessibility, and label availability. These dimensions provide a common basis for comparing the data used for pretraining, adaptation, and evaluation, as well as the reproducibility and transfer evidence supported by each study.

Across the 66 dataset entries summarized in Table~\ref{tab:wfm_unified_datasets}, several patterns emerge. First, synthetic and real data are nearly balanced: 33 entries (50.0\%) are synthetic, 31 (47.0\%) are real, and 2 (3.0\%) combine both sources. Second, dataset reuse across the WFM pipeline is common, with 30 entries (45.5\%) used for both PT and FT/Eval, compared with 19 (28.8\%) used only for FT/Eval, 12 (18.2\%) only for PT, and 5 (7.6\%) only for evaluation. Third, channel-oriented representations dominate the dataset landscape, with 35 CSI-, CIR-, or channel-based entries (53.0\%), compared with 22 I/Q-related (33.3\%) and 9 spectrogram-related entries (13.6\%). Finally, accessibility remains a major limitation: only 22 entries (33.3\%) are explicitly public, while the remainder are private, unspecified, constructed, or generated. 

\begin{table*}[!t]
\centering
\scriptsize
\setlength{\tabcolsep}{2.0pt}
\renewcommand{\arraystretch}{0.96}
\begin{threeparttable}
\caption{Unified summary of datasets used by physical-layer wireless foundation models.}
\label{tab:wfm_unified_datasets}
\begin{tabular}{p{4cm} p{2.4cm} p{1.5cm} p{3cm} c c c c p{2.5cm}}
\toprule
\textbf{Dataset} & \textbf{Task(s)} & \textbf{Stage} & \textbf{Modality} & \textbf{R/S} & \textbf{Access} & \textbf{Origin} & \textbf{Labels} & \textbf{Used in} \\
\midrule


RF-S
& SID
& PT
& Spectrogram
& R
& N/S
& C
& Unlabeled
& \cite{aboulfotouh20256gwavesfm,aboulfotouh2025multimodal,mohamed2026latentwave} \\

RF Signal Class.~\cite{zahid2024commrad}
& SID
& FT/Eval
& Spectrogram
& R
& Pub.
& U
& Labeled
& \cite{aboulfotouh20256gwavesfm,aboulfotouh2025multimodal,mohamed2026latentwave} \\

WiFi-CSI~\cite{wang2022caution}
& SID
& PT
& CSI
& R
& N/S
& U
& Unlabeled
& \cite{aboulfotouh20256gwavesfm} \\

RADIOML2016.10A~\cite{oshea2016radioml}
& SID
& FT/Eval
& Raw IQ
& S
& Pub.
& U
& Labeled
& \cite{mashaal2025iqfm,cheraghinia2025lightweightedge,chu2026wirelessjepa,zhou2025spectrumfm} \\

RADIOML2016.10B~\cite{oshea2016radioml}
& SID
& FT/Eval
& Raw IQ
& S
& Pub.
& U
& Labeled
& \cite{zhou2025spectrumfm} \\

RADIOML2016.04C~\cite{oshea2016radioml}
& SID
& FT/Eval
& Raw IQ
& S
& Pub.
& U
& Labeled
& \cite{zhou2025spectrumfm} \\

GNSS Jamming~\cite{moralesferre2021rawiq}
& SID
& FT/Eval
& Raw IQ
& S
& Pub.
& G
& Labeled
& \cite{chu2026wirelessjepa} \\

WiFi Protocol Room A
& SID
& FT/Eval
& Raw IQ
& R
& N/S
& C
& Labeled
& \cite{chu2026wirelessjepa} \\

5G NR Interference
& SID
& FT/Eval
& Raw IQ
& R
& Priv.
& C
& Labeled
& \cite{chu2026wirelessjepa} \\

LWM-Spectro corpus
& SID
& PT
& I/Q spectrogram
& S
& N/S
& C
& Unlabeled
& \cite{kim2026lwmspectro} \\

LTE SNR/Doppler set
& SID
& FT/Eval
& I/Q spectrogram
& S
& N/S
& G
& Labeled
& \cite{kim2026lwmspectro} \\

Phoenix mixed eval.
& SID
& FT/Eval
& I/Q spectrogram
& S
& N/S
& G
& Labeled
& \cite{kim2026lwmspectro} \\

Phoenix unseen scenario
& SID
& FT/Eval
& I/Q spectrogram
& S
& N/S
& G
& Labeled
& \cite{kim2026lwmspectro} \\

TechRec / Ghent IQ~\cite{fontaine2019lowcomplexity}
& SID; SSM
& Both
& IQ timeseries
& R
& Pub.
& U
& Mixed
& \cite{cheraghinia2025unifiedfm,cheraghinia2025lightweightedge,zhou2025spectrumfm} \\

USRP B200mini IQ~\cite{subray2023realworld}
& SID; SSM
& Both
& IQ timeseries
& R
& Pub.
& U
& Mixed
& \cite{cheraghinia2025unifiedfm,cheraghinia2025lightweightedge} \\

Sub-GHz IoT IQ~\cite{fontaine2020multiband}
& SID
& Both
& IQ timeseries
& R
& Pub.
& U
& Mixed
& \cite{cheraghinia2025unifiedfm,cheraghinia2025lightweightedge} \\

OTA MIMO IQ testbed
& SID; AoA
& Both
& Raw IQ
& R
& Priv.
& C
& Labeled
& \cite{mashaal2025iqfm,chu2026wirelessjepa,aboulfotouh2025multimodal} \\

SDR multi-ant. IQ corpus
& SID; AoA
& Both
& Raw IQ
& R
& Priv.
& C
& Labeled
& \cite{kanu2025ssradiorep} \\

RML2018.01A~\cite{deepsig2018radioml}
& SID; SSM
& PT
& Raw IQ
& S
& Pub.
& U
& Labeled
& \cite{zhou2025spectrumfm} \\

SpectrumFM collected spectrum set
& SID; SSM
& PT
& Raw IQ
& R
& Priv.
& C
& N/S
& \cite{zhou2025spectrumfm} \\

POWDER RFF~\cite{reusmuns2020powder}
& RFF
& Both
& Raw IQ / spectrogram
& R
& Pub.
& U
& Mixed
& \cite{mashaal2025iqfm,chu2026wirelessjepa,aboulfotouh2025multimodal,mohamed2026latentwave} \\

WiFo-MUD demod. set
& Dem
& Both
& Wireless signal + CSI
& M
& N/S
& C
& Labeled
& \cite{yang2026wifomud} \\

MUSE multi-task set
& Dem; CER; Loc; Prc
& Both
& Scene + wireless signal
& S
& N/S
& C
& Labeled
& \cite{zheng2025musefm} \\


5G-CSI~\cite{pan2022cfrcsi}
& SID; CER
& PT
& CSI
& R
& Pub.
& U
& Unlabeled
& \cite{aboulfotouh20256gwavesfm,mohamed2026latentwave} \\

MIMO-OFDM CE~\cite{hoydis2022sionna}
& CER
& FT/Eval
& CSI / OFDM channels
& S
& Pub.
& G
& Labeled
& \cite{aboulfotouh20256gwavesfm} \\

BERT4MIMO TDL CSI corpus
& CER; CP
& Both
& CSI matrices
& S
& N/S
& G
& Unlabeled
& \cite{catak2025bert4mimo} \\

SynthSoM-U2G
& CER
& Both
& Sensing data + channel
& S
& N/S
& C
& Paired
& \cite{han2025wicomg} \\

3GPP channel sets
& CER; CP; FB; Loc
& Both
& CSI
& S
& N/S
& G
& Mixed
& \cite{jiang2026csimae,alikhani2026lwmtemporal} \\

DeepMIMO~\cite{alkhateeb2019deepmimo}
& CER; CP; CDiag; FB; Loc; BM
& Both
& CSI / channel tensors
& S
& Pub.
& G
& Mixed
& \cite{alikhani2024lwm,yang2025wirelessgpt,guler2025wimae,wang2025filterandattend,raviv2026wimamba,jiang2025csiclip,pan2025lwlm,alikhani2026lwmtemporal,zhao2026sifo,chen2026spamae,mohamed2026latentwave,jiang2026csiclippp} \\

ChannelGPT CSI time-series set
& CP
& Both
& CSI time series
& S
& N/S
& C
& Paired
& \cite{yu2024channelgpt} \\

ChannelGPT env.-channel set
& CER; CP
& Both
& Env. map + CSI
& S
& N/S
& C
& Paired
& \cite{yu2024channelgpt} \\

Traciverse
& CER; CP; EnvR
& PT
& CSI / wireless channels
& S
& N/S
& G
& Unlabeled
& \cite{yang2025wirelessgpt} \\

Sionna RT~\cite{hoydis2022sionna}
& CER; CP; CDiag; Loc; EnvR; BM
& Both
& CSI / wireless channels
& S
& N/S
& G
& Mixed
& \cite{yang2025wirelessgpt,jiang2025csiclip,jiang2026csiclippp} \\

WINNER II CE set
& CER
& FT/Eval
& CSI
& S
& N/S
& G
& Labeled
& \cite{yang2025wirelessgpt} \\

WiFo CSI sets (D1--D16)
& CP
& PT
& CSI
& S
& N/S
& G
& Unlabeled
& \cite{liu2024wifo,zhang2025tinywifo} \\

WiFo CSI test sets (D17--D19)
& CP
& Eval
& CSI
& S
& N/S
& G
& Paired
& \cite{liu2024wifo,zhang2025tinywifo} \\

QuaDRiGa channel sets~\cite{jaeckel2014quadriga}
& CER; CP; CDiag; FB; Loc; EnvR; BM; Prc
& Both
& CSI / channel tensors
& S
& N/S
& G
& Mixed
& \cite{liu2024wifo,wen2025icwlm,sheng2025multitaskfm,wen2026wifoe,zhang2026hetercsi,yang2026comhymba} \\

LH-CDF
& FB
& Both
& CSI
& M
& N/S
& C
& Mixed
& \cite{liu2025wifocf} \\

M3SC~\cite{m3sc}
& CER; CP; Loc; AoA; BM
& Both
& Sensing data + CSI
& S
& N/S
& U
& Mixed
& \cite{zhang2026wifom2,liu2026wifomisac} \\

SynthSoM~\cite{synthsom}
& CER; CP; Loc; AoA; BM
& Both
& Sensing data + CSI
& S
& N/S
& U
& Mixed
& \cite{zhang2026wifom2,liu2026wifomisac} \\

SynthSoM-Twin~\cite{synthsomtwin}
& CER; CP; BM
& Eval
& Sensing data + CSI
& S
& N/S
& U
& Labeled
& \cite{zhang2026wifom2} \\

DeepSense-6G~\cite{deepsense6g}
& CER; CP; BM
& Eval
& Sensing data + CSI
& R
& Pub.
& U
& Labeled
& \cite{zhang2026wifom2,liu2026wifomisac} \\

ViWi~\cite{viwi}
& CER; CP; BM
& Eval
& Sensing data + CSI
& S
& Pub.
& U
& Labeled
& \cite{zhang2026wifom2} \\


5G CIR industrial scenario
& Loc
& Both
& CIR fingerprints
& R
& N/S
& C
& Paired
& \cite{aboulfotouh2024ssradio} \\

5G CIR corridor scenario
& Loc
& Both
& CIR fingerprints
& R
& N/S
& C
& Paired
& \cite{aboulfotouh2024ssradio} \\

QuaDRiGa 5G CIR indoor set~\cite{jaeckel2017quadriga}
& Loc
& PT
& CIR fingerprints
& S
& N/S
& G
& Paired
& \cite{aboulfotouh2024ssradio} \\

5G NR Positioning~\cite{gao2022toward}
& Loc
& FT/Eval
& CSI
& R
& N/S
& U
& Labeled
& \cite{aboulfotouh20256gwavesfm,aboulfotouh2025multimodal,mohamed2026latentwave} \\

UWB LOS/NLOS CIR~\cite{fontaine2024transfer}
& CDiag; Loc
& Both
& CIR
& R
& Pub.
& U
& Labeled
& \cite{cheraghinia2025unifiedfm,cheraghinia2025lightweightedge} \\

RadioFM 5G CIR
& Loc
& Both
& CIR
& R
& N/S
& U
& Mixed
& \cite{ott2024radiofm} \\

KUL CSI set~\cite{bast2020csi}
& Loc
& Both
& CSI
& R
& Pub.
& U
& Paired
& \cite{ott2025spartran} \\

FH-IIS CIR set~\cite{stahlke2023fh}
& Loc
& Both
& CIR fingerprints
& R
& Pub.
& U
& Paired
& \cite{ott2025spartran} \\

HSD~\cite{yang2022efficientfi}
& Act
& FT/Eval
& CSI
& R
& Pub.
& U
& Labeled
& \cite{aboulfotouh2024radio6gfm} \\

Human Activity Sensing~\cite{yang2022efficientfi}
& Act
& Both
& CSI
& R
& N/S
& U
& Mixed
& \cite{aboulfotouh20256gwavesfm,aboulfotouh2025multimodal,mohamed2026latentwave} \\



mmWave radar corpus
& Act
& Both
& FMCW radar
& R
& N/S
& U
& Mixed
& \cite{serbetci2025pilotwsensing} \\

SoM pathloss corpus
& EnvR
& Both
& Pathloss / env. maps
& S
& N/S
& C
& Paired
& \cite{sun2025wicopg} \\

AM-FM WiFi CSI corpus
& Act
& PT
& CSI
& R
& Priv.
& C
& Unlabeled
& \cite{zhu2026amfm} \\

WiFi sensing benchmarks
& Act; Loc
& FT/Eval
& CSI
& R
& N/S
& U
& Labeled
& \cite{zhu2026amfm} \\

CSI-Bench~\cite{zhu2025csibench}
& Act; Loc
& Both
& WiFi CSI
& R
& Pub.
& C
& Mixed
& \cite{luo2026csijepa} \\


DeepBeam~\cite{klautau2021mimodata}
& BM
& FT/Eval
& Raw IQ / beam
& R
& Pub.
& U
& Labeled
& \cite{mashaal2025iqfm} \\


Real-time Radio Dataset (RRD)
& SSM
& PT
& Raw IQ $\rightarrow$ spectrogram
& R
& Priv.
& C
& Unlabeled
& \cite{aboulfotouh2024ssradio,aboulfotouh2024radio6gfm} \\

NR--LTE Segmentation Dataset
& SSM
& FT/Eval
& Spectrogram
& S
& N/S
& G
& Labeled
& \cite{aboulfotouh2024ssradio,aboulfotouh2024radio6gfm} \\

ICARUS~\cite{roy2023icarus}
& SSM
& FT/Eval
& Raw IQ
& R
& Pub.
& U
& Labeled
& \cite{aboulfotouh2025multimodal} \\

SpectrumFM anomaly set
& SSM
& FT/Eval
& Raw IQ
& R
& N/S
& C
& Labeled
& \cite{zhou2025spectrumfm} \\

ARM-Omni
& SSM; EnvR
& Both
& RSS / aerial radio maps
& S
& N/S
& G
& Mixed
& \cite{gao2026farm} \\

FM-RME datasets (D1--D6)
& SSM; EnvR
& PT
& 4D spectrum / PSD maps
& S
& N/S
& C
& Unlabeled
& \cite{yang2026fmrme} \\

FM-RME test set (D7)
& SSM; EnvR
& Eval
& 4D spectrum / PSD maps
& S
& N/S
& C
& Unlabeled
& \cite{yang2026fmrme} \\

\bottomrule
\end{tabular}

\begin{tablenotes}[flushleft]
\footnotesize
\item
\textbf{Modality:} Env. = environment; PSD = power spectral density. \textbf{Origin} U = unspecified/unclear, C = constructed/collected, or G = generated/simulated. 
\textbf{Data type:} R = real, S = synthetic/simulated and M = mixed. \textbf{Access:} Pub. = public; Priv. = private or not publicly released; N/S = not specified. 
\end{tablenotes}
\end{threeparttable}
\end{table*}

\section{Signal Recognition and Demodulation}
\label{sec:analysis_signal}

\textbf{Overview.} Signal recognition and demodulation WFMs focus on identifying, classifying, or recovering wireless signals from I/Q, spectrogram, and multimodal inputs, as summarized in Table~\ref{tab:wfm_sid_dem_overview}. I/Q remains the dominant modality, but mixed-modality inputs appear in 7 of the 12 models, showing a frequent use of enrich raw signal representations with channel or contextual information. Pretraining objectives are diverse: hybrid and masked-reconstruction strategies are the most common, with 3 models each, followed by contrastive and latent-prediction objectives with 2 models each. Architecturally, Transformer-based and hybrid designs are increasingly dominant, while pure CNNs remain in models where local I/Q structure is central. Evaluation shift remains limited: 3 models report OOD evaluation, 5 partial shift, 3 ID evaluation, and 1 ID$^{\dagger}$ setting outside a standard PT$\rightarrow$FT pipeline.

\subsection{Data Modalities and Dataset Design}


In SID/Dem WFMs, the choice of input data defines what signal information the model can use and what type of transfer can be expected. Raw I/Q is attractive because it stays close to the received waveform and avoids heavy preprocessing. \textit{IQFM}~\cite{mashaal2025iqfm} illustrates this direction by learning directly from I/Q observations for signal-understanding tasks. However, this choice also makes the model sensitive to heterogeneous systems and propagation conditions, which motivates alternatives such as the spectrogram-based design of \textit{LWM-Spectro}~\cite{kim2026lwmspectro}.

Spectrogram-based WFMs instead convert wireless signals into time--frequency inputs, which can make signal patterns easier to organize for recognition tasks~\cite{aboulfotouh20256gwavesfm,kim2026lwmspectro,mohamed2026latentwave}. \textit{Multimodal WFM}~\cite{aboulfotouh2025multimodal} extends this idea by combining spectrogram and I/Q views, testing whether time--frequency structure and waveform-level information can provide complementary cues for RF classification. However, spectrogram-based representations may hide information from the original complex signal, especially phase. This may be acceptable for signal identification, but it can be more limiting for demodulation, where symbol recovery depends more directly on phase and constellation structure.

Other modalities become relevant when the target task depends not only on the received waveform, but also on propagation or environmental context. Unified FM~\cite{cheraghinia2025unifiedfm} combines I/Q and CIR representations, allowing technology recognition to exploit both signal and channel-related information. The distinction is clearer in \textit{WiFo-MUD}~\cite{yang2026wifomud}: for multi-user demodulation, CSI is not auxiliary context, but part of the problem itself, because it describes how users' signals are mixed at the receiver.

Dataset design follows a similar trade-off. Controlled testbed data, as in \textit{IQFM}~\cite{mashaal2025iqfm}, make transfer easier to interpret but limit the diversity seen during pretraining. In contrast, heterogeneous aggregation exposes the model to variation across technologies, devices, locations, sampling rates, frequencies, and channel conditions~\cite{zhou2025spectrumfm,cheraghinia2025unifiedfm}. This broader exposure strengthens stress testing, but also makes it harder to attribute transfer gains to a specific factor, such as the input representation, dataset composition, or pretraining objective.


\subsection{Pretraining Objectives}


Pretraining objectives differ not only in downstream performance, but also in the signal assumptions they impose. Contrastive objectives show that I/Q encoders can support downstream adaptation with limited supervision, including few-shot, frozen, or fine-tuned settings~\cite{mashaal2025iqfm,kanu2025ssradiorep}. However, their transfer depends on the chosen augmentations: if a transformation removes phase, timing, amplitude, or modulation cues that are relevant to the target task, the learned invariance may become harmful. Thus, contrastive learning is viable for SID/Dem only when the imposed invariances remain physically meaningful.

\textit{WirelessJEPA}~\cite{chu2026wirelessjepa} contrasts masked latent prediction with the contrastive design of \textit{IQFM}~\cite{mashaal2025iqfm}: instead of learning from transformed versions of the same signal, the model predicts missing parts from the surrounding context in a learned representation space. \textit{LatentWave}~\cite{mohamed2026latentwave} extends this idea to spectrogram and CSI inputs, testing whether the same latent-prediction principle can also compete with direct masked reconstruction. These works are promising because they use masking without requiring the model to reconstruct the original signal explicitly. However, their gains are not uniform, since performance depends on the input modality, the masking strategy, and the downstream task.

Masked reconstruction follows another direction by forcing the model to recover missing signal structure from visible context. 
It appears in complete WFM pipelines such as \textit{6G WavesFM}~\cite{aboulfotouh20256gwavesfm} and \textit{Multimodal WFM}~\cite{aboulfotouh2025multimodal}, where reconstruction is used to learn structured representations from spectrogram, I/Q, or multimodal inputs. These works support masked reconstruction as a practical pretraining strategy, but they mainly validate full WFM designs instead of isolating reconstruction as the decisive factor. \textit{SpectrumFM}~\cite{zhou2025spectrumfm} provides clearer objective-level evidence by comparing masked reconstruction, next-slot prediction, and their combination, with the hybrid objective performing best. This suggests that SID/Dem transfer may benefit from combining local signal recovery with temporal or spectral prediction.

The objective becomes more task-specific when the problem moves from recognition to signal recovery. \textit{WiFo-MUD}~\cite{yang2026wifomud} uses diffusion-based modeling for multi-user demodulation, where the goal is not only to separate signal classes, but to recover transmitted symbols under channel mixing, interference, and SNR imbalance. In this case, the pretraining objective is closer to conditional signal recovery than to generic representation learning. \textit{MUSE-FM}~\cite{zheng2025musefm} represents another task-aware direction: its task-guided multi-objective formulation aligns the backbone with predefined communication functions, but its transfer evidence depends on how close new tasks remain to those seen during training.


\subsection{Adaptation Strategies}


Lightweight adaptation is most convincing when the downstream task is recognition-oriented because the model usually only needs to decide which class a signal belongs to, rather than recover the transmitted symbols or reconstruct the full received signal. In this setting, pretraining is useful if it has already grouped signals according to relevant patterns, such as modulation shape, technology-specific time--frequency structure, transmitter fingerprints, or spectrum occupancy.
\textit{IQFM}~\cite{mashaal2025iqfm}, \textit{WirelessJEPA}~\cite{chu2026wirelessjepa}, and \textit{LatentWave}~\cite{mohamed2026latentwave} support this interpretation through frozen, linear-probing, k-NN, or nearest-neighbor-style evaluations on recognition tasks. These protocols are useful because they test whether the pretrained features are already informative enough for class discrimination with minimal task-specific training. \textit{Multimodal WFM}~\cite{aboulfotouh2025multimodal} further shows that stronger adaptation can still be beneficial when recognition depends on cues that change across modalities, channels, SNR levels, devices, or datasets; in such cases, partial fine-tuning or LoRA can refine the representation instead of only fitting a shallow decision layer. \textit{Lightweight Edge FM}~\cite{cheraghinia2025lightweightedge}
provides a complementary efficiency-oriented case. It adapts a
patch-independent MLP encoder through a two-stage procedure,
consisting of linear probing followed by end-to-end fine-tuning, demonstrating that full task-specific adaptation
can remain inexpensive when the pretrained backbone is substantially
smaller. Moreover, the model is fine-tuned on technology classes
excluded from pretraining, providing evidence that a compact
representation can be extended to previously unseen signal classes
with limited additional complexity.

This evidence should not be directly generalized to demodulation. The reviewed works do not show that frozen encoders, linear probes, or LoRA necessarily fail for Dem; rather, they show that the adaptation problem changes when the target moves from class separation to symbol recovery. In demodulation and multi-user inference, the model must handle channel mixing, interference, noise, and user-specific signal overlap, which may not be captured by a small classifier head. \textit{MUSE-FM}~\cite{zheng2025musefm} illustrates an intermediate case, where adaptation depends not only on learned features, but also on task identity, scenario information, and communication context. \textit{WiFo-MUD}~\cite{yang2026wifomud} is even more distinct because multi-user demodulation relies on diffusion-based recovery, channel coverage, and explicit noise modeling, making it difficult to compare with standard frozen-head or LoRA-based recognition pipelines.


\subsection{Evaluation Shift}


In SID/Dem, shift labels are hard to compare because they may refer to very different changes, familiar settings mainly test whether pretrained features remain useful for recognition, whereas related but distinct settings provide stronger, though still partial, evidence of transfer. Representative cases include \textit{SpectrumFM}~\cite{zhou2025spectrumfm}, which aggregates heterogeneous signal conditions, \textit{Multimodal WFM}~\cite{aboulfotouh2025multimodal}, which evaluates recognition with complementary I/Q and spectrogram views, and \textit{LatentWave}~\cite{mohamed2026latentwave}, which evaluates RF classification after spectrogram--CSI pretraining. These settings broaden evaluation beyond familiar conditions, but they often change several factors at once, such as the dataset, modality, signal regime, or task setup. As a result, they show that the model can tolerate broader variation, but they do not always reveal whether transfer is driven by the input modality, the pretraining objective, the architecture, or the dataset composition.

\textit{IQFM}~\cite{mashaal2025iqfm} and \textit{WirelessJEPA}~\cite{chu2026wirelessjepa} provide a useful example of this evaluation practice because they compare familiar settings with external recognition targets within the raw-I/Q family. This comparison shows that OOD transfer is task-dependent: \textit{WirelessJEPA} improves over the contrastive \textit{IQFM} baseline in most OOD cases, but the gain is not uniform across all targets~\cite{chu2026wirelessjepa}. This reinforces the main point that shifted evaluation should not only report whether transfer occurs, but also indicate where it succeeds and where it breaks. For SID, a model may generalize to one external recognition task while losing strength when another dataset changes the signal structure in a different way.

Demodulation requires a separate interpretation of shift. In SID, OOD evaluation tests whether signal classes remain recognizable across datasets or conditions. In Dem, it tests whether transmitted symbols can still be recovered when the communication setting changes. \textit{WiFo-MUD}~\cite{yang2026wifomud} evaluates this second case through changes in user configuration, channel conditions, SNR imbalance, modulation, antenna settings, and interference. Its evidence is therefore valuable for Dem, but not directly comparable with SID models evaluated through classification accuracy.

SID evaluations need clearer familiar-versus-shifted protocols that isolate the source of distribution change, while Dem evaluations need stronger evidence of recovery under changed channel, interference, and user configurations.

\subsection{Key Insights and Open Challenges}

\textbf{SID and Dem provide different types of transfer evidence.}
For SID, transfer mainly means that signal classes remain separable when datasets, signal conditions, or adaptation regimes change~\cite{aboulfotouh20256gwavesfm,chu2026wirelessjepa}. For Dem, the evidence is stronger only when transmitted symbols can still be recovered under changed channel, interference, user, SNR, modulation, or antenna conditions~\cite{yang2026wifomud}. Therefore, SID and Dem should not be interpreted as equivalent evidence of foundation-model generalization: one tests robust recognition, while the other tests structured signal recovery.

\textbf{Current gains are difficult to attribute to a single design choice.}
The reviewed models differ simultaneously in modality, pretraining objective, backbone, dataset composition, and adaptation protocol. As a result, it remains unclear whether performance gains come from the wireless representation itself, the learning objective, the amount and diversity of data, or the downstream task setup. This is especially important for comparing I/Q, spectrogram-based, masked, latent-prediction, contrastive, and diffusion-based approaches, which encode different assumptions about which signal structure should be preserved.

\textbf{Stronger evaluation requires isolating the source of shift.}
Heterogeneous datasets expose models to broader variation across devices, protocols, frequencies, sampling rates, locations, and channel conditions~\cite{zhou2025spectrumfm,cheraghinia2025unifiedfm}. This improves stress testing, but it can also hide why transfer succeeds or fails. SID/Dem evaluations should therefore separate the effects of dataset shift, signal-condition shift, task shift, and adaptation strategy, instead of reporting only aggregate ID/OOD performance

\begin{table*}[t]
\centering
\footnotesize
\setlength{\tabcolsep}{3.0pt}
\renewcommand{\arraystretch}{1.08}
\begin{threeparttable}
    \caption{Overview of WFMs for signal recognition and demodulation tasks.}
\label{tab:wfm_sid_dem_overview}
\begin{tabular}{p{2.5cm} p{2.6cm} p{2cm} p{1.9cm} p{4.3cm} p{1.7cm} c}
\toprule
\textbf{Model} 
& \textbf{PT Modality/Data}
& \textbf{PT Objective} 
& \textbf{Backbone} 
& \textbf{FT Task/Data}
& \textbf{Adaptation} 
& \textbf{Eval. Shift} \\
\midrule

6G WavesFM~\cite{aboulfotouh20256gwavesfm}
& \makecell[l]{Spectrogram$^{*}$\\CSI~\cite{wang2022caution,pan2022cfrcsi}}
& Masked Rec.
& ViT
& RF cls. -- Spectrogram~\cite{zahid2024commrad}
& \makecell[l]{Frozen FT\\Partial FT\\PEFT}
& Partial \\

IQFM~\cite{mashaal2025iqfm}
& IQ$^{*}$
& Contrastive
& CNN
& Modulation cls. -- IQ$^{*}$; IQ~\cite{oshea2016radioml}
& \makecell[l]{Frozen FT\\PEFT}
& ID + OOD \\

Multimodal WFM~\cite{aboulfotouh2025multimodal}
& \makecell[l]{Spectrogram$^{*}$\\IQ$^{*}$}
& Masked Rec.
& Multimodal ViT
& RF cls. -- Spectrogram~\cite{zahid2024commrad}
& \makecell[l]{Frozen FT\\Partial FT\\PEFT}
& Partial \\

Unified FM~\cite{cheraghinia2025unifiedfm}
& \makecell[l]{IQ~\cite{fontaine2019lowcomplexity,subray2023realworld,fontaine2020multiband}\\CIR~\cite{fontaine2024transfer}}
& Masked Rec.
& Transformer
& Technology rec. -- IQ~\cite{fontaine2019lowcomplexity,subray2023realworld,fontaine2020multiband}
& \makecell[l]{Partial FT\\Frozen FT}
& ID \\

SSRadioRep~\cite{kanu2025ssradiorep}
& IQ$^{*}$
& Contrastive
& CNN
& Modulation cls. -- IQ$^{*}$
& \makecell[l]{Frozen FT\\Full FT}
& ID \\

Lightweight Edge FM~\cite{cheraghinia2025lightweightedge}
& \makecell[l]{IQ~\cite{fontaine2019lowcomplexity,subray2023realworld,fontaine2020multiband,oshea2016radioml}\\CIR~\cite{fontaine2024transfer}}
& Hybrid
& MLP
& \makecell[l]{Technology rec. -- IQ~\cite{fontaine2019lowcomplexity,subray2023realworld,fontaine2020multiband}\\Modulation cls. -- IQ~\cite{oshea2016radioml}}
& \makecell[l]{Frozen FT\\Full FT}
& ID \\

SpectrumFM~\cite{zhou2025spectrumfm}
& \makecell[l]{IQ~\cite{deepsig2018radioml,fontaine2019lowcomplexity}\\IQ$^{*}$}
& Hybrid
& CNN-Transformer
& \makecell[l]{Modulation cls. -- IQ~\cite{oshea2016radioml}\\Technology cls. -- IQ~\cite{fontaine2019lowcomplexity}}
& \makecell[l]{PEFT\\Frozen FT}
& Partial \\

LWM-Spectro~\cite{kim2026lwmspectro}
& Spectrogram$^{*}$
& Hybrid
& MoE Transformer
& Modulation cls. -- Spectrogram$^{*}$
& \makecell[l]{Frozen FT\\MT-FT}
& Partial \\

WirelessJEPA~\cite{chu2026wirelessjepa}
& IQ$^{*}$
& Latent Prediction
& CNN-JEPA
& \makecell[l]{Modulation cls. -- IQ$^{*}$; IQ~\cite{oshea2016radioml}\\Jamming cls. -- IQ~\cite{moralesferre2021rawiq}\\Protocol cls. -- IQ$^{*}$\\Interference cls. -- IQ$^{*}$}
& Frozen FT
& ID + OOD \\

MUSE-FM~\cite{zheng2025musefm}
& \makecell[l]{Scene$^{*}$\\W. signal$^{*}$}
& Task-driven Supervision
& Transformer
& \makecell[l]{MIMO det. -- Scene$^{*}$ + W. signal$^{*}$\\Decoding -- Scene$^{*}$ + W. signal$^{*}$}
& \makecell[l]{MT-FT\\Zero-shot}
& ID$^{\dagger}$ \\

WiFo-MUD~\cite{yang2026wifomud}
& \makecell[l]{Rx signal$^{*}$\\CSI~\cite{jaeckel2014quadriga,alkhateeb2019deepmimo,synthsom}}
& Generative
& \makecell[l]{Diffusion\\Transformer}
& \makecell[l]{MU demod. -- Rx signal$^{*}$ +\\ CSI~\cite{jaeckel2014quadriga,alkhateeb2019deepmimo,synthsom}}
& Zero-shot
& ID + OOD$^{\dagger}$ \\

LatentWave~\cite{mohamed2026latentwave}
& \makecell[l]{Spectrogram$^{*}$\\CSI~\cite{pan2022cfrcsi,wang2022caution}}
& Latent Prediction
& Transformer-JEPA
& RF cls. -- Spectrogram~\cite{zahid2024commrad}
& Frozen FT
& Partial \\

\bottomrule
\end{tabular}

\begin{tablenotes}[flushleft]
\footnotesize
\item \textbf{Abbreviations:} Abbreviations and shorthand terms used in this table are defined in Table~\ref{tab:abbreviations}.
\item \textbf{Data notation:} $^{*}$ indicates private, author-collected, author-generated, or not separately referenced data.
\item \textbf{Eval. Shift:} ID = same source as PT; Partial = related but distinct source; OOD = clearly different source/distribution.
\item $^{\dagger}$ The work does not follow a fully standard PT$\rightarrow$FT pipeline.
\end{tablenotes}
\end{threeparttable}
\end{table*}

\section{Channel Representation Learning}
\label{sec:analysis_channel}



\textbf{Overview.} Channel representation WFMs form one of the largest groups in the survey, centered on CSI/CIR inputs, reconstruction-oriented pretraining, and Transformer-style backbones, as summarized in Table~\ref{tab:wfm_crl_overview}. CSI or CIR appears in most models, while some models also incorporate maps, sensing data, spectrograms, RSRP measurements, or scene-aware wireless signals to incorporate spatial and contextual cues. Masked reconstruction is the dominant pretraining objective, appearing in 15 of the 30 models, followed by hybrid objectives with 5 models; other objectives appear more selectively. Architecturally, Transformer-based variants dominate, with MLP and state-space designs appearing only in isolated cases. Evaluation shift remains mostly partial: 18 models report partial-only shift, 6 are strictly in-distribution, 4 report OOD-only settings, 1 combines ID and OOD evaluation, and 1 follows an ID$^{\dagger}$ setting outside a standard PT$\rightarrow$FT pipeline.

\subsection{Data Modalities and Dataset Design}

Channel WFMs differ first in how they represent the channel input. Frequency-domain CSI is the dominant choice because it directly captures the channel response required for estimation, prediction, feedback, and diagnostics~\cite{liu2024wifo,aboulfotouh20256gwavesfm}. However, CSI is not a standardized input: its structure and reliability depend on acquisition, system configuration, and propagation conditions. \textit{Filter-and-Attend}~\cite{wang2025filterandattend} shows that degraded pilot-estimated CSI may require correction before feature extraction, meaning that representation quality depends partly on how CSI is acquired. \textit{HeterCSI}~\cite{zhang2026hetercsi} and \textit{LWM-Temporal}~\cite{alikhani2026lwmtemporal} extend this issue to heterogeneous CSI configurations and sparse temporal structure. Thus, channel pretraining should expose the model to the temporal, frequency, antenna, and scenario variations that determine how CSI changes in practice.

Some WFMs instead exploit CSI and CIR as complementary frequency- and delay-domain views of the same channel. For example, \textit{CSI-CLIP}~\cite{jiang2025csiclip} uses CSI and CIR as paired channel views to learn aligned channel representations. This is useful because CIR makes the multi-path structure more explicit by showing how signal energy arrives over different delays, which can support tasks related to propagation paths, timing, or channel diagnostics. This differs from works that use CIR time series directly as the main input for localization- or diagnostic-oriented tasks~\cite{cheraghinia2025unifiedfm}. \textit{CSI-CLIP++}~\cite{jiang2026csiclippp} extends this paired-view direction by further emphasizing CSI--CIR consistency, making the relationship between frequency-domain and delay-domain channel representations part of the pretraining design. Other channel WFMs incorporate scene, map, geometry, sensing, or location information to relate channel behavior to layout, blockage, and user position~\cite{yu2024channelgpt,zheng2025musefm,yazdnian2026multimodalfm}. Such context can provide information unavailable from CSI alone, but requires additional data to be available and aligned across scenarios.

Finally, scaling channel pretraining is useful only when it increases the diversity of represented propagation and system conditions. \textit{WiFo}~\cite{liu2024wifo} illustrates this by pretraining across heterogeneous QuaDRiGa domains with different space--time--frequency CSI patterns. This reduces dependence on a single configuration, although transfer remains constrained by which dimensions of heterogeneity are actually covered during pretraining.


    
\subsection{Pretraining and Model Architectures}
    
One key point is how to convert channel measurements into a form that can be processed without losing its physical structure. This is nontrivial because CSI is complex-valued and its dimensions depend on the wireless system used to collect it. Channel WFMs typically map CSI into tensors, patches, or tokens before applying the backbone, contributing to the prevalence of Transformer-based designs. \textit{6G WavesFM}~\cite{aboulfotouh20256gwavesfm}
uses patch-based channel embeddings with positional encodings, while \textit{WiFo}~\cite{liu2024wifo} extends this representation to space--time--frequency CSI through 3D patching. In the massive-MIMO setting, \textit{BERT4MIMO}~\cite{catak2025bert4mimo} combines temporal and feature embeddings with BERT-style masking to reconstruct the real and imaginary components of high-dimensional CSI, allowing attention to capture dependencies across subcarriers, antennas, and time. These designs show that input representation determines which physical correlations the backbone can learn.

Pretraining objectives are likewise shaped by channel-specific constraints. Masked reconstruction is common because it learns channel structure from incomplete or compressed observations~\cite{alikhani2024lwm,jiang2026csimae}. Its effectiveness, however, depends on CSI quality: pretraining on clean observations may fail to capture degradations arising during practical acquisition~\cite{wang2025filterandattend}.
Recent works refine this reconstruction paradigm by adding channel-specific structure to the masking process. \textit{SPA-MAE}~\cite{chen2026spamae}, for example, introduces physics-guided constraints into masked reconstruction, illustrating that masking can better preserve spatial, frequency, and physical channel structure. Other objectives target dependencies not captured by reconstruction alone. Contrastive learning can align complementary channel views or improve representation discriminability: \textit{CSI-CLIP}~\cite{jiang2025csiclip}, for example, aligns CSI and CIR across frequency- and delay-domain representations. \textit{ContraWiMAE}~\cite{guler2025wimae} adds a contrastive component to masked reconstruction, improving the linear separability of the learned channel representations while preserving the reconstruction-oriented pretraining signal. 

Prediction-oriented and generative objectives address cases where the pretraining design is aligned with channel forecasting or synthesis. For example, \textit{Multi-Task Prediction FM}~\cite{sheng2025multitaskfm} uses temporal prediction to capture channel evolution from historical wireless time series, while \textit{ChannelGPT}~\cite{yu2024channelgpt} generates channel-related outputs conditioned on environmental context. \textit{WiCo-MG}~\cite{han2025wicomg} extends this generative direction to cross-modal channel modeling by learning to generate multipath representations from RGB sensing observations. Its two-stage architecture first aligns sensing and channel information in related latent spaces and then uses a frequency-aware shared--routed mixture-of-experts Transformer to learn the mapping between them.

Beyond the learning objective, channel WFMs must also remain computationally feasible as the channel representation grows. As the number of antennas, users, subcarriers, and time windows increases, full attention becomes expensive. \textit{WiMamba}~\cite{raviv2026wimamba} replaces full attention with state-space modeling to reduce memory and latency, while \textit{Tiny-WiFo}~\cite{zhang2025tinywifo} targets model-size constraints. These works show that channel backbones must balance representation quality with scalability and deployment cost.

\subsection{Adaptation Under Channel Shift}

Channel adaptation is often more constrained than classification because downstream outputs may depend on channel dimensions, compression rates, or target modalities. Prediction and reconstruction remain relatively close to the input representation, whereas estimation, CSI feedback, and sensing-based tasks often require more task-specific mappings. For example, in channel prediction, \textit{WiFo}~\cite{liu2024wifo} works on a single task, but different antenna layouts, subcarrier grids, time windows, and propagation scenarios turn that same task into many different problems in practice. This motivates the WFM setting: standard prediction methods usually need a separate model for each configuration. WiFo addresses this by pretraining one model on 16 simulated CSI datasets and transferring it to new configurations without fine-tuning. However, this represents a partial shift, since test configurations reuse simulated data from pretraining instead of a different CSI source or measurement setup.
\textit{HeterCSI}~\cite{zhang2026hetercsi} goes further by directly targeting differences in CSI size and scenario. It groups CSI samples of similar size and uses masking so the model does not learn artifacts to avoid learning padding artifacts. This allows zero-shot evaluation across 12 datasets. It also reports lower prediction error (NMSE) than WiFo's zero-shot results, supporting the idea that training on diverse CSI reduces the need to retrain per configuration. However, CSI feedback introduces additional constraints because outputs must match specific channel dimensions, feedback rates, and compression requirements.
\textit{WiFo-CF}~\cite{liu2025wifocf} addresses this challenge through heterogeneous pretraining across users and feedback rates, together with an MoE architecture that supports multiple configurations without requiring a separate network for each one. The model achieves strong performance under both ID and OOD settings, including evaluations on real-world measurements in addition to simulated data. \textit{SiFo}~\cite{zhao2026sifo} follows a different form of parameter-free adaptation by constructing a calibration memory from a small set of target-site users whose signal-strength measurements are paired with reference CSI. For each new user, the model retrieves the most similar calibration users and uses their stored information to guide CSI recovery.  \textit{ICWLM}~\cite{wen2025icwlm} also avoids parameter updates, but relies on in-context adaptation: input--output examples provided at inference time specify the channel mapping required for the current configuration.

Multi-task and multimodal WFMs introduce a broader challenge: reconciling heterogeneous output spaces. \textit{6G WavesFM}~\cite{aboulfotouh20256gwavesfm} shares about 80\% of its parameters across tasks through a common ViT backbone and lightweight task-specific adaptation. \textit{WirelessGPT}~\cite{yang2025wirelessgpt} provides a useful contrast: channel estimation uses parameter-efficient fine-tuning, whereas prediction reuses a frozen backbone with a task-specific head. This suggests that outputs closer to the pretrained representation permit lighter adaptation, while estimation requires greater task-specific adjustment. \textit{MUSE-FM}~\cite{zheng2025musefm} handles heterogeneous input--output formats through a prompt-guided encoder--decoder, while \textit{WiFo-MiSAC}~\cite{liu2026wifomisac} separates shared sensing--communication information from modality-specific features through mixture-of-experts routing, avoiding a fully shared representation across modalities.

\subsection{Key Insights and Open Challenges}

\textbf{Channel WFMs must handle non-ideal CSI:}
The same CSI input can represent very different conditions depending on whether it is clean, pilot-estimated, degraded, compressed, or constrained by feedback overhead. This makes it insufficient to report channel transfer only on ideal or homogeneous CSI, because the representation may fail for reasons tied to CSI acquisition or feedback constraints instead of to propagation structure alone~\cite{wang2025filterandattend,liu2025wifocf}. Stronger evaluations should therefore vary CSI quality, dimensionality, and feedback constraints explicitly.

\textbf{The value of a pretraining objective depends on the downstream channel task:}
Different objectives preserve different types of channel information. Masked reconstruction favors recoverable CSI structure, while contrastive learning emphasizes alignment or separability; neither is automatically useful for tasks requiring different properties, such as temporal prediction or constrained CSI feedback~\cite{jiang2026csimae,jiang2025csiclip}. Combining objectives is therefore not necessarily better; the contribution of each objective should be verified through controlled ablations showing whether it improves the target downstream task under the same data, backbone, and adaptation setting.


\textbf{Channel tasks do not adapt equally:}
The adaptation bottleneck changes with the channel output required by the downstream task. Prediction-oriented models can reduce retraining when pretraining captures relevant space--time--frequency variation~\cite{liu2024wifo}, or when the current channel mapping can be inferred from examples~\cite{wen2025icwlm}. By contrast, channel estimation and CSI feedback require task-specific output mapping, because the model must satisfy CSI dimensions, feedback rates, or reconstruction targets~\cite{jiang2026csimae,liu2025wifocf}.

\begin{table*}[t]
\centering
\footnotesize
\setlength{\tabcolsep}{3.0pt}
\renewcommand{\arraystretch}{1.08}
\begin{threeparttable}
\caption{Overview of WFMs for channel representation learning tasks.}
\label{tab:wfm_crl_overview}
\begin{tabular}{p{2.5cm} p{2.6cm} p{2cm} p{1.9cm} p{4.3cm} p{1.7cm} c}
\toprule
\textbf{Model} 
& \textbf{PT Modality/Data}
& \textbf{PT Objective} 
& \textbf{Backbone} 
& \textbf{FT Task/Data}
& \textbf{Adaptation} 
& \textbf{Eval. Shift} \\
\midrule

ChannelGPT~\cite{yu2024channelgpt}
& \makecell[l]{CSI$^{*}$\\Env. map$^{*}$}
& Generative
& Transformer
& \makecell[l]{Chnl. gen. -- CSI$^{*}$ + Env. map$^{*}$\\Chnl. pred. -- CSI$^{*}$}
& PEFT
& ID \\

WiFo~\cite{liu2024wifo}
& CSI~\cite{jaeckel2014quadriga}
& Masked Rec.
& Transformer
& Chnl. pred. -- CSI~\cite{jaeckel2014quadriga}
& Zero-shot
& Partial \\

LWM~\cite{alikhani2024lwm}
& CSI~\cite{alkhateeb2019deepmimo}
& Masked Rec.
& Transformer
& Chnl. diag. -- CSI~\cite{alkhateeb2019deepmimo}
& Frozen FT
& Partial \\

6G WavesFM~\cite{aboulfotouh20256gwavesfm}
& \makecell[l]{Spectrogram$^{*}$\\CSI~\cite{wang2022caution,pan2022cfrcsi}}
& Masked Rec.
& ViT
& Chnl. est. -- OFDM~\cite{hoydis2022sionna}
& \makecell[l]{Frozen FT\\Partial FT\\PEFT}
& Partial \\

WirelessGPT~\cite{yang2025wirelessgpt}
& \makecell[l]{CSI$^{*}$\\CSI~\cite{hoydis2022sionna}}
& Masked Rec.
& Transformer
& \makecell[l]{Chnl. est. -- CSI$^{*}$\\Chnl. pred. -- CSI~\cite{jaeckel2014quadriga}}
& \makecell[l]{PEFT\\Frozen FT}
& Partial \\

Unified FM~\cite{cheraghinia2025unifiedfm}
& \makecell[l]{IQ~\cite{fontaine2019lowcomplexity,subray2023realworld,fontaine2020multiband}\\CIR~\cite{fontaine2024transfer}}
& Masked Rec.
& Transformer
& Chnl. diag. -- CIR~\cite{fontaine2024transfer}
& \makecell[l]{Partial FT\\Frozen FT}
& ID \\

MUSE-FM~\cite{zheng2025musefm}
& \makecell[l]{Scene$^{*}$\\W. signal$^{*}$}
& Task-driven Supervision
& Transformer
& Chnl. est. -- Scene$^{*}$ + W. signal$^{*}$
& \makecell[l]{MT-FT\\Zero-shot}
& ID$^{\dagger}$ \\

CSI-CLIP~\cite{jiang2025csiclip}
& CSI~\cite{alkhateeb2019deepmimo}
& Contrastive
& Dual-branch
& Chnl. diag. -- CSI~\cite{alkhateeb2019deepmimo}
& Frozen FT
& Partial \\

ICWLM~\cite{wen2025icwlm}
& CSI~\cite{jaeckel2014quadriga}
& Task-driven Supervision
& Transformer
& Chnl. pred. -- CSI~\cite{jaeckel2014quadriga}
& Zero-shot
& Partial \\

ContraWiMAE~\cite{guler2025wimae}
& CSI~\cite{alkhateeb2019deepmimo}
& Hybrid
& Transformer
& \makecell[l]{Chnl. est. -- CSI~\cite{alkhateeb2019deepmimo}\\Chnl. diag. -- CSI~\cite{alkhateeb2019deepmimo}}
& Frozen FT
& Partial \\

Filter-and-Attend~\cite{wang2025filterandattend}
& CSI~\cite{alkhateeb2019deepmimo}
& Masked Rec.
& Transformer
& \makecell[l]{Chnl. est. -- CSI~\cite{alkhateeb2019deepmimo}\\Chnl. pred. -- CSI~\cite{alkhateeb2019deepmimo}}
& Frozen FT
& Partial \\

WiFo-CF~\cite{liu2025wifocf}
& \makecell[l]{CSI$^{*}$\\CSI~\cite{jaeckel2014quadriga}}
& Masked Rec.
& MoE Transformer
& CSI feedback -- CSI$^{*}$
& Frozen FT
& ID + OOD \\

BERT4MIMO~\cite{catak2025bert4mimo}
& CSI$^{*}$
& Masked Rec.
& Transformer
& Chnl. pred. -- CSI$^{*}$
& Full FT
& ID \\

Tiny-WiFo~\cite{zhang2025tinywifo}
& CSI~\cite{jaeckel2014quadriga}
& Distillation
& Transformer
& Chnl. pred. -- CSI~\cite{jaeckel2014quadriga}
& Zero-shot
& Partial \\

Lightweight Edge FM~\cite{cheraghinia2025lightweightedge}
& \makecell[l]{IQ~\cite{fontaine2019lowcomplexity,subray2023realworld,fontaine2020multiband,oshea2016radioml}\\CIR~\cite{fontaine2024transfer}}
& Hybrid
& MLP
& Chnl. diag. -- CIR~\cite{fontaine2024transfer}
& \makecell[l]{Frozen FT\\Full FT}
& ID \\

WiCo-MG~\cite{han2025wicomg}
& \makecell[l]{Sensing$^{*}$\\CSI$^{*}$}
& Generative
& Transformer
& Chnl. est. -- Sensing$^{*}$ + CSI$^{*}$
& Frozen FT
& ID \\

Multi-Task Prediction FM~\cite{sheng2025multitaskfm}
& CSI~\cite{jaeckel2014quadriga}
& Temporal Prediction
& Transformer
& Chnl. pred. -- CSI~\cite{jaeckel2014quadriga}
& MT-FT
& Partial \\

CSI-MAE~\cite{jiang2026csimae}
& CSI$^{*}$
& Masked Rec.
& Transformer
& \makecell[l]{Chnl. est. -- CSI$^{*}$\\CSI feedback -- CSI$^{*}$}
& \makecell[l]{Frozen FT\\Zero-shot}
& Partial \\

LWM-Spectro~\cite{kim2026lwmspectro}
& Spectrogram$^{*}$
& Hybrid
& MoE Transformer
& Chnl. diag. -- Spectrogram$^{*}$
& \makecell[l]{Frozen FT\\MT-FT}
& Partial \\

WiFo-M$^2$~\cite{zhang2026wifom2}
& \makecell[l]{Sensing~\cite{m3sc,synthsom}\\CSI~\cite{m3sc,synthsom}}
& Hybrid
& Transformer
& \makecell[l]{Chnl. est. -- Sensing~\cite{synthsomtwin,deepsense6g,viwi}\\Chnl. pred. -- CSI~\cite{synthsomtwin,deepsense6g,viwi}}
& Frozen FT
& OOD \\

MMFM4WCS~\cite{yazdnian2026multimodalfm}
& \makecell[l]{Map$^{*}$\\CSI$^{*}$}
& Masked Rec.
& Transformer
& Chnl. est. -- Map$^{*}$ + CSI$^{*}$
& Frozen FT
& ID \\

WiMamba~\cite{raviv2026wimamba}
& CSI~\cite{alkhateeb2019deepmimo}
& Masked Rec.
& SSM
& \makecell[l]{Chnl. est. -- CSI~\cite{alkhateeb2019deepmimo}\\Chnl. diag. -- CSI~\cite{alkhateeb2019deepmimo}}
& Frozen FT
& Partial \\

HeterCSI~\cite{zhang2026hetercsi}
& CSI~\cite{jaeckel2014quadriga}
& Masked Rec.
& Transformer
& \makecell[l]{Chnl. rec. -- CSI~\cite{jaeckel2014quadriga}\\Chnl. pred. -- CSI~\cite{jaeckel2014quadriga}}
& Zero-shot
& Partial \\

LWM-Temporal~\cite{alikhani2026lwmtemporal}
& CSI~\cite{alkhateeb2019deepmimo}
& Masked Rec.
& Transformer
& Chnl. pred. -- CSI~\cite{alkhateeb2019deepmimo}
& Partial FT
& Partial \\

WiFo-MiSAC~\cite{liu2026wifomisac}
& \makecell[l]{Sensing~\cite{m3sc,synthsom}\\CSI~\cite{m3sc,synthsom}}
& Hybrid
& MoE Transformer
& \makecell[l]{Chnl. est. -- Sen. + CSI~\cite{m3sc,synthsom}\\Chnl. pred. -- Sen. + CSI~\cite{m3sc,synthsom}}
& Zero-shot
& OOD \\

SiFo~\cite{zhao2026sifo}
& \makecell[l]{RSRP$^{*}$\\CSI~\cite{alkhateeb2019deepmimo}}
& Task-driven Supervision
& MLP
& CSI feedback -- RSRP$^{*}$ + CSI~\cite{alkhateeb2019deepmimo}
& Zero-shot$^{\dagger}$
& OOD \\

SPA-MAE~\cite{chen2026spamae}
& CSI~\cite{alkhateeb2019deepmimo}
& Masked Rec.
& Transformer
& \makecell[l]{Chnl. est. -- CSI~\cite{alkhateeb2019deepmimo}\\Chnl. diag. -- CSI~\cite{alkhateeb2019deepmimo}}
& Frozen FT
& Partial \\

ComHymba~\cite{yang2026comhymba}
& CSI~\cite{jaeckel2014quadriga}
& Masked Rec.
& SSM--Transformer
& \makecell[l]{Chnl. rec. -- CSI~\cite{jaeckel2014quadriga}\\Chnl. diag. -- CSI~\cite{jaeckel2014quadriga}}
& Frozen FT
& Partial \\

LatentWave~\cite{mohamed2026latentwave}
& \makecell[l]{Spectrogram$^{*}$\\CSI~\cite{pan2022cfrcsi}}
& \makecell[l]{Latent\\Prediction}
& Transformer
& Chnl. diag. -- CSI~\cite{alkhateeb2019deepmimo}
& Frozen FT
& Partial \\

CSI-CLIP++~\cite{jiang2026csiclippp}
& \makecell[l]{CSI~\cite{alkhateeb2019deepmimo}\\CIR~\cite{alkhateeb2019deepmimo}}
& Contrastive
& \makecell[l]{Dual-branch\\Transformer}
& Chnl. diag. -- CSI~\cite{alkhateeb2019deepmimo}
& Frozen FT
& OOD \\

\bottomrule
\end{tabular}

\begin{tablenotes}[flushleft]
\footnotesize
\item \textbf{Data notation:} $^{*}$ indicates private, author-collected, author-generated, or not separately referenced data.
\item \textbf{Eval. Shift:} ID = same source as PT; Partial = related but distinct source; OOD = clearly different source/distribution.
\item $^{\dagger}$ The work does not follow a fully standard PT$\rightarrow$FT pipeline; in SiFo, target-site adaptation is calibration-aided but does not update model parameters.
\end{tablenotes}
\end{threeparttable}
\end{table*}

\section{RF Sensing and Localization}
\label{sec:analysis_rfsensing}

\textbf{Overview.} RF sensing and localization WFMs use wireless observations to infer spatial, environmental, or identity-related information, as summarized in Table~\ref{tab:wfm_rf_overview}. Their inputs are highly diverse, spanning CSI/CIR, I/Q, spectrograms, sensing data, and environmental or radio maps. Masked reconstruction is the dominant pretraining objective, appearing in 15 of the 30 models, followed by hybrid (5), contrastive (4), and masked latent-prediction (3) objectives, with other strategies appearing less frequently. Transformer-based backbones dominate, while CNNs and SSMs are less common. Evaluation remains mostly partial: 13 models report partial-only shift, 8 are strictly ID, 6 OOD-only, 2 combine ID and OOD evaluation, and 1 follows an ID$^{\dagger}$ setting outside a standard PT$\rightarrow$FT pipeline.


\subsection{Data Modalities}


Modality choice determines how explicitly spatial and propagation information is exposed to the model before learning begins. 

Waveform or signal-level inputs such as I/Q samples, spectrograms, contain useful phase, frequency, and temporal patterns, but the relationship between those patterns and the sensing task is not explicit in the signal itself. The model must learn which parts of the signal are informative for inferring the target sensing output.
\textit{IQFM}~\cite{mashaal2025iqfm} and WirelessJEPA~\cite{chu2026wirelessjepa} illustrate this setting by using raw I/Q representations for tasks including AoA estimation and RF fingerprinting~\cite{mashaal2025iqfm}.
Channel-structured inputs, especially CSI and CIR, reduce part of this representational gap because they are already organized around propagation geometry. For example, \textit{CSI-CLIP}~\cite{jiang2025csiclip} supports positioning through CSI--CIR alignment, while \textit{CSI-MAE}~\cite{jiang2026csimae} uses masked CSI autoencoding for cross-scenario localization. However, these inputs still require the model to extract spatial meaning from the channel measurements. \textit{CSI-JEPA}~\cite{luo2026csijepa} further evaluates CSI representations on CSI collected from real WiFi sensing scenarios rather than only simulated channel data, across activity, localization, and user-related tasks.


A recent work~\cite{aboulfotouh2025multimodal} bridges these modality groups using a ViT-style backbone with masked wireless modeling, to support different RF sensing and localization inputs, such as CIR for UWB indoor/industrial localization, CSI for positioning and human activity recognition, I/Q for fingerprinting, and spectrograms for RF classification. A contrasting case is \textit{Pilot-WSensing}~\cite{serbetci2025pilotwsensing}: it uses FMCW radar and shows that WFMs pretrained on other modalities do not necessarily perform well on radar. This is likely due to the unique chirp structure, range-Doppler representations, and temporal-spatial characteristics that differ from communication-centric RF modalities. 
\textit{WiFo-MiSAC}~\cite{liu2026wifomisac} adds a stronger multimodal sensing case, built on synchronized CSI, radar, and map data with more than 1B complex CSI entries and over 200k CSI--radar--map triplets. Given this heterogeneity, it uses a shared-specific mixture-of-experts design to separate common and modality-specific information, reducing cross-modal interference from undifferentiated fusion.



Furthermore, other models incorporate spatial or environmental context more explicitly. For instance, \textit{MMFM4WCS}~\cite{yazdnian2026multimodalfm} combines partial CSI with explicit scene-geometry inputs, represented as building-height/occupancy maps, and learns cross-modal physical representations for tasks such as localization and MIMO precoding. \textit{MUSE-FM}~\cite{zheng2025musefm} is another work that explicitly adds environmental context as a multimodal input and uses scene graphs as prior knowledge about environment and channel distributions. Another example is \textit{WiCo-PG}~\cite{sun2025wicopg}, which generates pathloss maps from UAV imagery and carrier-frequency information.   


These richer inputs can improve sensing or localization performance, but they also change how results should be interpreted, as the gains may be coming from stronger RF representation learning, or from the additional spatial/environmental information already present in the input.



\subsection{Pretraining and Model Architectures}


Masked reconstruction is the dominant pretraining objective in RF sensing WFMs, but the choice of what to mask reflects different assumptions about where useful RF structure lives: in propagation delays, time--frequency patterns, or spatial field distributions. What varies is what gets masked and why. In \textit{RadioFM}~\cite{ott2024radiofm}, masking and dropping CIR inputs encourage the model to learn propagation structure useful for 5G indoor localization. In \textit{6G-RadioFM}~\cite{aboulfotouh2024radio6gfm}, masked spectrogram modeling applied with a ViT backbone targets time--frequency RF patterns for human activity sensing and spectrogram segmentation. \textit{FM-RME}~\cite{yang2026fmrme} extends this principle to spatial RF fields by reconstructing partial PSD and spectrum-map observations across space, time, and frequency simultaneously.

Some works push further by embedding RF-specific assumptions directly into the reconstruction objective. Rather than treating missing components as generic tokens to recover, \textit{SpaRTran}~\cite{ott2025spartran} introduces a sparsity bias motivated by radio propagation, encouraging representations in which dominant paths or components are captured instead of treating missing observations as generic tokens. This illustrates how wireless inductive biases can turn reconstruction into a more physically informed learning objective.



Contrastive pretraining poses a different requirement: the chosen views must preserve the RF cues needed by the target task. This is why IQFM uses task-aware augmentations, and SSRadioRep reports that carefully designed augmentations are important for learning transferable radio representations~\cite{mashaal2025iqfm,kanu2025ssradiorep}. For example, fingerprinting depends on device-specific signal patterns, so augmentations that erase those patterns can weaken the task. Localization and AoA estimation depend on timing, phase, and antenna relationships, so views that distort these properties can also be harmful. The key is choosing augmentations that keep the RF information needed for the sensing task. On the other hand, some works respond to the limitations of single objectives by combining them. \textit{LWLM}~\cite{pan2025lwlm} introduces a hybrid objective for localization that combines masked channel modeling, domain-transformation invariance, and position-invariant contrastive learning. These components respectively capture CSI structure, consistency across channel representations, and location-relevant information, improving ToA/AoA and single-/multi-BS localization, including label-limited and unseen-BS settings. 
Beyond hybrid objectives, masked latent prediction offers another way to train sensing-oriented WFMs by predicting hidden regions in representation space instead of reconstructing the original signal. \textit{WirelessJEPA}~\cite{chu2026wirelessjepa} applies this idea to multi-antenna I/Q signals arranged as antenna--time grids: hiding antennas tests spatial reasoning, while hiding time segments tests temporal reasoning. The model is evaluated on six downstream tasks and reports stronger OOD performance than contrastive baselines. \textit{CSI-JEPA}~\cite{luo2026csijepa} applies the same principle to time--subcarrier regions in real WiFi CSI, reporting substantial gains across seven sensing tasks and up to 98\% label savings. \textit{LatentWave}~\cite{mohamed2026latentwave} further shows that the choice of hidden region changes the downstream behavior: frequency masking favors positioning and beam prediction, while region masking is better for RF classification. Temporal prediction introduces another learning signal for localization-oriented sensing. \textit{Multi-Task Prediction FM}~\cite{sheng2025multitaskfm} uses CSI and location series to model localization as a sequential prediction problem instead of as a single-shot inference task. This makes it relevant to RF sensing because the learned representation is encouraged to capture temporal channel and mobility dynamics, although its evidence remains partial because transfer is evaluated within related CSI-based localization settings.


Architecturally, Transformer and ViT-style backbones dominate RF sensing WFMs because they accommodate tokenized CSI, CIR, spectrograms, maps, and scene-aware inputs within a common framework~\cite{ott2024radiofm,aboulfotouh2024radio6gfm}. This flexibility makes attention-based models suitable for heterogeneous sensing modalities, but it does not imply that attention is always the most efficient choice. \textit{WiMamba}~\cite{raviv2026wimamba} provides a useful counterpoint by replacing attention with a bidirectional Mamba backbone based on selective state-space models, aiming to reduce latency and memory cost for long CSI token sequences while preserving structured channel dependencies.


\subsection{Model Adaptation and Efficiency}

Efficient adaptation is closely tied to the amount of new labeled data or reference measurements required in the target environment. For localization, this usually means collecting new reference positions, fingerprints, or labeled channel measurements. \textit{RadioFM}~\cite{ott2024radiofm} addresses this by reaching comparable localization accuracy with ten times less reference data, reducing the cost of fingerprinting campaigns. \textit{LWLM}~\cite{pan2025lwlm} follows the same direction by improving label-limited fine-tuning and unseen-BS localization settings. \textit{AM-FM}~\cite{zhu2026amfm} extends this direction to WiFi sensing by pretraining a single backbone on 9.2 million unlabeled CSI samples across diverse devices and environments and transferring it to nine downstream tasks, reducing downstream labeling requirements.

A related efficiency problem arises in radio-map reconstruction, where cost lies not only in labels but also in dense spatial measurements. \textit{FARM}~\cite{gao2026farm} estimates aerial radio maps from sparse RSS observations and evaluates zero-shot transfer under three OOD shifts: unseen altitude/coverage settings, carrier frequencies, and antenna patterns. It consistently outperforms RadioUNet across these shifts, showing pretrained representations can reduce dependence on dense target-site measurements under changing deployment conditions.

A second consideration is the compute and memory cost of adapting to multiple tasks. \textit{6G WavesFM}~\cite{aboulfotouh20256gwavesfm} shares one pretrained backbone across sensing, communication, and localization using task-specific heads and LoRA, reducing the need for separate models. Its reported convergence speedup of up to 5$\times$ further suggests that broader pretraining can reduce downstream training cost. \textit{CSI-MAE}~\cite{jiang2026csimae} uses lightweight decoder fine-tuning and zero-shot transfer to reduce training overhead under cross-scenario reuse. \textit{WiMamba}~\cite{raviv2026wimamba} addresses inference efficiency by replacing full attention with state-space modeling for long CSI sequences. \textit{ComHymba}~\cite{yang2026comhymba} follows a related direction by combining windowed attention with state-space modules, reporting gains across channel reconstruction, environmental sensing, and beam management, with up to 3.3$\times$ inference speedup over Transformer backbones. 

\subsection{Evaluation Shift}


In RF sensing and localization, evaluation shift should test whether a WFM remains useful when the sensing situation changes, not only when the data are similar to pretraining. This distinction matters because sensing performance can change when the physical setting changes, including the environment layout, device placement, user behavior, or sensing modality. Therefore, the most informative shift in this category is not simply a new dataset, but a new sensing situation.

Most reviewed works still provide limited evidence for this type of shift. \textit{RadioFM}~\cite{ott2024radiofm} shows that pretraining can support indoor localization with fewer reference data, but the evaluation remains close to the original sensing setup. \textit{CSI-CLIP}~\cite{jiang2025csiclip} provides a partial-shift case through positioning evaluation, reporting a 22\% reduction in mean positioning error. \textit{CSI-CLIP++}~\cite{jiang2026csiclippp} extends this line by testing CSI--CIR pretraining across different environments, carrier frequencies, and data scales, and by including cross-simulator positioning transfer on Sionna RT. This makes the shift evidence stronger than in the original \textit{CSI-CLIP}, but it is still mainly evidence of channel-based positioning transfer, not a controlled study of changes in room layout, device placement, user behavior, or obstacles. Other channel-based models also report transfer across related scenarios or configurations, but they do not systematically isolate changes in the physical sensing environment~\cite{yang2025wirelessgpt,jiang2026csimae}.

A smaller set of works gives stronger evidence, but each tests a different kind of shift. For instance, \textit{6G-RadioFM}~\cite{aboulfotouh2024radio6gfm}, evaluates CSI-based activity recognition after spectrogram-based pretraining, introducing both modality and sensing-task shift. 
\textit{AM-FM}~\cite{zhu2026amfm} evaluates sensing across broader WiFi conditions, making it closer to practical RF sensing variability. \textit{IQFM}~\cite{mashaal2025iqfm} and \textit{WirelessJEPA}~\cite{chu2026wirelessjepa} are useful in a different way: they include ID and OOD comparisons, which make it possible to see whether performance changes under shifted conditions. However, these works still lack a systematic analysis of controlled changes in room geometry, user behavior, device placement, obstacles, or sensing layout.


\subsection{Key Insights and Open Challenges}




\textbf{JEPA behavior is strongly shaped by the prediction target:}
JEPA-based WFMs show that latent prediction does not produce a task-agnostic representation by default. Different masking geometries emphasize different wireless dependencies and can favor different downstream tasks, indicating that the usefulness of the learned representation depends strongly on what physical structure is selected as the prediction target~\cite{chu2026wirelessjepa,mohamed2026latentwave}.

\textbf{Multimodal gains remain difficult to attribute:}
RF sensing can benefit from complementary I/Q, CSI, spectrogram, radar, or environmental inputs, but their individual contribution remains difficult to isolate~\cite{aboulfotouh2025multimodal}. Modality ablations and missing-modality evaluations are therefore needed to distinguish gains from stronger representation learning from those obtained simply by providing more informative inputs.

\textbf{Adaptation efficiency has multiple dimensions:}
For RF localization, efficiency is not only determined by the number of trainable parameters, but also by the amount of labeled target-domain data and the computational cost of adaptation and inference~\cite{ott2024radiofm,pan2025lwlm}. These dimensions should be reported separately to avoid treating parameter efficiency as a complete measure of adaptation cost.

\textbf{Physical-environment shift remains under-evaluated:}
RF sensing generalization is strongly affected by changes in layout, device placement, user behavior, and sensing geometry, yet such factors are rarely isolated in current evaluations~\cite{aboulfotouh2024radio6gfm,zhu2026amfm}. Dataset, modality, and task shifts should therefore be distinguished from controlled changes in the physical sensing environment.

\begin{table*}[t]
\centering
\footnotesize
\setlength{\tabcolsep}{3.0pt}
\renewcommand{\arraystretch}{1.08}
\begin{threeparttable}
\caption{Overview of WFMs for RF sensing and localization tasks.}
\label{tab:wfm_rf_overview}
\begin{tabular}{p{2.5cm} p{2.6cm} p{2cm} p{1.9cm} p{4.3cm} p{1.7cm} c}
\toprule
\textbf{Model} 
& \textbf{PT Modality/Data}
& \textbf{PT Objective} 
& \textbf{Backbone} 
& \textbf{FT Task/Data}
& \textbf{Adaptation} 
& \textbf{Eval. Shift} \\
\midrule

RadioFM~\cite{ott2024radiofm}
& CIR$^{*}$
& Masked Rec.
& Transformer
& Indoor loc. -- CIR$^{*}$
& \makecell[l]{Partial FT\\Frozen FT}
& ID \\

6G-RadioFM~\cite{aboulfotouh2024radio6gfm}
& Spectrogram$^{*}$
& Masked Rec.
& ViT
& Activity rec. -- CSI~\cite{yang2022efficientfi}
& Frozen FT
& OOD \\

6G WavesFM~\cite{aboulfotouh20256gwavesfm}
& \makecell[l]{Spectrogram$^{*}$\\CSI~\cite{wang2022caution,pan2022cfrcsi}}
& Masked Rec.
& ViT
& \makecell[l]{Activity rec. -- CSI~\cite{yang2022efficientfi}\\Localization -- CSI~\cite{gao2022toward}}
& \makecell[l]{Frozen FT\\Partial FT\\PEFT}
& Partial \\

Multimodal WFM~\cite{aboulfotouh2025multimodal}
& \makecell[l]{Spectrogram$^{*}$\\IQ$^{*}$}
& Masked Rec.
& Multimodal ViT
& \makecell[l]{Activity rec. -- CSI~\cite{yang2022efficientfi}\\Positioning -- CSI~\cite{gao2022toward}\\RF fingerprinting -- IQ~\cite{reusmuns2020powder}}
& \makecell[l]{Frozen FT\\Partial FT\\PEFT}
& Partial \\

WirelessGPT~\cite{yang2025wirelessgpt}
& \makecell[l]{CSI$^{*}$\\CSI~\cite{hoydis2022sionna}}
& Masked Rec.
& Transformer
& \makecell[l]{Activity rec. -- CSI$^{*}$\\Env. recon. -- CSI$^{*}$}
& \makecell[l]{Frozen FT\\PEFT}
& Partial \\

IQFM~\cite{mashaal2025iqfm}
& IQ$^{*}$
& Contrastive
& CNN
& \makecell[l]{AoA -- IQ$^{*}$\\RF fingerprinting -- IQ~\cite{reusmuns2020powder}}
& \makecell[l]{Frozen FT\\PEFT}
& ID + OOD \\

Unified FM~\cite{cheraghinia2025unifiedfm}
& \makecell[l]{IQ~\cite{fontaine2019lowcomplexity,subray2023realworld,fontaine2020multiband}\\CIR~\cite{fontaine2024transfer}}
& Masked Rec.
& Transformer
& \makecell[l]{Localization -- CIR~\cite{fontaine2024transfer}\\Range corr. -- CIR~\cite{fontaine2024transfer}}
& \makecell[l]{Partial FT\\Frozen FT}
& ID \\

MUSE-FM~\cite{zheng2025musefm}
& \makecell[l]{Scene$^{*}$\\W. signal$^{*}$}
& Task-driven Supervision
& Transformer
& Localization -- Scene$^{*}$ + W. signal$^{*}$
& \makecell[l]{MT-FT\\Zero-shot}
& ID$^{\dagger}$ \\

CSI-CLIP~\cite{jiang2025csiclip}
& CSI~\cite{alkhateeb2019deepmimo}
& Contrastive
& Dual-branch
& Positioning -- CSI~\cite{alkhateeb2019deepmimo}
& Frozen FT
& Partial \\

Filter-and-Attend~\cite{wang2025filterandattend}
& CSI~\cite{alkhateeb2019deepmimo}
& Masked Rec.
& Transformer
& Localization -- CSI~\cite{alkhateeb2019deepmimo}
& Frozen FT
& Partial \\

LWLM~\cite{pan2025lwlm}
& CSI~\cite{alkhateeb2019deepmimo}
& Hybrid
& Transformer
& \makecell[l]{ToA -- CSI~\cite{alkhateeb2019deepmimo}\\AoA -- CSI~\cite{alkhateeb2019deepmimo}\\Single-/multi-BS loc. -- CSI~\cite{alkhateeb2019deepmimo}}
& Frozen FT
& Partial \\

WiFo-CF~\cite{liu2025wifocf}
& \makecell[l]{CSI$^{*}$\\CSI~\cite{jaeckel2014quadriga}}
& Masked Rec.
& MoE Transformer
& Indoor loc. -- CSI$^{*}$
& Frozen FT
& ID \\

SSRadioRep~\cite{kanu2025ssradiorep}
& IQ$^{*}$
& Contrastive
& CNN
& AoA -- IQ$^{*}$
& \makecell[l]{Frozen FT\\Full FT}
& ID \\

SpaRTran~\cite{ott2025spartran}
& \makecell[l]{CIR~\cite{stahlke2023fh}\\CSI~\cite{bast2020csi}}
& Masked Rec.
& Transformer
& Localization -- CIR~\cite{stahlke2023fh} + CSI~\cite{bast2020csi}
& Frozen FT
& ID \\

WiCo-PG~\cite{sun2025wicopg}
& \makecell[l]{Pathloss$^{*}$\\Env. map$^{*}$}
& Generative
& Transformer
& Env. recon. -- Pathloss$^{*}$ + Env. map$^{*}$
& Frozen FT
& ID \\

Multi-Task Prediction FM~\cite{sheng2025multitaskfm}
& \makecell[l]{CSI~\cite{jaeckel2014quadriga}\\Location series$^{*}$}
& Temporal Prediction
& Transformer
& Localization -- CSI~\cite{jaeckel2014quadriga}
& \makecell[l]{MT-FT\\Zero-shot}
& Partial \\

Pilot WSensing FM~\cite{serbetci2025pilotwsensing}
& FMCW radar$^{*}$
& Hybrid 
& Transformer
& Activity rec. -- FMCW radar$^{*}$
& Frozen FT
& ID \\


CSI-MAE~\cite{jiang2026csimae}
& CSI$^{*}$
& Masked Rec.
& Transformer
& Localization -- CSI$^{*}$
& \makecell[l]{Frozen FT\\Zero-shot}
& Partial \\

WirelessJEPA~\cite{chu2026wirelessjepa}
& IQ$^{*}$
& Latent Prediction
& CNN-JEPA
& \makecell[l]{AoA -- IQ$^{*}$\\RF fingerprinting -- IQ~\cite{reusmuns2020powder}}
& Frozen FT
& ID + OOD \\

MMFM4WCS~\cite{yazdnian2026multimodalfm}
& \makecell[l]{Map$^{*}$\\CSI$^{*}$}
& Masked Rec.
& Transformer
& Localization -- Map$^{*}$ + CSI$^{*}$
& Frozen FT
& ID \\

AM-FM~\cite{zhu2026amfm}
& CSI$^{*}$
& Hybrid
& Transformer
& \makecell[l]{Activity rec. -- CSI$^{*}$\\Localization -- CSI$^{*}$}
& Frozen FT
& OOD \\

FM-RME~\cite{yang2026fmrme}
& \makecell[l]{PSD map$^{*}$\\Spectrum map$^{*}$}
& Masked Rec.
& Geometry-aware Transformer
& Env. recon. -- PSD map$^{*}$ + Spectrum map$^{*}$
& Zero-shot
& Partial \\

WiMamba~\cite{raviv2026wimamba}
& CSI~\cite{alkhateeb2019deepmimo}
& Masked Rec.
& SSM (Mamba)
& Localization -- CSI~\cite{alkhateeb2019deepmimo}
& Frozen FT
& Partial \\

WiFo-MiSAC~\cite{liu2026wifomisac}
& \makecell[l]{Sensing~\cite{m3sc,synthsom}\\CSI~\cite{m3sc,synthsom}}
& Hybrid
& MoE Transformer
& \makecell[l]{Distance est. -- Sen. + CSI~\cite{m3sc,synthsom}\\AoA est. -- Sen. + CSI~\cite{m3sc,synthsom}}
& Zero-shot
& OOD \\

CSI-JEPA~\cite{luo2026csijepa}
& CSI~\cite{zhu2025csibench}
& \makecell[l]{Latent\\Prediction}
& Transformer
& \makecell[l]{Activity rec. -- CSI~\cite{zhu2025csibench}\\Localization/sensing -- CSI~\cite{zhu2025csibench}}
& Frozen FT
& OOD \\

FARM~\cite{gao2026farm}
& \makecell[l]{RSS maps$^{*}$\\Aerial radio maps$^{*}$}
& Hybrid
& Transformer
& Env. recon. -- Aerial radio maps$^{*}$
& Zero-shot
& OOD \\

SPA-MAE~\cite{chen2026spamae}
& CSI~\cite{alkhateeb2019deepmimo}
& Masked Rec.
& Transformer
& \makecell[l]{Positioning -- CSI~\cite{alkhateeb2019deepmimo}\\LoS/NLoS -- CSI~\cite{alkhateeb2019deepmimo}}
& Frozen FT
& Partial \\

ComHymba~\cite{yang2026comhymba}
& CSI~\cite{jaeckel2014quadriga}
& Masked Rec.
& SSM--Transformer
& Env. sensing -- CSI~\cite{jaeckel2014quadriga}
& Frozen FT
& Partial \\

LatentWave~\cite{mohamed2026latentwave}
& \makecell[l]{Spectrogram$^{*}$\\CSI~\cite{pan2022cfrcsi}}
& \makecell[l]{Latent\\Prediction}
& Transformer
& \makecell[l]{Positioning -- CSI~\cite{gao2022toward}\\LoS/NLoS -- CSI~\cite{alkhateeb2019deepmimo}}
& Frozen FT
& Partial \\

CSI-CLIP++~\cite{jiang2026csiclippp}
& \makecell[l]{CSI~\cite{alkhateeb2019deepmimo}\\CIR~\cite{alkhateeb2019deepmimo}}
& Contrastive
& \makecell[l]{Dual-branch\\Transformer}
& Positioning -- CSI~\cite{alkhateeb2019deepmimo}
& Frozen FT
& Partial \\

\bottomrule
\end{tabular}

\begin{tablenotes}[flushleft]
\footnotesize
\item \textbf{Data notation:} $^{*}$ indicates private, author-collected, author-generated, or not separately referenced data.
\item \textbf{Eval. Shift:} ID = same source as PT; Partial = related but distinct source; OOD = clearly different source/distribution.
\item $^{\dagger}$ The work does not follow a fully standard PT$\rightarrow$FT pipeline.
\end{tablenotes}
\end{threeparttable}
\end{table*}

\section{Beam Management}
\label{sec:analysis_beam}

\textbf{Overview.} Beam management WFMs rely primarily on channel-aware and spatially informative inputs that capture propagation geometry and directional transmission, as summarized in Table~\ref{tab:wfm_beam_overview}. CSI-centered inputs dominate, either alone or combined with CIR, sensing, maps, spectrograms, or scene context, while I/Q-only inputs are less common. Masked reconstruction is the leading pretraining objective (7 of 16 models), followed by contrastive and hybrid approaches (3 each). Evaluation is predominantly partial-shift, reported by 10 models, compared with 2 ID, 3 OOD, and 1 ID$^{\dagger}$ setting.

\subsection{Data Modalities and Data Design}

In beam management and precoding WFMs, beam indices or precoders are usually treated as downstream targets, not as pretraining modalities, because they are tied to a codebook, antenna configuration, or optimization criterion. In contrast, CSI and CIR provide reusable channel structure that can support beam-related decisions beyond a single beam-label space. This leads to a first design strategy, \emph{channel-centric} where the model is pretrained mainly from measured or simulated channel representations and is later adapted to beam prediction or precoding. For example, \textit{LWM}~\cite{alikhani2024lwm} pretrained a task-agnostic channel model whose features are reused for downstream taks, such as sub-6~GHz to mmWave beam prediction. The strength of this design is that the input is already close to the physical quantity used for beam selection. However, this design also inherits a channel-distribution dependence: transfer depends on whether the pretrained representation covers the target channel conditions and configurations. This limitation is explicit in \textit{CSI-CLIP}~\cite{jiang2025csiclip}, which reports gains on beam management across scenarios and frequencies under compatible CSI configurations, but notes weaker generalization when the number of transmit/receive antennas or subcarriers differs from the training configuration.

A second design strategy is \emph{environment-aware}, where the model incorporates contextual information about the propagation environment without requiring the representation to be learned from a single channel observation alone. The key difference is not simply the use of additional modalities, but the construction of paired channel--context data. \textit{MMFM4WCS}~\cite{yazdnian2026multimodalfm} makes this assumption explicit by pretraining with CSI, a static 3D environment representation, and relative user-location information, aiming to learn physics-aware representations that transfer to tasks such as hybrid MIMO precoding. \textit{WiFo-M$^2$}~\cite{zhang2026wifom2}, learns out-of-band channel-aware features from paired multimodal sensing and CSI through contrastive pretraining to support PHY actions such as beam prediction when direct channel observations are incomplete, delayed, or configuration-dependent. 
The strength of this design is that environmental context may improve robustness when the channel observation alone is incomplete. Its limitation is that the benefit is conditional: if the contextual modality is weakly aligned with the channel or unavailable at deployment time, the added complexity does not guarantee better transfer.


\subsection{Pretraining and Model Architectures}

Beam management is a stricter test of channel representations than generic channel tasks because the model must preserve the channel differences that change the beam decision. This becomes visible when pretrained channel embeddings are reused for beam prediction, as in \textit{LWM}~\cite{alikhani2024lwm}: the representation is useful only if it retains the information needed to select the correct beam. \textit{CSI-CLIP}~\cite{jiang2025csiclip} provides a clear warning in this direction: CSI--CIR alignment reduces positioning error by 22\%, but improves beam-management accuracy by only 1\% over supervised methods. However, \textit{CSI-CLIP++}~\cite{jiang2026csiclippp} shows that this limitation is not inherent to CSI--CIR contrastive pretraining. By scaling the CSI--CIR alignment setup and evaluating across broader DeepMIMO settings, it reports beam-prediction Top-1 gains of up to 19.31 percentage points. This suggests one possible remedy: an alignment objective can become more useful for beam prediction when it is exposed to enough beam-relevant channel variation, although this still depends on whether the training setup covers the distinctions induced by the target codebook.

\textit{ContraWiMAE}~\cite{guler2025wimae} follows a more beam-oriented route. Instead of relying only on channel reconstruction or generic alignment, it combines masked channel modeling with a contrastive component designed to improve downstream separability. Its beam-selection results show why this matters under scarce labels and harder beam decisions. For codebook size 32, \textit{WiMAE} reaches 39.9\% top-1 accuracy with only 1\% of the training data, outperforming \textit{LWM} and raw-channel baselines; adding the contrastive component in \textit{ContraWiMAE} further improves linear separability as beam-class complexity increases. The solution here is not simply to scale pretraining, but to shape the representation so that beam-relevant differences are easier to recover with limited labeled data.

Precoding extends this issue from beam selection to system-level control. \textit{MMIMO-Prc-FM}~\cite{emery2025precodingfm} shows that masked CSI pretraining can support downstream precoding through a frozen backbone, although the evaluation remains tied to the original data and system setting. \textit{WiFo-E}~\cite{wen2026wifoe} addresses a broader source of difficulty: end-to-end FDD precoding models trained for one configuration can fail when the number of antennas, users, pilot length, or feedback budget changes. Its sparse MoE Transformer is introduced to reduce task interference and share knowledge across heterogeneous configurations, with reported gains over per-configuration training and generalization to unseen settings. This points to a third remedy for beam-related WFMs: the architecture must account for configuration variability, not only for the channel representation itself.


\subsection{Evaluation Shift}

Beam-management evaluations reveal a tension between stronger OOD evidence and label comparability. Since the optimal beam is defined by a codebook, antenna geometry, channel representation, and deployment setup, changing these factors too aggressively can make the downstream label itself change meaning. For this reason, several works test meaningful but controlled shifts: they move beyond the exact pretraining setting, while keeping the beam task physically and label-wise comparable.

This is the role of channel-centric evaluations such as \textit{SPA-MAE}~\cite{chen2026spamae} and \textit{ComHymba}~\cite{yang2026comhymba}. These works stress the pretrained representation through unseen scenarios, cross-band settings, different SNR levels, or broader simulated channel conditions, while keeping the beam-label construction compatible with the underlying channel and codebook assumptions. Such evaluations are not weak by default: they test whether channel features remain useful when beam prediction becomes harder, without making the target label incomparable.

A stronger form of shift appears when the model must recover beam-relevant information from a different sensing context, not only from a related CSI setting. \textit{WiFo-M$^2$}~\cite{zhang2026wifom2} is the clearest example: it learns out-of-band channel-aware features by aligning multimodal sensing with CSI during pretraining, then evaluates beam prediction in unseen scenarios from SynthSoM-Twin, DeepSense-6G, and ViWi. The important point is not only that the datasets differ, but that the model can use environmental sensing as a proxy for beam-relevant channel structure when direct CSI is unavailable or incomplete. In the measured DeepSense-6G scenario, the image-based variant with a frozen pretrained backbone reaches 91.0\%, 100\%, and 100\% Top-1/3/5 beam accuracy, showing that the learned sensing-to-channel representation can remain useful under a substantially different evaluation context.

\subsection{Adaptation Strategies}

In beam management, adaptation should show whether the pretrained representation is already aligned with the beam decision. This matters because beam labels are not generic classes; they indicate which beam or precoder the PHY system should use. \textit{IQFM}~\cite{mashaal2025iqfm} illustrates this in beam prediction. With 50 labeled samples per class, LoRA reaches 52.6\% accuracy, outperforming both the supervised baseline at 42.5\% and linear probing at 35.5\%. With 500 samples per class, LoRA rises to 94.1\%, while the supervised baseline reaches 89.5\% and linear probing remains much lower at 42.7\%. This suggests that the pretrained I/Q representation contains beam-related information, but this information is not directly accessible through a simple linear head; additional adaptation is needed to align it with the codebook-dependent beam decision.

\textit{ContraWiMAE}~\cite{guler2025wimae} shows the same issue from another angle. Under linear probing, \textit{ContraWiMAE} outperforms \textit{LWM} by 42.3 percentage points, suggesting that its pretrained representation exposes beam-relevant structure more directly. However, when both representations are evaluated with a stronger ResNet-Wide downstream head, the gap narrows to 6.4 percentage points. This indicates that a more expressive downstream model can recover part of the beam/codebook mapping even from a weaker representation. For beam management, the key evidence is therefore not only final accuracy, but how much adaptation is needed before the representation becomes useful for beam selection.

\subsection{Key Insights and Open Challenges}

\textbf{Beam transfer depends on system compatibility:}
Beam selection and precoding depend directly on channel conditions, but the corresponding labels are tied to the system setup. As a result, CSI/CIR-based WFMs can transfer to beam-related tasks only when the antenna/user configuration, channel dimensions, codebook, or deployment scenario remain compatible with the pretraining setting. Environment-aware inputs can help when CSI is missing, delayed, or unreliable by adding sensing, location, or scene information about the propagation context. However, this benefit depends on having that context available at inference time and aligned with the channel samples and beam labels~\cite{zhang2026wifom2}.

\textbf{Channel similarity does not guarantee beam discrimination:}
Beam selection is not determined by channel similarity alone, but by how channel states map to a specific codebook or precoding rule. Two channels that are close in a CSI/CIR embedding may still require different beams, while different-looking channels may share the same optimal codeword. This is why CSI--CIR alignment can improve spatial or channel representation quality without necessarily producing beam-discriminative embeddings~\cite{jiang2025csiclip,jiang2026csiclippp}. Beam-oriented pretraining, broader alignment settings, or downstream adaptation are therefore needed to expose the channel differences that actually change the PHY action.

\textbf{Beam OOD is constrained by label-system dependence:}
In beam management, the beam label is tied to the codebook and system setup, so changing the antenna configuration or channel dimensions too much can change the meaning of the optimal beam. This is why many works keep the downstream beam setup compatible with pretraining, even when testing shifted data or tasks~\cite{alikhani2024lwm,guler2025wimae}. Stronger OOD evidence is still possible by learning sensing-to-channel representations that remain useful for beam management across unseen scenarios, including settings where direct CSI is unavailable or incomplete~\cite{zhang2026wifom2}.

\begin{table*}[!t]
\centering
\footnotesize
\setlength{\tabcolsep}{3.0pt}
\renewcommand{\arraystretch}{1.08}
\begin{threeparttable}
\caption{Overview of WFMs for beam management tasks.}
\label{tab:wfm_beam_overview}
\begin{tabular}{p{2.5cm} p{2.3cm} p{2cm} p{1.9cm} p{4.3cm} p{1.7cm} c}
\toprule
\textbf{Model} 
& \textbf{PT Modality/Data}
& \textbf{PT Objective} 
& \textbf{Backbone} 
& \textbf{FT Task/Data}
& \textbf{Adaptation} 
& \textbf{Eval. Shift} \\
\midrule

LWM~\cite{alikhani2024lwm}
& CSI~\cite{alkhateeb2019deepmimo}
& Masked Rec.
& Transformer
& Beam pred. -- CSI~\cite{alkhateeb2019deepmimo}
& Frozen FT
& Partial \\

IQFM~\cite{mashaal2025iqfm}
& IQ$^{*}$
& Contrastive
& CNN
& Beam pred. -- IQ/beam~\cite{klautau2021mimodata}
& \makecell[l]{Frozen FT\\PEFT}
& OOD \\

MUSE-FM~\cite{zheng2025musefm}
& \makecell[l]{Scene$^{*}$\\W. signal$^{*}$}
& Task-driven Supervision
& Transformer
& MU prec. -- Scene$^{*}$ + W. signal$^{*}$
& \makecell[l]{MT-FT\\Zero-shot}
& ID$^{\dagger}$ \\

CSI-CLIP~\cite{jiang2025csiclip}
& CSI~\cite{alkhateeb2019deepmimo}
& Contrastive
& Dual-branch
& Beam mgmt. -- CSI~\cite{alkhateeb2019deepmimo}
& Frozen FT
& Partial \\

ICWLM~\cite{wen2025icwlm}
& CSI~\cite{jaeckel2014quadriga}
& Task-drive Supervision
& Transformer
& Precoding -- CSI~\cite{jaeckel2014quadriga}
& Zero-shot
& Partial \\

ContraWiMAE~\cite{guler2025wimae}
& CSI~\cite{alkhateeb2019deepmimo}
& Hybrid
& Transformer
& Beam mgmt. -- CSI~\cite{alkhateeb2019deepmimo}
& Frozen FT
& Partial \\

MMIMO-Prc-FM~\cite{emery2025precodingfm}
& CSI$^{*}$
& Masked Rec.
& Transformer
& Precoding -- CSI$^{*}$
& Frozen FT
& ID \\



WiFo-M$^2$~\cite{zhang2026wifom2}
& \makecell[l]{Sensing~\cite{m3sc,synthsom}\\CSI~\cite{m3sc,synthsom}}
& Hybrid
& Transformer
& Beam mgmt. -- Sensing~\cite{synthsomtwin,deepsense6g,viwi} + CSI~\cite{synthsomtwin,deepsense6g,viwi}
& Frozen FT
& OOD \\

WiFo-E~\cite{wen2026wifoe}
& CSI~\cite{jaeckel2014quadriga}
& Masked Rec.
& Transformer
& Precoding -- CSI~\cite{jaeckel2014quadriga}
& Frozen FT
& Partial \\

MMFM4WCS~\cite{yazdnian2026multimodalfm}
& \makecell[l]{Map$^{*}$\\CSI$^{*}$}
& Masked Rec.
& Transformer
& Precoding -- Map$^{*}$ + CSI$^{*}$
& Frozen FT
& ID \\

WiMamba~\cite{raviv2026wimamba}
& CSI~\cite{alkhateeb2019deepmimo}
& Masked Rec.
& SSM (Mamba)
& Beam mgmt. -- CSI~\cite{alkhateeb2019deepmimo}
& Frozen FT
& Partial \\

WiFo-MiSAC~\cite{liu2026wifomisac}
& \makecell[l]{Sensing~\cite{m3sc,synthsom}\\CSI~\cite{m3sc,synthsom}}
& Hybrid
& MoE Transformer
& Beam pred. -- Sen. + CSI~\cite{m3sc,synthsom}
& Zero-shot
& OOD \\

SPA-MAE~\cite{chen2026spamae}
& CSI~\cite{alkhateeb2019deepmimo}
& Masked Rec.
& Transformer
& Beam pred. -- CSI~\cite{alkhateeb2019deepmimo}
& Frozen FT
& Partial \\

ComHymba~\cite{yang2026comhymba}
& CSI~\cite{jaeckel2014quadriga}
& Masked Rec.
& SSM--Transformer
& Beam mgmt. -- CSI~\cite{jaeckel2014quadriga}
& Frozen FT
& Partial \\

LatentWave~\cite{mohamed2026latentwave}
& \makecell[l]{Spectrogram$^{*}$\\CSI~\cite{pan2022cfrcsi}}
& \makecell[l]{Masked Latent\\Prediction}
& Transformer
& Beam pred. -- CSI~\cite{alkhateeb2019deepmimo}
& Frozen FT
& Partial \\

CSI-CLIP++~\cite{jiang2026csiclippp}
& \makecell[l]{CSI~\cite{alkhateeb2019deepmimo}\\CIR~\cite{alkhateeb2019deepmimo}}
& Contrastive
& \makecell[l]{Dual-branch\\Transformer}
& Beam pred. -- CSI~\cite{alkhateeb2019deepmimo}
& Frozen FT
& OOD \\

\bottomrule
\end{tabular}

\begin{tablenotes}[flushleft]
\footnotesize
\item \textbf{Abbreviations:} Abbreviations and shorthand terms used in this table are defined in Table~\ref{tab:abbreviations}.
\item \textbf{Data notation:} $^{*}$ indicates private, author-collected, author-generated, or not separately referenced data.
\item \textbf{Eval. Shift:} ID = same source as PT; Partial = related but distinct source; OOD = clearly different source/distribution.
\end{tablenotes}
\end{threeparttable}
\end{table*}

\section{Spectrum Sensing and Monitoring}
\label{sec:analysis_spectrum}

\textbf{Overview.} Spectrum sensing and monitoring WFMs form a smaller, task-specific group focused on spectrum occupancy, temporal dynamics, and radio-environment structure, as summarized in Table~\ref{tab:wfm_ssm_overview}. Spectrograms appear in 3 of the 6 models, mainly for forecasting and segmentation, while others use I/Q, PSD/spectrum maps, or radio maps. Masked reconstruction is the dominant pretraining objective (4 of 6 models), while \textit{SpectrumFM} and \textit{FARM} use hybrid objectives incorporating temporal prediction or generative modeling. Architectures remain diverse, spanning RNN, ViT, CNN--Transformer, geometry-aware Transformer, and diffusion Transformer designs. Partial-shift evaluation is the most common pattern: 3 models report partial-only shift, 1 combines ID and partial evaluation, and 2 report OOD settings.

\subsection{Spectrum Views, Pretraining, and Architectures}

Spectrum WFMs do not learn a single common notion of spectrum monitoring. Their learned behavior depends strongly on the input representation: spectrogram-based models focus on time--frequency occupancy, I/Q-based models stay closer to signal-level dynamics, and spectrum-map models represent spatial radio-environment structure.
\textit{SSRadio}~\cite{aboulfotouh2024ssradio} and \textit{6G-RadioFM}~\cite{aboulfotouh2024radio6gfm} represent spectrum activity as time--frequency occupancy patterns, which naturally supports forecasting and segmentation but keeps the learned behavior tied to a spectrogram-based view of spectrum monitoring. 

A different formulation appears when the objective is not only to recover occupancy structure, but also to model temporal spectrum dynamics and anomalous activity from received samples. \textit{SpectrumFM}~\cite{zhou2025spectrumfm} uses raw I/Q data for spectrum sensing and anomaly detection, but its contribution is better understood through the coupling between representation, architecture, and objective. Its CNN--Transformer architecture follows this choice: the CNN captures local signal patterns, while self-attention captures longer-range dependencies. Its hybrid objective, combining masked reconstruction with next-slot prediction, further encourages the model to learn both missing signal content and temporal spectrum dynamics. The reported results support this design, including an AUC of 0.97 for spectrum sensing at $-4$~dB SNR 
and an anomaly-detection AUC gain of more than 10\% over the adversarial autoencoder baseline~\cite{zhou2025spectrumfm}.

A third formulation treats spectrum monitoring as spatial radio-environment estimation, where the target is a spatial--temporal--spectral field rather than an occupancy mask or anomaly label. \textit{FM-RME}~\cite{yang2026fmrme} uses PSD and spectrum maps with a geometry-aware module that preserves propagation patterns across position and orientation, together with masked recovery of space--time--frequency structure. The pretrained model also transfers without fine-tuning to an unseen radio-map dataset with different temporal and spectral parameters, outperforming the compared baselines.

\subsection{Adaptation Strategies}

Adaptation results in spectrum WFMs reveal a recurring gap between learning spectrum activity and learning spectrum semantics. In the spectrogram view, \textit{SSRadio}~\cite{aboulfotouh2024ssradio} freezes the 5-layer ConvLSTM backbone learned with masked spectrogram modeling and fine-tunes only the task head for forecasting and segmentation. This leads to faster convergence, but the tuned FM still falls slightly below the from-scratch baseline in forecasting. For segmentation, the authors report stronger difficulty for the NR class and attribute it to distribution differences between the pretraining and segmentation datasets, as well as the mismatch between regression-based pretraining and classification-based segmentation. This suggests that masked spectrogram features can capture time--frequency activity, but may not be sufficiently class-discriminative for technology-level segmentation.

A clearer version of the same issue appears in I/Q-based interference tasks. \textit{Multimodal WFM}~\cite{aboulfotouh2025multimodal} shows that interference detection is already strong with linear probing, reaching 96.40\%, and improves only modestly with LoRA, reaching 99.60\%. Interference classification is more diagnostic: performance increases from 58.25\% with linear probing to 66.93\% with LoRA. This indicates that the pretrained representation captures the presence of interference more readily than the structure needed to separate interference types, so adaptation is most valuable when the downstream task requires finer decision boundaries.

Spectrum-map modeling tests a different form of reuse: whether propagation structure can transfer without task-specific tuning. \textit{FM-RME}~\cite{yang2026fmrme} applies radio-map estimation to an unseen dataset with different temporal and spectral parameters, without fine-tuning, and reports stronger performance than the compared baselines. This suggests that geometry-aware pretraining and spatial--temporal--spectral masked recovery can support zero-shot transfer when the representation captures reusable spectrum-propagation structure.

\subsection{Evaluation Shift}

Spectrum sensing and monitoring WFMs still provide limited OOD evidence. Most works evaluate transfer within a single spectrum view instead of across broad spectrum conditions. For instance, \textit{SSRadio}~\cite{aboulfotouh2024ssradio} and \textit{6G-RadioFM}~\cite{aboulfotouh2024radio6gfm} remain within spectrogram-based forecasting or segmentation, while \textit{SpectrumFM}~\cite{zhou2025spectrumfm} evaluates I/Q-based sensing and anomaly detection. These results are useful, but they mainly show controlled transfer inside the representation view used by each model.

A different form of partial shift appears in \textit{FM-RME}~\cite{yang2026fmrme}, where the shift concerns radio-map estimation rather than interference recognition. The model is evaluated on an unseen dataset with different temporal and spectral parameters, testing whether learned spectrum-propagation structure transfers across radio-environment settings. This evidence is valuable, but it is not directly comparable with I/Q interference detection or spectrogram segmentation because the output itself is a spatial--temporal--spectral map.

Stronger OOD evidence appears when the downstream setting changes the data source, interference condition, or spectrum view more clearly. \textit{Multimodal WFM}~\cite{aboulfotouh2025multimodal} is pretrained on mixed wireless modalities but evaluated on over-the-air LTE I/Q recordings with DSSS interference for interference detection and classification. This tests whether a representation learned from heterogeneous wireless inputs remains useful under a shifted I/Q interference setting. The results indicate that detection transfers more easily than classification, suggesting that OOD spectrum evaluation should distinguish simple interference presence from the harder problem of identifying interference types.


\subsection{Key Insights and Open Challenges}

\textbf{Geometry-aware design can support zero-shot spatial transfer:}
Spectrum-map and radio-map tasks are strongly shaped by spatial propagation patterns, making geometry-aware design particularly relevant for this family of WFMs. Geometry-aware feature extraction has supported zero-shot radio-map estimation across changed radio-environment settings~\cite{yang2026fmrme}, suggesting that spatial propagation relationships learned during pretraining can remain useful beyond the original map configuration.


\textbf{Not all spectrum tasks test the same transfer difficulty:}
In spectrum WFMs, strong transfer on one downstream task should not be interpreted as evidence of general spectrum reuse. Detecting spectral activity or interference can be easier than distinguishing interference types, where additional task-specific adaptation provides clearer benefits~\cite{aboulfotouh2025multimodal}. Segmentation introduces another level of difficulty because the model must localize structured time--frequency activity under distribution shift, making frozen spectrogram transfer less reliable~\cite{aboulfotouh2024ssradio}. This suggests that spectrum evaluation should separate simple activity detection from tasks that require richer time--frequency or interference-level interpretation.

\textbf{OOD evidence remains limited for spectrum WFMs:}
Current spectrum WFMs show useful controlled transfer, but most evaluations remain close to the spectrum view used by each model. The clearest OOD case appears when a model pretrained with mixed wireless modalities is evaluated on real over-the-air I/Q recordings with DSSS interference~\cite{aboulfotouh2025multimodal}, but this remains an isolated example. Systematic OOD evaluation should vary emitter types, frequency bands, and interference patterns independently to identify what spectrum structure actually transfers.

\begin{table*}[t]
\centering
\footnotesize
\setlength{\tabcolsep}{3.0pt}
\renewcommand{\arraystretch}{1.08}
\begin{threeparttable}
\caption{Overview of WFMs for spectrum sensing and monitoring tasks.}
\label{tab:wfm_ssm_overview}
\begin{tabular}{p{2.5cm} p{2.3cm} p{2cm} p{1.9cm} p{4.3cm} p{1.7cm} c}
\toprule
\textbf{Model} 
& \textbf{PT Modality/Data}
& \textbf{PT Objective} 
& \textbf{Backbone} 
& \textbf{FT Task/Data}
& \textbf{Adaptation} 
& \textbf{Eval. Shift} \\
\midrule

SSRadio~\cite{aboulfotouh2024ssradio}
& Spectrogram$^{*}$
& Masked Rec.
& RNN
& \makecell[l]{Spectrum forecast. -- Spectrogram$^{*}$\\Spectrum seg. -- Spectrogram$^{*}$}
& Frozen FT
& ID + Partial \\

6G-RadioFM~\cite{aboulfotouh2024radio6gfm}
& Spectrogram$^{*}$
& Masked Rec.
& ViT
& Spectrum seg. -- Spectrogram$^{*}$
& Frozen FT
& Partial \\

Multimodal WFM~\cite{aboulfotouh2025multimodal}
& \makecell[l]{Spectrogram$^{*}$\\IQ$^{*}$}
& Masked Rec.
& ViT
& Interference det./cls. -- IQ~\cite{roy2023icarus}
& \makecell[l]{Frozen FT\\Partial FT\\PEFT}
& OOD \\

SpectrumFM~\cite{zhou2025spectrumfm}
& \makecell[l]{IQ~\cite{deepsig2018radioml,fontaine2019lowcomplexity}\\IQ$^{*}$}
& Hybrid
& CNN-Transformer
& \makecell[l]{Spectrum sensing -- IQ~\cite{oshea2016radioml}\\Anomaly det. -- IQ$^{*}$}
& \makecell[l]{PEFT\\Frozen FT}
& Partial \\

FM-RME~\cite{yang2026fmrme}
& \makecell[l]{PSD map$^{*}$\\Spectrum map$^{*}$}
& Masked Rec.
& Geometry-aware Transformer
& \makecell[l]{Radio map est. -- PSD map$^{*}$ +\\Spectrum map$^{*}$}
& Zero-shot
& Partial \\

FARM~\cite{gao2026farm}
& \makecell[l]{RSS map$^{*}$\\Aerial radio map$^{*}$}
& Hybrid
& \makecell[l]{Diffusion\\Transformer}
& Radio map est. -- Aerial radio map$^{*}$
& Zero-shot
& OOD \\

\bottomrule
\end{tabular}

\begin{tablenotes}[flushleft]
\footnotesize
\item \textbf{Abbreviations:} Abbreviations and shorthand terms used in this table are defined in Table~\ref{tab:abbreviations}.
\item \textbf{Eval. Shift} indicates whether the downstream evaluation uses data from the same source as PT (ID), a closely related but distinct source (Partial), or a clearly different source/distribution (OOD). Combined labels indicate multiple downstream settings.
\end{tablenotes}
\end{threeparttable}
\end{table*}

\section{Multi-Task Physical-Layer WFMs}
\label{sec:multitask_phy_wfms}

The cross-category PHY group includes models whose evaluation or design spans multiple physical-layer task families. This group is important because it tests a stronger version of the WFM idea: whether a pretrained wireless representation can support different PHY functions, rather than only improve one task under one modality or dataset. However, not all cross-category evidence has the same meaning. Some models mainly demonstrate broad downstream evaluation, while others explicitly design a shared backbone, task-conditioned interface, or physical-context mechanism for multi-task reuse.

A first group consists of models that use multiple downstream tasks mainly as a test of representation breadth. \textit{IQFM}~\cite{mashaal2025iqfm}, \textit{SSRadioRep}~\cite{kanu2025ssradiorep}, and \textit{WirelessJEPA}~\cite{chu2026wirelessjepa} evaluate whether a pretrained encoder can support different signal, sensing, or beam-related tasks with limited adaptation. These works are useful because they expose the representation to more than one downstream function. However, they should be interpreted as \emph{cross-task evaluation} rather than full multi-task WFM design, since the main contribution remains the learned encoder and its adaptation behavior, not a unified task interface.

A second group more directly targets multi-task reuse through shared architectures or task-conditioned adaptation. \textit{6G WavesFM}~\cite{aboulfotouh20256gwavesfm} is a clear example because it shares a pretrained backbone across communication, sensing, localization, and signal-classification tasks, using task-specific heads and parameter-efficient adaptation. \textit{MUSE-FM}~\cite{zheng2025musefm} follows a different route by conditioning the model on task descriptions and scenario information, while \textit{ICWLM}~\cite{wen2025icwlm} uses in-context input--output examples to adapt the same model to different PHY mappings. These designs move closer to the foundation-model objective because they address not only representation reuse, but also how heterogeneous PHY tasks are specified to a common model.

A third group uses physical, environmental, or temporal context to connect tasks that would otherwise appear separate. \textit{MMFM4WCS}~\cite{yazdnian2026multimodalfm} combines CSI, scene geometry, and user-location information to learn representations that transfer across communication and spatial tasks. \textit{WirelessGPT}~\cite{yang2025wirelessgpt} and \textit{Multi-Task Prediction FM}~\cite{sheng2025multitaskfm} also point toward broader reuse by modeling multiple wireless variables or prediction tasks within a common framework. The value of these approaches is that they incorporate system context or temporal structure, but their generality still depends on whether the learned representation transfers beyond related configurations, scenarios, or task formats.

Overall, task count alone is a weak signal of foundation-model generality. What is still missing across all three groups is controlled evidence isolating the source of transfer: none of the surveyed works report ablations that separate gains due to the shared representation from gains due to task-specific conditioning or context injection. Establishing this distinction is arguably the key open question for validating cross-category PHY WFMs.

\section{Future Directions and Recommendations}
\label{sec:open_directions}


Based on the comparative analysis conducted in this survey, this section identifies future research directions for the development and evaluation of WFMs and gives some recommendations.



\subsection{OOD Evaluation for WFM Robustness}

OOD evaluation should become a central component of WFM assessment. Across the surveyed literature, strong OOD evidence remains limited: many works show useful transfer on familiar data sources or partial shifts, while fewer evaluate downstream settings that clearly depart from the pretraining conditions. This distinction is especially important in wireless, where a change may stress the model but still remain close to the original setting. Effective OOD evaluation should separate downstream data from pretraining data along a major wireless factor, such as the signal source or protocol; the propagation environment or physical layout; the antenna, hardware, or beam system; or the interference and spectrum distribution. This would clarify whether a pretrained representation remains reusable beyond nearby conditions, which is central to assessing WFM robustness.

\subsection{Predictive Pretraining Objectives}

Predictive pretraining is emerging as a promising direction for WFMs, as it can guide the encoder toward wireless-specific structure. In the JEPA paradigm, the masking pattern can act as an inductive bias: antenna, temporal, frequency, and spatial masks emphasize different wireless dependencies~\cite{chu2026wirelessjepa,mohamed2026latentwave}. Future work should move beyond manually design masking rules toward adaptive target selection, where the model learns which regions provide the most informative prediction targets for each wireless observation, potentially combining multiple geometries instead of relying on a fixed masking strategy.

Temporal prediction provides a complementary direction for learning wireless dynamics. Current approaches can forecast variables such as channel conditions, angle, or traffic over time~\cite{zhou2025spectrumfm}. However, these variables are often predicted separately, even though they are physically related. For example, user movement can change the channel and, consequently, the beam that should be selected. Future predictive WFMs should therefore learn not only how each variable evolves, but also how changes in one wireless variable affect the others.




\subsection{Wireless-Specific Architectures}

Most WFMs still rely on Transformer or ViT-based backbones because they are flexible and easy to adapt across tokenized wireless inputs. However, recent models suggest that wireless data may benefit from more specialized architectures. For example, state-space models have been explored to improve efficiency in long wireless sequences~\cite{raviv2026wimamba}, while MoE-based designs address heterogeneity across feedback or precoding configurations~\cite{liu2025wifocf,wen2026wifoe}. Future architectures should therefore be evaluated not only by accuracy, but also by whether their structural bias matches wireless-specific constraints such as temporal structure, spatial organization, and configuration variability.





\subsection{Efficient and Deployable WFMs}

A WFM can show strong downstream accuracy and still be impractical for wireless deployment. Many PHY tasks require fast inference, limited memory, and low-cost adaptation, especially at the edge or under real-time constraints. Future evaluations should therefore report deployment-related costs together with downstream performance, including latency, memory usage, trainable parameters, and adaptation time. This would clarify whether a WFM is feasible beyond offline evaluation.

\subsection{Multimodal, Environment-Aware, and Attributable Transfer}

Multimodal and environment-aware WFMs are important because many PHY tasks depend on the propagation context behind the received signal. Models such as \textit{MUSE-FM} move toward task- and context-aware transfer~\cite{zheng2025musefm}, while \textit{FM-RME} shows how geometry-aware inputs can support radio-map estimation~\cite{yang2026fmrme}. However, adding maps, scenes, sensing data, or physical context also makes transfer harder to attribute: gains may come from reusable RF representations, added context, or the adaptation protocol. Future studies should include modality-controlled ablations, frozen-probing baselines, and fixed adaptation budgets. This would clarify whether multimodal WFMs improve the reusable representation itself or mainly the complete downstream pipeline.



\subsection{RRM WFMs}

Most reviewed WFMs remain centered on physical-layer representation learning. A natural next step is to extend these models toward radio resource management (RRM), where the objective is not only to estimate or classify wireless states, but to support decisions such as scheduling, power control, handover, spectrum sharing, beam allocation, and interference coordination. This transition is challenging because RRM depends on traffic demand, mobility, fairness, latency, energy constraints, and multi-cell interactions. Future RRM WFMs should therefore combine wireless representation learning with decision-oriented evaluation and closed-loop simulation.

\subsection{Wireless World Models}

An emerging research direction is the development of Wireless World Models (WWMs). The motivation comes from a limitation observed across current WFMs: many models learn useful wireless representations, but their transfer is often evaluated under controlled changes. In real deployments, however, wireless conditions change because users move, links become blocked, interference appears, spectrum occupancy varies, and system configurations are updated. WWMs aim to address this limitation by learning how wireless conditions change over time. Instead of focusing only on the current observation, a WWM would use temporal and physical context to capture how channels, interference, spectrum activity, or beam quality evolve. This type of representation could be reused in tasks where future or changing conditions matter, such as channel prediction, beam adaptation, spectrum monitoring, and radio resource management.


\section{Conclusion}
\label{sec:conclusion}

This survey systematically analyzed wireless foundation models (WFMs) for physical-layer applications, focusing on how wireless representations are pretrained, adapted, and evaluated across major PHY task families. By comparing pretraining objectives, backbone architectures, adaptation strategies, downstream tasks, and evaluation shifts, we identified the main design patterns and limitations of the rapidly evolving WFM landscape.

Our analysis shows growing evidence that representations learned from wireless data can be reused across tasks, datasets, and operating conditions, but the strength of this evidence varies substantially across task families. Signal recognition, demodulation, channel representation learning, RF sensing, beam management, and spectrum monitoring test different forms of transfer, making performance gains and OOD results difficult to compare directly. Moreover, many studies simultaneously vary the pretraining data, modality, objective, architecture, and adaptation strategy, making it difficult to isolate the source of transfer gains. Broader downstream evaluation alone is therefore insufficient to establish strong foundation-model behavior; evaluations must also determine what knowledge transfers, under which distribution shifts, and with how much task-specific adaptation.

Progress toward more general WFMs will require broader wireless data, controlled and comparable evaluation protocols, and clearer attribution of gains to the learned representation itself. Wireless-specific inductive biases, multimodal and physics-aware representations, efficient adaptation, and deployment constraints remain important directions, but should be assessed under benchmarks that systematically test reuse across meaningfully different wireless tasks and operating conditions. Current results are promising, yet stronger evidence is still needed before a shared wireless representation can be considered broadly reusable across the diversity of environments, devices, and system configurations expected in future wireless networks.

\balance 
\bibliographystyle{IEEEtran}
\bibliography{references}
\end{document}